\documentclass[english]{article}
\usepackage[T1]{fontenc}
\usepackage[latin9]{inputenc}
\usepackage{geometry}
\usepackage{babel}
\usepackage{booktabs}
\usepackage{amsmath}
\usepackage{amssymb}
\usepackage{graphicx}
\usepackage{rotating}
\usepackage{setspace}
\usepackage{wasysym}
\PassOptionsToPackage{normalem}{ulem}
\usepackage{ulem}
\usepackage[unicode=true,
 bookmarks=true,bookmarksnumbered=false,bookmarksopen=false,
 breaklinks=false,pdfborder={0 0 0},pdfborderstyle={},backref=false,colorlinks=false]
 {hyperref}
\hypersetup{pdftitle={Quasi-Experimental Shift-Share Designs},
 pdfauthor={Kirill Borusyak, Peter Hull, and Xavier Jaravel}}

\makeatletter

\providecommand{\tabularnewline}{\\}

\newenvironment{lyxlist}[1]
	{\begin{list}{}
		{\settowidth{\labelwidth}{#1}
		 \setlength{\leftmargin}{\labelwidth}
		 \addtolength{\leftmargin}{\labelsep}
		 }}
	{\end{list}}
\ifx\proof\undefined
\newenvironment{proof}[1][\protect\proofname]{\par
	\normalfont\topsep6\p@\@plus6\p@\relax
	\trivlist
	\itemindent\parindent
	\item[\hskip\labelsep\scshape #1]\ignorespaces
}{%
	\endtrivlist\@endpefalse
}
\providecommand{\proofname}{Proof}
\fi

\usepackage{changepage}
\usepackage{xcolor}
\hypersetup{
    colorlinks,
    linkcolor={black},
    citecolor={black},
    urlcolor={black}
}

\usepackage{minitoc}
\usepackage[authordate, backend=biber, uniquename=false, uniquelist=false, noibid, maxcitenames=2, mincitenames=1, footmarkoff]{biblatex-chicago}

\DeclareFieldFormat{citehyperref}{%
  \DeclareFieldAlias{bibhyperref}{noformat}
  \bibhyperref{#1}}

\DeclareFieldFormat{textcitehyperref}{%
  \DeclareFieldAlias{bibhyperref}{noformat}
  \bibhyperref{%
    #1%
    \ifbool{cbx:parens}
      {\bibcloseparen\global\boolfalse{cbx:parens}}
      {}}}

\savebibmacro{cite}
\savebibmacro{textcite}

\renewbibmacro*{cite}{%
  \printtext[citehyperref]{%
    \restorebibmacro{cite}%
    \usebibmacro{cite}}}

\renewbibmacro*{textcite}{%
  \ifboolexpr{
    ( not test {\iffieldundef{prenote}} and
      test {\ifnumequal{\value{citecount}}{1}} )
    or
    ( not test {\iffieldundef{postnote}} and
      test {\ifnumequal{\value{citecount}}{\value{citetotal}}} )
  }
    {\DeclareFieldAlias{textcitehyperref}{noformat}}
    {}%
  \printtext[textcitehyperref]{%
    \restorebibmacro{textcite}%
    \usebibmacro{textcite}}}

\AtBeginBibliography{

}

\AtEveryBibitem{%
  \clearfield{issn} 
  \clearfield{doi} 
  \clearfield{isbn} 
  \clearfield{eprint} 
  \clearfield{number} 

  \ifentrytype{online}{}{
    \clearfield{url}
  }
}

\usepackage{pdflscape}

\makeatother

\usepackage[style=chicago-authordate]{biblatex}
\begin{document}
\title{Quasi-Experimental Shift-Share Research Designs}
\author{\vspace{1.5cm}
}
\author{Kirill Borusyak\\
UCL and CEPR\and Peter Hull\\
U Chicago and NBER\and Xavier Jaravel~~\\
LSE and CEPR\thanks{Contact: k.borusyak@ucl.ac.uk, hull@uchicago.edu, and x.jaravel@lse.ac.uk. We are grateful to Rodrigo Ad\~{a}o, Joshua Angrist, David Autor, Moya Chin, Andy Garin, Ed Glaeser, Paul Goldsmith-Pinkham, Larry Katz, Michal Koles\'{a}r, Gabriel Kreindler, Jack Liebersohn, Eduardo Morales, Jack Mountjoy, J\"{o}rn-Steffen Pischke, Brendan Price, Isaac Sorkin, Jann Spiess, Itzchak Tzachi Raz, various seminar participants, and five anonymous referees for helpful comments. We thank David Autor, David Dorn, and Gordon Hanson, as well as Paul Goldsmith-Pinkham, Isaac Sorkin, and Henry Swift, for providing replication code and data.}}
\date{\vspace{0.25cm}
December 2020}

\maketitle
\vspace{0.25cm}

\begin{abstract}
\begin{singlespace} \noindent\begin{adjustwidth*}{0.75cm}{0.75cm} {\normalsize Many
studies use shift-share (or ``Bartik'') instruments, which average
a set of shocks with exposure share weights. We provide a new econometric
framework for shift-share instrumental variable (SSIV) regressions
in which identification follows from the quasi-random assignment of
shocks, while exposure shares are allowed to be endogenous. The framework
is motivated by an equivalence result: the orthogonality between a
shift-share instrument and an unobserved residual can be represented
as the orthogonality between the underlying shocks and a shock-level
unobservable. SSIV regression coefficients can similarly be obtained
from an equivalent shock-level regression, motivating shock-level
conditions for their consistency. We discuss and illustrate several
practical insights of this framework in the setting of \textcite{AutorDorn2001},
estimating the effect of Chinese import competition on manufacturing
employment across U.S. commuting zones.}\end{adjustwidth*} \end{singlespace} \vfill{}
\thispagestyle{empty}\newpage
\global\long\def\expec#1{\mathbb{E}\left[#1\right]}%
\global\long\def\var#1{\mathrm{Var}\left[#1\right]}%
\global\long\def\cov#1{\mathrm{Cov}\left[#1\right]}%
\global\long\def\one{\mathbf{1}}%
\global\long\def\diag{\operatorname{diag}}%
\global\long\def\tr{\operatorname{tr}}%
\global\long\def\plim{\operatorname*{plim}}%
\setcounter{page}{1}\dosecttoc\addtocontents{toc}{\protect\setcounter{tocdepth}{0}}
\end{abstract}

\section{Introduction}

A large and growing number of empirical studies use shift-share instruments:
weighted averages of a common set of shocks, with weights reflecting
heterogeneous shock exposure. In many settings, such as those of \textcite{Bartik1991},
\textcite{Blanchard1992} and \textcite{AutorDorn2001}, a regional
instrument is constructed from shocks to industries with local industry
employment shares measuring the shock exposure. In other settings,
researchers may combine shocks across countries, socio-demographic
groups, or foreign markets to instrument for treatments at the regional,
individual, or firm level.\footnote{Observations in shift-share designs may, for example, represent regions
impacted by immigration shocks from different countries \parencite{Card2001,Peri2016},
firms differentially exposed to foreign market shocks \parencite{Hummels2014,Berman2011},
product groups purchased by different types of consumers \parencite{Jaravel2017},
groups of individuals facing different national income trends \parencite{Boustan2013},
or countries differentially exposed to the U.S. food aid supply shocks
\parencite{nunnqian14}. We present a taxonomy of existing shift-share
designs, and how they relate to our framework, in Section \ref{subsec:Taxonomy}.}

The claim for instrument validity in shift-share instrumental variable
(SSIV) regressions must rely on some assumptions about the shocks,
exposure shares, or both. This paper develops a novel framework for
understanding such regressions as leveraging exogenous variation in
the shocks, allowing the variation in exposure shares to be endogenous.
Our approach is motivated by an equivalence result: the orthogonality
between a shift-share instrument and an unobserved residual can be
represented as the orthogonality between the underlying shocks and
a shock-level unobservable. Given a first stage, it follows that the
instrument identifies a parameter of interest if and only if the shocks
are uncorrelated with this unobservable, which captures the average
unobserved determinants of the original outcome among observations
most exposed to a given shock. SSIV regression coefficients can similarly
be obtained from an equivalent IV regression estimated at the level
of shocks. In this regression the outcome and treatment variables
are first averaged, using exposure shares as weights, to obtain shock-level
aggregates. The shocks then directly instrument for the aggregated
treatment. Importantly, these equivalence results only rely on the
structure of the shift-share instrument and thus apply to outcomes
and treatments that are not typically computed at the level of shocks.

We use these equivalence results to derive two conditions sufficient
for SSIV consistency. First, we assume shocks are as-good-as-randomly
assigned as if arising from a natural experiment. This is enough for
the shift-share instrument to be valid: i.e. for the shocks to be
uncorrelated with the relevant unobservables in expectation. Second,
we assume that a shock-level law of large numbers applies\textemdash that
the instrument incorporates many sufficiently independent shocks,
each with sufficiently small average exposure. Instrument relevance
further holds when individual units are mostly exposed to only a small
number of shocks, provided those shocks affect treatment. Our two
quasi-experimental conditions are similar to ones imposed in other
settings where the underlying shocks are directly used as instruments,
bringing SSIV to familiar econometric territory.\footnote{For example, \textcite{AADHP2016} study the impact of import competition
from China on U.S. industry employment using industry (i.e. shock-level)
regressions with shocks constructed similarly to those underlying
the regional shift-share instrument used in \textcite{AutorDorn2001}.
Our framework shows that both studies can rely on similar econometric
assumptions, though the economic interpretations of the estimates
differ.} 

We extend our quasi-experimental approach to settings where shocks
are as-good-as-randomly assigned only conditionally on shock-level
observables, to SSIVs with exposure shares that do not add up to a
constant for each observation, and to panel data. For conditional
random assignment, we show that quasi-experimental shock variation
can be isolated with regression controls that have a shift-share structure.
Namely, it is enough to control for an exposure-weighted sum of the
relevant shock-level confounders. Relatedly, in SSIVs with ``incomplete
shares,'' where the sum of exposure shares varies across observations,
it is important to control for the sum of exposure shares as the exposure-weighted
sum of a constant. In panel data, we show that the SSIV estimator
can be consistent both with many shocks per period and with many periods.
We also show that unit fixed effects only isolate variation in shocks
over time when exposure shares are time-invariant. In other extensions
we show how SSIV with multiple endogenous variables can be viewed
quasi-experimentally and how multiple sets of quasi-random shocks
can be combined with new overidentified shock-level IV procedures.

Our framework also bears practical tools for SSIV inference and testing.
\textcite{Adao} show that conventional standard errors in SSIV regressions
may be invalid because observations which similar exposure shares
are likely to have correlated residuals. They are also the first to
propose a solution to this inference problem in a framework based
on ours, with identifying variation in shocks. We present a convenient
alternative based on our equivalence result: estimating SSIV coefficients
at the level of identifying variation (shocks) can yield asymptotically
valid standard errors. The validity of this solution requires an additional
assumption on the structure of the included controls (producing standard
errors that are conservative otherwise). However, it offers several
practical features: it can be implemented with standard statistical
software, extended to various forms of shock dependence (e.g. autocorrelation),
and computed in some settings where the estimator of \textcite{Adao}
fails (e.g. when there are more shocks than observations). Appropriate
measures of first-stage relevance and valid falsification tests of
shock exogeneity can also be obtained with conventional shock-level
procedures. Monte-Carlo simulations confirm the accuracy of our asymptotic
approximations in moderately-sized samples of shocks, and that the
finite-sample properties of SSIV are similar to those of conventional
shock-level IV regressions which use the same shocks as instruments.

We illustrate the practical insights from our framework in the setting
of \textcite{AutorDorn2001}, who estimate the effect of increased
Chinese import penetration on manufacturing employment across U.S.
commuting zones. We find supporting evidence for the interpretation
of their SSIV as leveraging quasi-random variation in industry-specific
Chinese import shocks. This application uses a new Stata package,
\emph{ssaggregate}, which we have developed to help practitioners
implement the appropriate shock-level analyses.\footnote{\label{fn:ssagg}This Stata package creates the shock-level aggregates
used in the equivalent regression. Users can install this package
with the command \emph{ssc install ssaggregate.} See the associated
help file and this paper's replication archive at \href{https://github.com/borusyak/shift-share}{https://github.com/borusyak/shift-share}
for more details.}

Our quasi-experimental approach is not the only framework for SSIV
identification and consistency. In related work, \textcite{GPSS}
formalize a different approach based on the exogeneity of the exposure
shares, imposing no explicit assumption of shock exogeneity. This
framework is motivated by a different equivalence result: the SSIV
coefficient also coincides with a generalized method of moments estimator,
with exposure shares as multiple excluded instruments. Though exposure
exogeneity is a sufficient condition for SSIV identification (and,
as such, implies our shock-level orthogonality condition), we focus
on plausible conditions under which it is not necessary.

We delineate two cases where identification via exogenous shocks is
attractive. In the first case, the shift-share instrument is based
on a set of shocks which can itself be thought of as an instrument.
Consider the \citeauthor{AutorDorn2001} (2013; hereafter ADH) shift-share
instrument, which combines industry-specific changes in Chinese import
competition (the shocks) with local exposure given by the lagged industrial
composition of U.S. regions (the exposure shares). In such a setting,
exogeneity of industry employment shares is difficult to justify \emph{a
priori} since unobserved industry shocks (e.g., automation or innovation
trends) are likely to affect regional outcomes through the same mixture
of exposure shares. Our approach, in contrast, allows researchers
to specify a set of shocks that are plausibly uncorrelated with such
unobserved factors. Consistent with this general principle, ADH attempt
to purge their industry shocks from U.S.-specific confounders by measuring
Chinese import growth outside of the United States. Similarly, \textcite{Hummels2014}
combine country-by-product changes in transportation costs to Denmark
(as shocks) with lagged firm-specific composition of intermediate
inputs and their sources (as shares). They argue these shocks are
``idiosyncratic,'' which our approach formalizes as ``independent
from relevant country-by-product unobservables.'' Other recent examples
of where our approach may naturally apply are found, for example,
in finance \parencite{Xu2018}, the immigration literature \parencite{Peri2016},
and studies of innovation \parencite{Stuen2012}.

In the second case, a researcher can think of quasi-experimental shocks
which are not observed directly but are instead estimated in-sample
in an initial step, potentially introducing biases. In the canonical
estimation of regional labor supply elasticities by \textcite{Bartik1991},
for example, the shocks are measured as national industry growth rates.
Such growth captures national industry labor demand shocks, which
one may be willing to assume are as-good-as-randomly assigned across
industries; however, industry growth rates also depend on unobserved
regional labor supply shocks. We show that our framework can still
apply to such settings by casting the industry employment growth rates
as noisy estimates of latent quasi-experimental demand shocks and
establishing conditions to ensure the supply-driven estimation error
is asymptotically ignorable. These conditions are weaker if the latent
shocks are estimated as leave-one-out averages. Although leave-one-out
shift-share IV estimates do not have a convenient shock-level representation,
we provide evidence that in the \textcite{Bartik1991} setting this
leave-out adjustment is unimportant.

Formally, our approach to SSIV relates to the analysis of IV estimators
with many invalid instruments by \textcite{Kolesar2015}. Consistency
in that setting follows when violations of individual instrument exclusion
restrictions are uncorrelated with their first-stage effects. For
quasi-experimental SSIV, the exposure shares can be thought of as
a set of invalid instruments (per the \textcite{GPSS} interpretation),
and our orthogonality condition requires their exclusion restriction
violations to be uncorrelated with the shocks. Despite this formal
similarity, we argue that shift-share identification is better understood
through the quasi-random assignment of a single instrument (shocks),
rather than through a large set of invalid instruments (exposure shares)
that nevertheless produce a consistent estimate. This view is reinforced
by our equivalence results, yields a natural shock-level identification
condition, and suggests new validations and extensions of SSIV.

Our analysis also relates to other recent methodological studies of
shift-share designs, including those of \textcite{Jaeger2017} and
\textcite{Broxterman2018}. The former highlights biases of SSIV due
to endogenous local labor market dynamics, and we show how their solution
can be implemented in our setting. The latter studies the empirical
performance of different shift-share instrument constructions. As
discussed above we also draw on the inferential framework of \textcite{Adao},
who derive valid standard errors in shift-share designs with a large
number of idiosyncratic shocks. More broadly, our paper adds to a
growing literature studying the causal interpretation of common research
designs, including work by \textcite{bj16}, \textcite{gb18}, \textcite{as_es},
and \textcite{csa_es} for event study designs; \textcite{dCdH18}
for two-way fixed effects regressions; \textcite{sloczynski_weights}
for regressions with other controls; and \textcite{hull18} for mover
designs.

The remainder of this paper is organized as follows. Section 2 introduces
the environment, derives our equivalence results, and motivates our
approach to SSIV identification and consistency. Section 3 establishes
the baseline quasi-experimental assumptions and Section 4 derives
various extensions. Section 5 discusses shock-level procedures for
valid SSIV inference and testing. Section 6 summarizes the types of
empirical settings where our framework may be applied and illustrates
its practical implications in the ADH setting. Section 7 concludes.

\section{\label{sec:Shocks-as-Instruments}Setting and Motivation}

We begin by presenting the SSIV setting and motivating our approach
to identification and consistency with two equivalence results. We
first show that population orthogonality of the shift-share instrument
can be recast at the shock level, motivating identification by exogenous
shocks when exposure shares are endogenous. We then derive a similar
shock-level equivalence result for the SSIV estimator, motivating
its consistency with many as-good-as-randomly assigned shocks.

\subsection{\label{subsec:setup}The Shift-Share IV Setting}

We observe an outcome $y_{\ell}$, treatment $x_{\ell}$, control
vector $w_{\ell}$ (which includes a constant) and shift-share instrument
$z_{\ell}$ for a set of observations $\ell=1,\dots,L$. We also observe
a set of regression weights $e_{\ell}>0$ with $\sum_{\ell}e_{\ell}=1$
($e_{\ell}=\frac{1}{L}$ covers the unweighted case). The instrument
can be written as
\begin{align}
z_{\ell} & =\sum_{n}s_{\ell n}g_{n},\label{eq:eq:bartik}
\end{align}
for a set of observed shocks $g_{n}$, $n=1,\dots,N$, and a set of
observed shares $s_{\ell n}\ge0$ defining the exposure of each observation
$\ell$ to each shock $n$. Initially we assume the sum of these exposure
weights is constant across observations, i.e. that $\sum_{n}s_{\ell n}=1$;
we relax this assumption in Section \ref{subsec:Incomplete-shares}.\footnote{Note that the shares are defined relative to the total across components
$n$. In practice shift-share instruments are sometimes presented
differently, with the shares defined relative to the total across
observations $\ell$ (see footnote \ref{fn:ADH-details} for an example
in the \textcite{AutorDorn2001} setting). We recommend that researchers
follow the representation in (\ref{eq:eq:bartik}) to apply our theoretical
results.} Although our focus is on shift-share IV, we note that the setup nests
shift-share reduced-form regressions, of $y_{\ell}$ on $z_{\ell}$
and $w_{\ell}$, when $x_{\ell}=z_{\ell}$.

We seek to estimate the causal effect or structural parameter $\beta$
in a linear model of
\begin{equation}
y_{\ell}=\beta x_{\ell}+w_{\ell}^{\prime}\gamma+\varepsilon_{\ell},\label{eq:causalmodel}
\end{equation}
where the residual $\varepsilon_{\ell}$ is defined to be orthogonal
with the control vector $w_{\ell}$.\footnote{Formally, given a linear causal or structural model of $y_{\ell}=\beta x_{\ell}+\epsilon_{\ell}$
we define $\gamma=\expec{\sum_{\ell}e_{\ell}w_{\ell}w_{\ell}^{\prime}}^{-1}\expec{\sum_{\ell}e_{\ell}w_{\ell}\epsilon_{\ell}}$
and $\varepsilon_{\ell}=\epsilon_{\ell}-w_{\ell}^{\prime}\gamma$
as the residual from this population projection, satisfying $\expec{\sum_{\ell}e_{\ell}w_{\ell}\varepsilon_{\ell}}=0$.
Defining a unique $\gamma$ requires an implicit maintained assumption
that $\expec{\sum_{\ell}e_{\ell}w_{\ell}w_{\ell}^{\prime}}$ is of
full rank, which holds when there is no perfect collinearity in the
control vector. We consider models with heterogeneous treatment effects
in Appendix \ref{sec:hetFX}; see footnote \ref{fn:hetfx} for a summary.\label{fn:model_derivation}} For example, we might be interested in estimating a classic model
of labor supply which relates observations of log wage growth $y_{\ell}$
and log employment growth $x_{\ell}$ across local labor markets $\ell$
by an inverse labor supply elasticity $\beta$. The residual $\varepsilon_{\ell}$
in equation (\ref{eq:causalmodel}) would then contain all local labor
supply shocks, such as those arising from demographic, human capital,
or migration changes, that are not systematically related to the observed
controls in $w_{\ell}$.\footnote{While this simple labor supply equation is only well-defined under
certain assumptions (for instance, it rules out wage bargaining and
profit sharing between firms and workers), it is a standard modeling
tool. We note that it is inconsequential whether wages or employment
are on the right-hand side of the second stage regression; we choose
wage growth as the outcome following the tradition of \textcite{Bartik1991}.} To estimate $\beta$ we require an instrument capturing variation
in local labor demand.

We consider a $z_{\ell}$ based on the introduction of new import
tariffs $g_{n}$ across different industries $n$, with $s_{\ell n}$
denoting location $\ell$'s lagged shares of industry employment.\footnote{Import tariffs affect import prices, consumer demand for domestic
products, and in turn labor demand. Recent studies illustrate that
one can obtain quasi-random identifying variation in import tariffs
in practice. Changes in import tariffs across industries have been
used for identification in both industry-level analyses (e.g. \textcite{Fajgelbaum2020})
and shift-share analyses of regional outcomes (e.g. \textcite{Kovak2013}).} In estimating $\beta$ we may weight observations by the overall
lagged regional employment, $e_{\ell}$. We return to this labor supply
example at several points to ground the following theoretic discussion.

It is worth highlighting that in studying this setting we do not impose
a typical assumption of independent and identically-distributed (\emph{iid})
data $\left\{ e_{\ell},z_{\ell},w_{\ell},x_{\ell},\varepsilon_{\ell}\right\} $,
as might arise from random sampling of potential observations. Relaxing
the usual \emph{iid }assumption is required for us to treat the $g_{n}$
as random variables, which generate dependencies of the instrument
(\ref{eq:eq:bartik}) across observations exposed to the same random
shocks. The non-\emph{iid }setting further allows for unobserved common
shocks, which may generate dependencies in the residual $\varepsilon_{\ell}$.

Given this non-\emph{iid} setting, we consider IV identification of
$\beta$ by the full-data moment condition
\begin{align}
\expec{\sum_{\ell}e_{\ell}z_{\ell}\varepsilon_{\ell}} & =0.\label{eq:identification}
\end{align}
This condition captures the orthogonality of the shift-share instrument
with the second-stage residual, in expectation over realizations of
$\left\{ e_{\ell},z_{\ell},\varepsilon_{\ell}\right\} $ for all $\ell=1,\dots,L$.
When such orthogonality holds the $\beta$ parameter is identified:
i.e., uniquely recoverable from observable moments, provided the instrument
has a first stage.\footnote{Formally, when equation (\ref{eq:identification}) holds the moment
condition $m(b,c)\equiv\expec{\sum_{\ell}e_{\ell}(z_{\ell},w_{\ell}^{\prime})^{\prime}(y_{\ell}-bx_{\ell}-w_{\ell}^{\prime}c)}=0$
has a unique solution of $(\beta,\gamma)$, provided $\expec{\sum_{\ell}e_{\ell}(z_{\ell},w_{\ell}^{\prime})^{\prime}(x_{\ell},w_{\ell}^{\prime})}$
is of full-rank.} The full-data orthogonality condition generalizes the conventional
condition of $\expec{z_{\ell}\varepsilon_{\ell}}=0$, which might
be considered in an \emph{iid }setting with fixed $e_{\ell}$.

The moment condition (\ref{eq:identification}) yields a natural estimator
of $\beta$: the coefficient on $x_{\ell}$ in an IV regression of
$y_{\ell}$ which instruments by $z_{\ell}$, controls for $w_{\ell}$,
and weights by $e_{\ell}$. By the Frisch-Waugh-Lovell theorem, this
SSIV estimator can be represented as a bivariate IV regression of
outcome and treatment residuals, or as the ratio of $e_{\ell}$-weighted
sample covariances between the instrument and the residualized outcome
and treatment:
\begin{equation}
\hat{\beta}=\frac{\sum_{\ell}e_{\ell}z_{\ell}y_{\ell}^{\perp}}{\sum_{\ell}e_{\ell}z_{\ell}x_{\ell}^{\perp}},\label{eq:beta_hat}
\end{equation}
where $v_{\ell}^{\perp}$ denotes the residual from an $e_{\ell}$-weighted
sample projection of a variable $v_{\ell}$ on the control vector
$w_{\ell}$. Note that by the properties of residualization, it is
enough to residualize $y_{\ell}$ and $x_{\ell}$ without also residualizing
the shift-share instrument $z_{\ell}$.

In our non-\emph{iid} setting, we study consistency and other asymptotic
properties of $\hat{\beta}$ by considering a sequence of data-generating
processes, indexed by $L$, for the complete data $\left\{ e_{\ell},s_{\ell n},g_{n},w_{\ell},x_{\ell},\varepsilon_{\ell}\right\} $,
for $\ell=1,\dots,L$, $n=1\dots,N$, and $N=N(L)$. Consistency,
for example, is defined as $\hat{\beta}\stackrel{p}{\to}\beta$ as
$L\to\infty$ along this sequence. We do not employ conventional sampling-based
asymptotic sequences (and corresponding laws of large numbers) as
these are generally inappropriate in a non-\emph{iid }setting where
both $z_{\ell}$ and $\varepsilon_{\ell}$ may exhibit non-standard
mutual dependencies. It is worth emphasizing that any assumptions
on the data-generating sequence are useful only to approximate the
finite-sample distribution of the SSIV estimator, not to define an
actual process for realizations of the data. For example, we will
consider below a sequence in which the number of shocks $N$ grows
with $L$, recognizing that in reality shift-share instruments are
constructed from a fixed set of shocks (e.g. tariffs across all industries)
along with a fixed number of observations (e.g. all local labor markets).
The assumption of growing $N$ should here be interpreted as a way
to capture the presence of a large number of shocks in a given set
of observations, such that the asymptotic sequence provides a good
approximation to the observed data.\footnote{This is similar to how \textcite{Bekker1994} uses a non-standard
asymptotic sequence to analyze IV estimators with many instruments:
``The sequence is designed to make the asymptotic distribution fit
the finite sample distribution better. It is completely irrelevant
whether or not further sampling will lead to samples conforming to
this sequence'' (p. 658).}

\subsection{A Shock-Level Orthogonality Condition\label{subsec:moment-equiv}}

We first build intuition for our approach to satisfying the IV moment
condition by showing that the structure of the shift-share instrument
allows equation (\ref{eq:identification}) to be rewritten as condition
on the orthogonality of shocks $g_{n}$. Namely, by exchanging the
order of summation across $\ell$ and $n$, we obtain
\begin{align}
\expec{\sum_{\ell}e_{\ell}z_{\ell}\varepsilon_{\ell}}=\expec{\sum_{\ell}e_{\ell}\sum_{n}s_{\ell n}g_{n}\varepsilon_{\ell}}=\expec{\sum_{n}s_{n}g_{n}\bar{\varepsilon}_{n}} & =0,\label{eq:id_rewrite}
\end{align}
where we define $s_{n}=\sum_{\ell}e_{\ell}s_{\ell n}$ and $\bar{\varepsilon}_{n}=\frac{\sum_{\ell}e_{\ell}s_{\ell n}\varepsilon_{\ell}}{\sum_{\ell}e_{\ell}s_{\ell n}}$.
Just as the left-hand side of this expression captures the orthogonality
of the instrument $z_{\ell}$ with the residual $\varepsilon_{\ell}$
when weighted by $e_{\ell}$, the right-hand side captures the orthogonality
of shocks $g_{n}$ and $\bar{\varepsilon}_{n}$ when weighted by $s_{n}$.
Since these two expressions are equivalent, equation (\ref{eq:id_rewrite})
shows that such shock orthogonality is necessary and sufficient condition
for the orthogonality of the shift-share instrument. As with $e_{\ell}$,
the shock-level weights are also non-negative and sum to one, since
$\sum_{n}s_{n}=\sum_{\ell}e_{\ell}\left(\sum_{n}s_{\ell n}\right)=1$.
The shock-level unobservables $\bar{\varepsilon}_{n}$ represent exposure-weighted
averages of the residuals $\varepsilon_{\ell}$.

The labor supply example is useful for unpacking this first equivalence
result. When $s_{\ell n}$ are lagged employment shares and $e_{\ell}$
are similarly lagged regional employment weights, the $s_{n}$ weights
are proportional to the lagged industry employment.\footnote{Without regression weights (i.e. $e_{\ell}=\frac{1}{L}$), $s_{n}$
is instead the average employment share of industry $n$ across locations.} Moreover, with $\varepsilon_{\ell}$ capturing unmeasured supply
shocks, $\bar{\varepsilon}_{n}$ is the average unobserved supply
shock among regions $\ell$ that are the most specialized in industry
$n$, in terms of their lagged employment $e_{\ell}s_{\ell n}$. Equation
(\ref{eq:id_rewrite}) then shows that for the shift-share instrument
$z_{\ell}$ to identify the labor supply elasticity $\beta$, the
industry demand shocks $g_{n}$ must be orthogonal with these industry-level
unobservables when weighted by industry size. For example, the industries
which experience a rise in import tariffs should not face systematically
different unobserved labor supply conditions (e.g., migration patterns)
in their primary markets.

Shock orthogonality is a necessary condition for SSIV identification
and is satisfied when, as in the preferred interpretation of \textcite{GPSS},
the exposure shares are exogenous, the data are \emph{iid}, and the
shocks are considered non-random.\footnote{Formally, in this framework $\expec{e_{\ell}s_{\ell n}\varepsilon_{\ell}}=0$
for each $(\ell,n)$, so $\expec{\sum_{n}s_{n}g_{n}\bar{\varepsilon}_{n}}=\sum_{n}g_{n}\sum_{\ell}\expec{e_{\ell}s_{\ell n}\varepsilon_{\ell}}=0$.} In practice, however, this approach to SSIV identification may be
untenable in many settings. In our labor supply example, the \textcite{GPSS}
approach to identification requires the (lagged) local employment
share of each industry to be a valid instrument in the labor supply
equation, i.e. uncorrelated with all unobserved labor supply shocks.
This assumption is unlikely to hold: changes in foreign immigration,
for example, are a type of local labor supply shock which is likely
related to the local industry composition (e.g., new immigrants may
prefer to settle in areas with larger clusters of specific industries,
such as high-tech, even conditionally on the prevailing wage). Formally,
whenever the second-stage error term has a component with the shift-share
structure, $\sum_{n}s_{\ell n}\nu_{n}$ for unobserved shocks $\nu_{n}$,
then the exposure shares will be mechanically endogenous even if the
$\nu_{n}$ and $g_{n}$ are uncorrelated (see Appendix \ref{subsec:appx-no-Unobserved-Shocks}
for a proof).\footnote{In the immigration example, $\nu_{n}$ is positive in high-tech industries
and negative in industries that do not attract immigrants. Note that
the same argument applies to migration flows within the U.S., which
can similarly make local labor supply shocks related to local industry
composition. Lagging local employment shares does not alleviate these
threats to identification in general.}

When shares are endogenous, equation (\ref{eq:id_rewrite}) suggests
that identification may instead follow from the exogeneity of shocks.
We formalize this approach in Section \ref{subsec:A1andA2}, by specifying
a quasi-experimental design in which the $g_{n}$ are as-good-as-randomly
assigned with respect to the other terms in the expression. We show
how this simple exogeneity can be relaxed with controls in Section
\ref{subsec:Conditional-Quasi-Random-Assignm}.

\subsection{Estimator Equivalence\label{subsec:numerical-equiv}}

We next build intuition for our approach to SSIV consistency by showing
that the estimate $\hat{\beta}$ is equivalently obtained as the coefficient
from a non-standard shock-level IV procedure, in which $g_{n}$ directly
serves as the instrument. This equivalence result suggests that the
large-sample properties of $\hat{\beta}$ can be derived from a law
of large numbers for the equivalent shock-level regression. An attractive
feature of this approach is that it does not rely on an assumption
of \emph{iid }observations, which can be untenable in the presence
of observed and unobserved $n$-level shocks. We instead place assumptions
on the assignment of the equivalent IV regression's instrument $g_{n}$,
similar to a more standard analysis of a randomized treatment in an
experimental settings (\cite{Abadie2017b}).

Formally, we have the following equivalence result:
\begin{lyxlist}{00.00.0000}
\item [{\textbf{Proposition}}] \textbf{1}\textbf{\emph{ }}The SSIV estimator
$\hat{\beta}$ equals the second-stage coefficient from a $s_{n}$-weighted
shock-level IV regression that uses the shocks $g_{n}$ as the instrument
in estimating
\begin{equation}
\bar{y}_{n}^{\perp}=\alpha+\beta\bar{x}_{n}^{\perp}+\bar{\varepsilon}_{n}^{\perp},\label{eq:ind_reg}
\end{equation}
where $\bar{v}_{n}=\frac{\sum_{\ell}e_{\ell}s_{\ell n}v_{\ell}}{\sum_{\ell}e_{\ell}s_{\ell n}}$
denotes an exposure-weighted average of variable $v_{\ell}$.
\end{lyxlist}
\begin{proof}
\noindent By definition of $z_{\ell}$, 
\begin{align}
\hat{\beta} & =\frac{\sum_{\ell}e_{\ell}\left(\sum_{n}s_{\ell n}g_{n}\right)y_{\ell}^{\perp}}{\sum_{\ell}e_{\ell}\left(\sum_{n}s_{\ell n}g_{n}\right)x_{\ell}^{\perp}}=\frac{\sum_{n}g_{n}\left(\sum_{\ell}e_{\ell}s_{\ell n}y_{\ell}^{\perp}\right)}{\sum_{n}g_{n}\left(\sum_{\ell}e_{\ell}s_{\ell n}x_{\ell}^{\perp}\right)}=\frac{\sum_{n}s_{n}g_{n}\bar{y}_{n}^{\perp}}{\sum_{n}s_{n}g_{n}\bar{x}_{n}^{\perp}}.\label{eq:shocks_instruments}
\end{align}
Furthermore $\sum_{n}s_{n}\bar{y}_{n}^{\perp}=\sum_{\ell}e_{\ell}\left(\sum_{n}s_{\ell n}\right)y_{\ell}^{\perp}=\sum_{\ell}e_{\ell}y_{\ell}^{\perp}=0$,
since $y_{\ell}^{\perp}$ is an $e_{\ell}$-weighted regression residual
and $\sum_{n}s_{\ell n}=1$. This and an analogous equality for $\bar{x}_{n}^{\perp}$
imply that (\ref{eq:shocks_instruments}) is a ratio of $s_{n}$-weighted
covariances, of $\bar{y}_{n}^{\perp}$ and $\bar{x}_{n}^{\perp}$
with $g_{n}$. Hence it is obtained from the specified IV regression.
\end{proof}
As with equation (\ref{eq:id_rewrite}), Proposition 1 exploits the
structure of the instrument to exchange orders of summation in the
expression for the SSIV estimator (\ref{eq:beta_hat}). This exchange
shows that SSIV estimates can also be thought to arise from variation
across shocks, rather than across observations. The equivalent IV
regression uses the shocks $g_{n}$ directly as the instrument and
shock-level aggregates of the original (residualized) outcome and
treatment, $\bar{y}_{n}^{\perp}$ and $\bar{x}_{n}^{\perp}$. Specifically,
$\bar{y}_{n}^{\perp}$ reflects the average residualized outcome of
the observations most exposed to the $n$th shock, while $\bar{x}_{n}^{\perp}$
is the same weighted average of residualized treatment. The regression
is weighted by $s_{n}$, representing each shock's average exposure
across the observations.\footnote{\label{fn:reduced-form-equiv}In the special case of reduced-form
shift-share regressions, Proposition 1 shows that the equivalent shock-level
procedure is still an IV regression, of $\bar{y}_{n}^{\perp}$ on
the transformed shift-share instrument $\bar{z}_{n}^{\perp}$, again
instrumented by $g_{n}$ and weighted by $s_{n}$.}

The fact that shift-share estimates can be equivalently obtained by
a shock-level IV procedure suggests a new approach to establishing
their consistency. Generally, IV regressions of the form of (\ref{eq:shocks_instruments})
will be consistent when the instrument (here, $g_{n}$) is as-good-as
randomly assigned, there is a large the number of observations (here,
$N$), the importance weights are sufficiently dispersed (here, that
the $s_{n}$ are not too skewed), and there is an asymptotic first
stage. Consistency is then guaranteed regardless of the correlation
structure of the residuals $\bar{\varepsilon}_{n}^{\perp}$, and thus
in the primitive residuals $\varepsilon_{\ell}$ and exposure shares
$s_{\ell n}$. We formalize this approach below.

Before proceeding, it is worth emphasizing that while the shock-level
IV regression from Proposition~1 motivates our approach to \emph{identification}
of $\beta$, it does not affect the \emph{interpretation} of the coefficient
as measuring an $\ell$-level relationship. The shock-level equation
(\ref{eq:ind_reg}), in which the outcome and treatment are unconventional
shock-level objects, does not have independent economic content. For
example, in the labor supply setting $\bar{y}_{n}$ is not industry
$n$'s wage growth; rather, it measures the average wage growth in
regions where industry $n$ employs the most workers. Thus, while
$\hat{\beta}$ can be computed at the industry level it estimates
the elasticity of regional, rather than industry, labor supply, and
could, for example, capture the kinds of local spillovers that a regression
of industry wages on industry employment cannot.\footnote{In Appendix \ref{subsec:abstract_model} we develop a stylized model
to illustrate how the SSIV coefficient can differ from a ``native''
shock-level IV coefficient in the presence of local spillovers or
treatment effect heterogeneity, though both parameters may be of interest.
Intuitively, in the labor supply case one may estimate a low regional
elasticity but a high elasticity of industry labor supply if, for
example, migration is constrained but workers are mobile across industries
within a region.} Furthermore, Proposition 1 holds even for the outcomes and treatments
which cannot be naturally computed at the shock level, e.g. when $n$
indexes industries and $y_{\ell}$ measures labor force non-participation,
as in \textcite{AutorDorn2001}.

\section{A Quasi-Experimental SSIV Framework\label{sec:QE_assignment}}

We now show how SSIV identification and consistency can be satisfied
by a quasi-experiment in which shocks are as-good-as-randomly assigned,
mutually uncorrelated, large in number, and sufficiently dispersed
in terms of their average exposure. Instrument relevance generally
holds in such settings when the exposure of individual observations
tends to be concentrated in a small number of shocks, and when those
shocks affect treatment. We then show how this framework is naturally
generalized to settings in which shocks are only conditionally quasi-randomly
assigned or exhibit some forms of mutual dependence, such as clustering.

\subsection{Quasi-Randomly Assigned and Mutually Uncorrelated Shocks\label{subsec:A1andA2}}

Our approach to SSIV consistency is based on a thought experiment
in which the shocks $g_{n}$ are as-good-as-randomly assigned conditional
on the shock-level unobservables $\bar{\varepsilon}_{n}$ and exposure
weights $s_{n}$. As motivated above, placing assumptions on this
assignment process (rather than on the sampling properties of observations)
has two key advantages. First, we do not rely on conventional assumptions
of independent or clustered data which are generally inconsistent
with the shift-share data structure when the shocks are considered
random variables. Second, in conditioning on $\bar{\varepsilon}=\left\{ \bar{\varepsilon}_{n}\right\} _{n}$
and $s=\left\{ s_{n}\right\} _{n}$ we place no restrictions on the
dependence between the $s_{\ell n}$ and $\varepsilon_{\ell}$, allowing
shock exposure to be endogenous. We first show that such endogeneity
need not pose problems for SSIV identification:
\begin{lyxlist}{00.00.0000}
\item [{\textbf{Proposition}}] \noindent \textbf{2 }The SSIV moment condition
(\ref{eq:identification}) is satisfied by the following condition:
\item [{\textbf{Assumption}}] \noindent \textbf{1}\emph{ (Quasi-random
shock assignment)}:\emph{ }\textbf{\emph{$\expec{g_{n}\mid\bar{\varepsilon},s}=\mu$}},\textbf{
}for all $n$.
\end{lyxlist}
\begin{proof}
\noindent By equation (\ref{eq:id_rewrite}) and the law of iterated
expectations, $\expec{\sum_{\ell}e_{\ell}z_{\ell}\varepsilon_{\ell}}=\expec{\sum_{n}s_{n}g_{n}\bar{\varepsilon}_{n}}=\mu\cdot\expec{\sum_{n}s_{n}\bar{\varepsilon}_{n}}$
under Assumption 1. Furthermore, since $\sum_{\ell}s_{\ell n}=1$
and $\expec{\sum_{\ell}e_{\ell}\varepsilon_{\ell}}=0$ by construction,
$\expec{\sum_{n}s_{n}\bar{\varepsilon}_{n}}=\expec{\left(\sum_{n}s_{\ell n}\right)\left(\sum_{\ell}e_{\ell}\varepsilon_{\ell}\right)}=0$.
\end{proof}
Proposition 2 shows that the shift-share instrument is valid, in that
the IV moment condition (\ref{eq:identification}) holds, when the
underlying shocks are as-good-as-randomly assigned: each $g_{n}$
has the same mean $\mu$, regardless of the realizations of the relevant
unobservables $\bar{\varepsilon}$ (and average exposures $s$). In
the labor supply example this assumption would mean that import tariffs
should not have been chosen strategically, based on labor supply trends,
or in a way that is correlated with such trends.\footnote{For example, if labor supply trends differ between regions specializing
in manufacturing vs. services, the import tariffs should apply to
both types of sectors. We discuss in Section \ref{subsec:Incomplete-shares}
how to apply our framework in the case where import tariffs only apply
to a subsector of the economy, e.g. in manufacturing only.}

It follows from Proposition 2 that $\beta$ is identified by Assumption
1 provided the instrument is relevant.\footnote{Appendix \ref{sec:hetFX} shows how SSIV identifies a convex average
of heterogeneous treatment effects (varying potentially across both
$\ell$ and $n$) under a stronger notion of as-good-as-random shock
assignment and a first-stage monotonicity condition. This can be seen
as generalizing both the IV identification result of \textcite{AGI2000}
to shift-share instruments, as well as the reduced-form shift-share
identification result in \textcite{Adao}.\label{fn:hetfx}} In practice, the existence of a non-zero first stage can be inferred
from the data; we discuss appropriate inferential techniques in Section
\ref{sec:InferenceDiagnostics}. To illustrate how such instrument
relevance might hold with quasi-experimental shocks, we consider a
simple first-stage model. Consider a setting without controls ($w_{\ell}=1$)
and where treatment is a share-weighted average of shock-specific
components: $x_{\ell}=\sum_{n}s_{\ell n}x_{\ell n}$, where $x_{\ell n}=\pi_{\ell n}g_{n}+\eta_{\ell n}$
with $\pi_{\ell n}\ge\bar{\pi}$ almost surely for some fixed $\bar{\pi}>0$.
In line with Assumption 1, suppose that the shocks are independent
mean-zero, given the full set of exposure shares $s_{\ell n}$ and
regression weights $e_{\ell}$ and the full set of $\pi_{\ell n}$
and $\eta_{\ell n}$, with variances that are bounded below by some
fixed $\bar{\sigma}_{g}^{2}>0$. Then the instrument first stage is
positive:
\begin{align}
\expec{\sum_{\ell}e_{\ell}z_{\ell}x_{\ell}} & =\expec{\sum_{\ell}e_{\ell}\left(\sum_{n}s_{\ell n}g_{n}\right)\left(\sum_{n}s_{\ell n}(\pi_{\ell n}g_{n}+\eta_{\ell n})\right)}\nonumber \\
 & \ge\bar{\pi}\bar{\sigma}_{g}^{2}\expec{\sum_{\ell}e_{\ell}\sum_{n}s_{\ell n}^{2}}>0.\label{eq:FirstStage}
\end{align}

Given identification, SSIV consistency follows from an appropriate
law of large numbers. Motivated by the estimator equivalence in Section
\ref{subsec:numerical-equiv}, we consider settings in which the effective
sample size of the shock-level IV regression (\ref{eq:ind_reg}) is
large and the observations of the effective instrument (shocks) are
mutually uncorrelated:
\begin{lyxlist}{00.00.0000}
\item [{\textbf{Assumption}}] \noindent \textbf{2}\emph{ (Many uncorrelated
shocks)}:\textbf{\emph{ }}\emph{$\expec{\sum_{n}s_{n}^{2}}\rightarrow0$}
and $\cov{g_{n},g_{n^{\prime}}\mid\bar{\varepsilon},s}=0$ for all
$(n,n^{\prime})$ with $n^{\prime}\neq n$.
\end{lyxlist}
\noindent The first part of Assumption 2 states that the expected
Herfindahl index of average shock exposure, $\expec{\sum_{n}s_{n}^{2}}$,
converges to zero as $L\rightarrow\infty$. This condition implies
that the number of observed shocks grows with the sample (since $\sum_{n}s_{n}^{2}\ge1/N$),
and can be interpreted as requiring a large effective sample for the
equivalent shock-level IV regression. An equivalent condition is that
the largest importance weight in this regression, $s_{n}$, becomes
vanishingly small.\footnote{\textcite{GPSS} propose a different measure of the importance of
a given $n$, termed ``Rotemberg weights.'' In Appendix \ref{sec:appx-Rotemberg}
we show the formal connection between $s_{n}$ and these weights,
and that the latter do not carry the sensitivity-to-misspecification
interpretation as they do in the exogenous shares view of \textcite{GPSS}.
Instead, the Rotemberg weight of shock $n$ measures the leverage
of $n$ in the equivalent shock-level IV regression from Proposition
1. Shocks may have large leverage either because of large $s_{n}$,
as would be captured by the Herfindahl index, or because the shocks
have a heavy-tailed distribution which is allowed by Assumption 2.} The second part of Assumption 2 states that the shocks are mutually
uncorrelated given the unobservables and $s_{n}$. Both of these conditions,
while novel for SSIV, would be standard assumptions to establish the
consistency of a conventional shock-level IV estimator with $g_{n}$
as the instrument and $s_{n}$ weights.

Assumptions 1 and 2 are the baseline assumptions of our quasi-experimental
framework. Given a standard relevance condition and additional regularity
conditions listed in Appendix \ref{sec:consistency}, they are sufficient
to establish SSIV consistency:\footnote{One high-level condition used in Proposition 3 (Assumption B2) is
that the control coefficient $\gamma$ is consistently estimated by
its sample analog, $\hat{\gamma}=\left(\sum_{\ell}e_{\ell}w_{\ell}w_{\ell}^{\prime}\right){}^{-1}\sum_{\ell}e_{\ell}w_{\ell}\epsilon_{\ell}$
(see footnote \ref{fn:model_derivation}). We discuss sufficient conditions
for this assumption in Appendix \ref{sec:controlconsistency}.}
\begin{lyxlist}{00.00.0000}
\item [{\textbf{Proposition}}] \noindent \textbf{3 }Suppose Assumptions
1 and 2 hold, $\sum_{\ell}e_{\ell}z_{\ell}x_{\ell}^{\perp}\xrightarrow{p}\pi$
with $\pi\neq0$, and Assumptions B1-B2 hold. Then $\hat{\beta}\xrightarrow{p}\beta$.
\end{lyxlist}
\begin{proof}
\noindent See Appendix \ref{sec:consistency}.
\end{proof}
As before, the relevance condition merits further discussion. In our
simple first-stage model, $\sum_{\ell}e_{\ell}z_{\ell}x_{\ell}^{\perp}$
converges to $\expec{\sum_{\ell}e_{\ell}z_{\ell}x_{\ell}}$ under
appropriate regularity conditions, which is bounded above zero by
a term proportional to $\expec{\sum_{\ell}e_{\ell}\sum_{n}s_{\ell n}^{2}}$.
Thus, in this case, SSIV relevance holds when the $e_{\ell}$-weighted
average of local exposure Herfindahl indices $\sum_{n}s_{\ell n}^{2}$
across observations does not vanish in expectation. In our running
labor supply example, where $x_{\ell n}$ is industry-by-region employment
growth, SSIV relevance generally arises from individual regions $\ell$
tending to specialize in a small number of industries $n$, provided
import tariffs have a non-vanishing effect on local industry employment.\footnote{\label{fn:no-Bartik-variation}Note that this precludes consideration
of an asymptotic sequence where $L$ remains finite as $N$ grows.
With $L$ (and also $e_{1},\dots,e_{L}$) fixed, Assumption 2 implies
$\sum_{\ell}e_{\ell}^{2}\expec{\sum_{n}s_{\ell n}^{2}}\to0$ and thus
$\var{z_{\ell}}=\var{\sum_{n}s_{\ell n}g_{n}}\to0$ for each $\ell$
if $\var{g_{n}}$ is bounded. If the instrument has asymptotically
no variation it cannot have a first stage, unless the $\pi_{\ell n}$
grow without bound. This result also highlights the role of picking
the shares which reflect the impact of $g_{n}$ on $x_{\ell}$. Here
when the shares are misspecified, i.e. when the treatment is constructed
from different shares $\tilde{s}_{\ell n}$ as $x_{\ell}=\sum_{n}\tilde{s}_{\ell n}x_{\ell n}$,
the first-stage is bounded by a term proportional to $\expec{\sum_{\ell}e_{\ell}\sum_{n}s_{\ell n}\tilde{s}_{\ell n}}$,
which can be arbitrarily small even if $\expec{\sum_{\ell}e_{\ell}\sum_{n}\tilde{s}_{\ell n}^{2}}\not\rightarrow0$.} Compare this to the Herfindahl condition in Assumption 2, which instead
states that the \emph{average }shares of industries across locations
become small. Both conditions may simultaneously hold when most regions
specialize in a small number of industries, differentially across
a large number of industries.\footnote{As an extreme example, suppose each region specializes on one industry
only: $s_{\ell n}=\one\left[n=n(\ell)\right]$ for some $n(\ell)$.
Then the average local concentration index $\sum_{\ell}e_{\ell}\sum_{n}s_{\ell n}^{2}$
equals one, while Assumption 2 holds when national industry composition
is asymptotically dispersed: for example, when $e_{\ell}=1/L$ and
$n(\ell)$ is drawn \emph{iid} across regions and uniformly over $1,\dots,N$.}

\subsection{Conditional Shock Assignment and Weak Shock Dependence\label{subsec:Conditional-Quasi-Random-Assignm}}

Proposition 3 establishes SSIV consistency when shocks have the same
expectation across $n$ and are mutually uncorrelated, but both requirements
are straightforward to relax. We next provide extensions that allow
the shock expectation to depend on observables and for weak mutual
dependence (such as clustering or serial correlation) of the residual
shock variation.

We first relax Assumptions 1 and 2 to only hold conditionally on a
vector of shock-level observables $q_{n}$ (that includes a constant).
For example, it may be more plausible that shocks are as-good-as-randomly
assigned within a set of observed clusters $c(n)\in\{1,\dots,C\}$
with non-random cluster-average shocks, in which case $q_{n}$ collects
$C-1$ cluster dummies and a constant. In the labor supply example,
this may allow import tariffs to vary systematically across groups
of industries with similar labor intensity, but be as-good-as-random
within each of those groups. In general, with $q=\left\{ q_{n}\right\} _{n}$,
we consider the following weakened version of Assumption 1:
\begin{lyxlist}{00.00.0000}
\item [{\textbf{Assumption}}] \noindent \textbf{3}\emph{ (Conditional quasi-random
shock assignment)}:\emph{ }\textbf{\emph{$\expec{g_{n}\mid\bar{\varepsilon},q,s}=q_{n}^{\prime}\mu$}}\emph{,}
for all\emph{ $n$.}
\end{lyxlist}
Similarly, we consider a weakened version of Assumption 2 which imposes
mutual uncorrelatedness on the residual $\tilde{g}_{n}=g_{n}-q_{n}^{\prime}\mu$:
\begin{lyxlist}{00.00.0000}
\item [{\textbf{Assumption}}] \noindent \textbf{4}\emph{ (Many uncorrelated
shock residuals)}:\textbf{\emph{ }}\emph{$\expec{\sum_{n}s_{n}^{2}}\rightarrow0$}
and $\cov{\tilde{g}_{n},\tilde{g}_{n^{\prime}}\mid\bar{\varepsilon},q,s}=0$
for all $(n,n^{\prime})$ with $n^{\prime}\neq n$.
\end{lyxlist}
In the shock cluster example, Assumption 4 applies with a small number
of clusters, each with its own random effect, as in that case a law
of large numbers may apply to the within-cluster residuals $\tilde{g}_{n}$
but not the original shocks $g_{n}$.

By a simple extension of the proof to Proposition 3, the SSIV estimator
is consistent when these conditions replace Assumptions 1 and 2 and
the residual shift-share instrument $\tilde{z}_{\ell}=\sum_{n}s_{\ell n}\tilde{g}_{n}$
replaces $z_{\ell}$. While this instrument is infeasible, since $\mu$
is unknown, the following result shows that SSIV regressions that
control for the exposure-weighted vector of shock-level controls,
$\tilde{w}_{\ell}=\sum_{n}s_{\ell n}q_{n}$, provide a feasible implementation:
\begin{lyxlist}{00.00.0000}
\item [{\textbf{Proposition}}] \noindent \textbf{4 }Suppose Assumptions
3 and 4 hold, $\sum_{\ell}e_{\ell}z_{\ell}x_{\ell}^{\perp}\xrightarrow{p}\pi$
with $\pi\neq0$, and Assumptions B1-B2 hold. Then $\hat{\beta}\xrightarrow{p}\beta$
provided $\tilde{w}_{\ell}$ is included in $w_{\ell}$.
\end{lyxlist}
\begin{proof}
\noindent See Appendix \ref{sec:consistency}.
\end{proof}
This result highlights a special role of controls with a shift-share
structure (i.e. $\sum_{n}s_{\ell n}q_{n}$): besides removing confounding
variation from the residual (as any $w_{\ell}$ would do), they can
also be viewed as removing such variation directly from the shocks
(i.e. implicitly using $\tilde{g}_{n}$ in place of $g_{n}$). In
particular, Proposition 4 shows that controlling for each observation's
individual exposure to each cluster $\sum_{n}s_{\ell n}\one\left[c\left(n\right)=c\right]$
isolates the within-cluster variation in shocks. This allows for a
thought experiment in which shocks are drawn quasi-randomly only within
observed clusters, but not across clusters with potentially different
shock means. Note that Proposition 3 is obtained as a special case
of Proposition 4, which sets $q_{n}=1$.

Even conditional on observables, mutual shock uncorrelatedness may
be undesirably strong. It is, however, straightforward to further
relax this assumption to allow for shock assignment processes with
weak mutual dependence, such as further clustering or autocorrelation.
In Appendix \ref{sec:consistency} we prove extensions of Proposition
4 which replace Assumption 4 with one of the following alternatives:
\begin{lyxlist}{00.00.0000}
\item [{\textbf{Assumption}}] \noindent \textbf{5}\emph{ (Many uncorrelated
shock clusters)}:\emph{ }There exists a partition of shocks into clusters
$c(n)$ such that $\expec{\sum_{c}s_{c}^{2}}\rightarrow0$ for $s_{c}=\sum_{n:\,c(n)=c}s_{n}$
and $\cov{\tilde{g}_{n},\tilde{g}_{n^{\prime}}\mid\bar{\varepsilon},q,s}=0$
for all $(n,n^{\prime})$ with $c(n)\ne c(n^{\prime})$;
\item [{\textbf{Assumption}}] \noindent \textbf{6}\emph{ (Many weakly correlated
shocks)}:\emph{ }For some sequence of numbers $B_{L}\ge0$ and a fixed
function $f(\cdot)$ satisfying $\sum_{r=1}^{\infty}f(r)<\infty$,
$B_{L}\expec{\sum_{n}s_{n}^{2}}\rightarrow0$ and $\left|\cov{\tilde{g}_{n},\tilde{g}_{n^{\prime}}\mid\bar{\varepsilon},q,s}\right|\le B_{L}\cdot f\left(\left|n^{\prime}-n\right|\right)$
for all $(n,n^{\prime})$. 
\end{lyxlist}
Assumption 5 relaxes Assumption 4 by allowing shock residuals to be
grouped within mutually mean-independent clusters $c(n)$, while placing
no restriction on their within-cluster correlation. At the same time,
the Herfindahl index assumption of Assumption 4 is strengthened to
hold for industry clusters, with $s_{c}$ denoting the average exposure
of cluster $c$. Assumption 6 takes a different approach, allowing
all nearby shock residuals to be mutually correlated provided their
covariance is bounded by a function $B_{L}\cdot f\left(\left|n'-n\right|\right)$.
This accommodates, for example, the case of first-order autoregressive
time series with the covariance bound declining at a geometric rate,
i.e. $f(r)=\delta^{r}$ for $\delta\in[0,1)$ and constant $B_{L}$.
With $B_{L}$ growing, stronger dependence of nearby shocks is also
allowed (see Appendix \ref{sec:consistency}).

\section{Extensions}

We now present several other extensions of our quasi-experimental
framework. Section \ref{subsec:Main-estimated-shocks} discusses how
our framework applies when the shocks are estimated within the sample,
as in the canonical \textcite{Bartik1991} study. Section \ref{subsec:Incomplete-shares}
explains the need for additional controls when the sum of exposure
shares vary across observations. Section \ref{subsec:Panels} considers
shift-share identification with panel data. Finally, Section \ref{subsec:Main-Multiple-shocks}
extends the framework to allow for multiple treatments and shift-share
instruments.

\subsection{Shift-Share Designs with Estimated Shocks\label{subsec:Main-estimated-shocks}}

In some shift-share designs, the shocks are equilibrium objects that
can be difficult to view as being quasi-randomly assigned. For example,
in the canonical \textcite{Bartik1991} estimation of the regional
labor supply elasticity, the shocks are national industry employment
growth rates. Such growth reflects labor demand shifters, which one
may be willing to assume are as-good-as-randomly assigned across industries.
However industry growth also aggregates regional labor supply shocks
that directly enter the residual $\varepsilon_{\ell}$. Here we show
how the quasi-experimental SSIV framework can still apply in such
cases, by viewing the $g_{n}$ as noisy estimates of some latent true
shocks $g_{n}^{\ast}$ (labor demand shifters, in the \textcite{Bartik1991}
example) that satisfy Assumption 1. We establish the conditions on
estimation noise (aggregated labor supply shocks, in \textcite{Bartik1991})
such that a feasible shift-share instrument estimator, perhaps involving
a leave-one-out correction as in \textcite{autorduggan03}, is asymptotically
valid.

We leave a more general treatment of this issue to Appendix \ref{subsec:appx-Estimated_shocks}
and for concreteness focus on the \textcite{Bartik1991} example.
The industry growth rates $g_{n}$ can be written as weighted averages
of the growth of each industry in each region: $g_{n}=\sum_{\ell}\omega_{\ell n}g_{\ell n}$,
where the weights $\omega_{\ell n}$ are the lagged shares of industry
employment located in region $\ell$, with $\sum_{\ell}\omega_{\ell n}=1$
for each $n$. In a standard model of regional labor markets, $g_{\ell n}$
includes (to first-order approximation) an industry labor demand shock
$g_{n}^{*}$ and a term that is proportional to the regional supply
shock $\varepsilon_{\ell}$.\footnote{Appendix \ref{subsec:appx-labor-demand-model} presents such a model,
showing that $g_{\ell n}$ also depends on the regional average of
$g_{n}^{\ast}$ (via local general equilibrium effects) and on idiosyncratic
region-specific demand shocks. Both of these are uncorrelated with
the error term in the model and thus do not lead to violations of
Assumption 1; we abstract away from this detail here.} We suppose that the demand shocks are as-good-as-randomly assigned
across industries, such that the infeasible SSIV estimator which uses
$z_{\ell}^{\ast}=\sum_{n}s_{\ell n}g_{n}^{\ast}$ as an instrument
satisfies our quasi-experimental framework. The asymptotic bias of
the feasible SSIV estimator which uses $z_{\ell}=\sum_{n}s_{\ell n}g_{n}$
then depends on the large-sample covariance between the labor supply
shocks $\varepsilon_{\ell}$ and an aggregate of the supply shock
``estimation error,'' 
\begin{equation}
\psi_{\ell}=z_{\ell}-z_{\ell}^{\ast}\propto\sum_{n}s_{\ell n}\sum_{\ell^{\prime}}\omega_{\ell^{\prime}n}\varepsilon_{\ell^{\prime}}.\label{eq:psi_nonLOO}
\end{equation}

Two insights follow from considering the bias term $\sum_{\ell}e_{\ell}\psi_{\ell}\varepsilon_{\ell}$.
First, part of the covariance between $\psi_{\ell}$ and $\varepsilon_{\ell}$
is mechanical, since $\varepsilon_{\ell}$ enters $\psi_{\ell}$.
In fact, if supply shocks are spatially uncorrelated this is the only
source of bias from using $z_{\ell}$ rather than $z_{\ell}^{*}$
as an instrument. This motivates the use of a leave-one-out (LOO)
shock estimator, $g_{n,-\ell}=\sum_{\ell^{\prime}\ne\ell}\omega_{\ell^{\prime}n}g_{\ell^{\prime}n}/\sum_{\ell^{\prime}\ne\ell}\omega_{\ell^{\prime}n}$,
and the feasible instrument $z_{\ell}^{LOO}=\sum_{n}s_{\ell n}g_{n,-\ell}$
to remove this mechanical covariance.\footnote{This problem of mechanical bias is similar to that of two-stage least
squares with many instruments \parencite{Bound1995a}, and the solution
is similar to the jackknife instrumental variable estimate approach
of \textcite{JIVE}.} Conversely, if the regional supply shocks $\varepsilon_{\ell}$ are
spatially correlated a LOO adjustment may not be sufficient to eliminate
mechanical bias in the feasible SSIV instrument, though more restrictive
split-sample methods (e.g. those estimating shocks from distant regions)
may suffice.

Second, in settings where there are many regions contributing to each
shock estimate even the mechanical part of $\sum_{\ell}e_{\ell}\psi_{\ell}\varepsilon_{\ell}$
may be ignorable, such that the conventional non-LOO shift-share instrument
$z_{\ell}$ (which, unlike $z_{\ell}^{LOO}$, has a convenient shock-level
representation per Proposition 1) is asymptotically valid when $z_{\ell}^{LOO}$
is.\footnote{\textcite{Adao} derive the corrected standard errors for LOO SSIV
and find that they are in practice very close to the non-LOO ones,
in which case the SSIV standard errors we derive in the next section
are approximately valid even when the LOO correction is used.} In Appendix \ref{subsec:appx-Estimated_shocks}, we derive a heuristic
for this case, under the assumption of spatially-independent supply
shocks. In a special case when each region is specialized in a single
industry and there are no importance weights, the key condition is
$L/N\rightarrow\infty$, or that the average number of regions specializing
in the typical industry is large. With incomplete specialization or
weights, the corresponding condition requires the typical industry
to be located in a much larger number of regions than the number of
industries that a typical region specializes in.

To illustrate the preceding points in the data, Appendix \ref{subsec:appx-Estimated_shocks}
replicates the setting of \textcite{Bartik1991} with and without
a LOO estimator, using data from \textcite{GPSS}. We find that in
practice the LOO correction does not matter for the SSIV estimate,
consistent with the findings of \textcite{GPSS} and \textcite{Adao},
and especially so when the regression is estimated without regional
employment weights. Our framework provides a explanation for this:
the heuristic statistic we derive is much larger without importance
weights. These findings imply that if, in the canonical \textcite{Bartik1991}
setting, one is willing to assume quasi-random assignment of the underlying
industry demand shocks and that the regional supply shocks are spatially-uncorrelated,
one can interpret the uncorrected SSIV estimator as leveraging demand
variation in large samples, as some of the literature has done (e.g.
\textcite{Suarez2016}).

\subsection{SSIVs with Incomplete Shares \label{subsec:Incomplete-shares}}

While we have so far assumed the sum of exposure shares is constant,
in practice this $S_{\ell}=\sum_{n}s_{\ell n}$ may vary across observations
$\ell$. For example, in the labor supply setting, the quasi-experiment
in tariffs may only cover manufacturing industries, while the lagged
manufacturing employment shares of $s_{\ell n}$ may be measured relative
to total employment in region $\ell$. In this case $S_{\ell}$ equals
the lagged total share of manufacturing employment in region $\ell$.
The \textcite{AutorDorn2001} setting is another example of this scenario,
as we discuss below.\footnote{We note that this scenario applies to quasi-experiments in which shocks
are impossible (\emph{ex ante}) for some industries. In contrast,
if all industries were equally likely to receive tariffs but only
some did \emph{ex post}, the set of $n$ should include all industries,
with $S_{\ell}=1$, and zero tariffs captured by $g_{n}=0$ for some
$n$. In such a case, however, it is unlikely that all manufacturing
industries receive the tariffs by chance when no non-manufacturing
industries do.}

Our framework highlights a potential problem with such ``incomplete
share'' settings: even if Assumptions 1 and 2 hold, the SSIV estimator
will generally leverage non-experimental variation in $S_{\ell}$
in addition to quasi-experimental variation in shocks. To see this
formally, note that one can always return to the complete shares setting
by rewriting the shift-share instrument with the ``missing'' (e.g.,
non-manufacturing) shock included: $z_{\ell}=s_{\ell0}g_{0}+\sum_{n>0}s_{\ell n}g_{n}$,
where $g_{0}=0$ and $s_{\ell0}=1-S_{\ell}$, yielding $\sum_{n=0}^{N}s_{\ell n}=1$
for all $\ell$. The previous quasi-experimental framework then applies
to this expanded set of shocks $g_{0},\dots,g_{N}$. But since $g_{0}=0$,
Proposition 3 requires in this case that $\expec{g_{n}\mid s,\bar{\varepsilon}}=0$
for $n>0$ as well; that is, that the expected shock to each manufacturing
industry is the same as the ``missing'' non-manufacturing shock
of zero. Otherwise, even when manufacturing shocks are random, regions
with higher manufacturing shares $S_{\ell}$ will tend to have systematically
different values of the instrument $z_{\ell}$, leading to bias when
these regions also have different unobservables.\footnote{Formally, if Assumptions 1 and 2 hold for all $n>0$ we have from
the proof to Proposition 3 that $\sum_{\ell}e_{\ell}z_{\ell}\varepsilon_{\ell}=\sum_{n=0}^{N}s_{n}g_{n}\bar{\varepsilon}_{n}=\expec{\sum_{n=0}^{N}s_{n}(g_{n}-\mu)\bar{\varepsilon}_{n}}+o_{p}(1)=-\mu\expec{s_{0}\bar{\varepsilon}_{0}}+o_{p}(1)$.
If $\mu\neq0$ and the missing industry share is large ($s_{0}\not\xrightarrow{p}0$)
this can only converge to zero when $\expec{s_{0}\bar{\varepsilon}_{0}}=\expec{\sum_{\ell}e_{\ell}s_{\ell0}\varepsilon_{\ell}}$
does, i.e. when $S_{\ell}$ is exogenous.}

Cast in this way, the incomplete shares issue has a natural solution
via Assumption 3: to control for the sum of exposure shares. Formally,
one can allow the missing and non-missing shocks to have different
means by conditioning on the indicator $\mathbf{1}[n>0]$ in the $q_{n}$
vector. By Proposition 4, the SSIV estimator allows for such conditional
quasi-random assignment when the control vector $w_{\ell}$ contains
the exposure-weighted average of $\mathbf{1}[n>0]$, which here is
$\sum_{n=0}^{N}s_{\ell n}\one\left[n>0\right]=S_{\ell}$.\footnote{By effectively ``dummying out'' the missing industry, SSIV regressions
that control for $S_{\ell}$ further require a weaker Herfindahl condition:
$\expec{\sum_{n=1}^{N}s_{n}^{2}}\rightarrow0$, allowing the non-manufacturing
industry share $s_{0}$ to stay large.} Thus, in the labor supply example, quasi-experimental variation in
manufacturing shocks is isolated provided one controls for a region's
lagged manufacturing share $S_{\ell}$. More generally, the control
$\sum_{n=1}^{N}s_{\ell n}q_{n}$ which, per Proposition 4, allows
the shock mean to depend on observables $q_{n}$ for $n>0$ changes
the interpretation in the incomplete shares case: it is a exposure-weighted
sum, rather than average. 

\subsection{Panel Data\label{subsec:Panels}}

In practice, SSIV regressions are often estimated with panel data,
where the outcome $y_{\ell t}$, treatment $x_{\ell t}$, exposure
shares $s_{\ell nt}$, and shocks $g_{nt}$ are additionally indexed
by time periods $t=1,\dots,T$.\footnote{Exposure shares are typically lagged and sometimes fixed in a pre-period.
Our subscript $t$ notation indicates that these shares are used to
construct the instrument for period $t$, not that they are measured
in that period.} In such settings a time-varying instrument $z_{\ell t}=\sum_{n}s_{\ell nt}g_{nt}$
is used, and the controls $w_{\ell t}$ may include unit- or period-specific
fixed effects.

It is straightforward to apply the preceding quasi-experimental framework
to the panel case with a simple relabeling: $\tilde{\ell}=\left(\ell,t\right)$
for the $LT$ observations and $\tilde{n}=\left(n,\tau\right)$ for
the $NT$ shocks (where $\tau$ also indexes time periods). With
exposure shares redefined as $\tilde{s}_{\tilde{\ell}\tilde{n}}=s_{\ell nt}\mathbf{1}[t=\tau]$
(i.e. by definition zero for $t\neq\tau$), the instrument can be
rewritten as $z_{\tilde{\ell}}=\sum_{\tilde{n}}\tilde{s}_{\tilde{\ell}\tilde{n}}g_{\tilde{n}}$,
mirroring the cross-sectional case.

With this relabeling, standard intuitions for panel consistency readily
translate into shift-share designs. In short panels or repeated cross-sections
(i.e. with fixed $T$) the SSIV estimator can be consistent if $L,N\to\infty$
and the cross-sectional conditions of Proposition 4 hold.\footnote{Arbitrary serial correlation of shocks can be allowed here via Assumption
5, with $n$ defining a cluster. One complication arises when an increasing
number of unit fixed effects are included in the control vector, violating
Assumption B2 in Proposition 4. In Appendix \ref{sec:paneldata} we
show how this incidental parameter problem can be solved  by imposing
shock exogeneity with respect to demeaned residuals.} Alternatively, consistency of the estimator can follow from a long
time series of shocks ($T\to\infty$) that have weak serial dependence,
even if $L$ and $N$ are small. This case accommodates, in particular,
shift-share designs that leverage purely time-series shocks ($N=1$,
$T\to\infty$), as in \textcite{nunnqian14}.\footnote{\textcite{nunnqian14} estimate the impact of U.S. food aid on civil
conflict, using variation in U.S. wheat production (a single ``shock''
per period) over a long time horizon ($T=36$ years), interacted with
a country\textquoteright s tendency to receive US food aid (the ``exposure
shares'' for $L=125$ countries). Our approach may also be appropriate
in settings where $LT$ and $NT$ are large despite moderate $N$
and $T$. \textcite{Berman2017}, for example, leverage price changes
for $N=14$ minerals over $T=14$ years in a very large cross-section
of spatial cells.}

One subtlety of panel SSIV regressions concerns the role of fixed
effect (FE) controls. Unit fixed effects play a dual role in conventional
panels (with exogenous shocks varying at the same level as the observations):
they purge both time-invariant unobservables ($\frac{1}{T}\sum_{\tau}\varepsilon_{\ell\tau}$)
from the residual and the time-invariant component of the shocks ($\frac{1}{T}\sum_{\tau}g_{n\tau}$).
While the first role directly extends to the FEs of cross-sectional
units $\ell$ in the shift-share case, the second role only does when
exposure shares are fixed across periods, i.e. when $s_{\ell nt}\equiv s_{\ell n0}$.\footnote{Shift-share IV settings with panel data and time-invariant shares
include, for example, \textcite{Berman2011}, \textcite{Berman2017},
\textcite{Hummels2014}, \textcite{Imbert2019}, and \textcite{nunnqian14}.} Similarly, while period FEs always purge period-specific unobservables
($\frac{1}{L}\sum_{\ell}\varepsilon_{\ell t}$), in SSIV designs they
only isolate within-period shock variation when the exposure shares
add up to one. With incomplete shares, in contrast, period FEs need
to be interacted with the sum of exposure shares $S_{\ell}$.\footnote{Both results follow from Proposition 4. The exposure-weighted sums
of shock-level unit FEs, which isolate the time-varying component
of $g_{nt}$, are only absorbed by observation-level FEs when the
exposure shares are time-invariant. Similarly, if the $q_{n}$ include
shock-level period FEs, the corresponding exposure-weighted sums equal
$\sum_{\tilde{n}}\tilde{s}_{\tilde{\ell}\tilde{n}}\mathbf{1}[\tau=\bar{t}]=S_{\ell}\mathbf{1}[t=\bar{t}]$,
which simplifies to period FEs when $S_{\ell}=1$.}

Finally, we note that while fixing exposure shares may have the advantage
of isolating cleaner time-varying shock variation, it may also have
an efficiency cost: lagging the shares by many periods is likely to
make the shift-share instrument less predictive of treatment (see
Appendix \ref{subsec:SSIV-Relevance-panels} for a formal argument).
If a researcher wants to update the shares to maximize the first stage,
but also isolate the shock variation over time (which is not achieved
by controlling for unit $\ell$ fixed effects with time-varying shares),
she may instead use the first-differenced specification. That is,
estimate $\Delta y_{\ell t}=\beta\Delta x_{\ell t}+\gamma^{\prime}\Delta w_{\ell t}+\Delta\varepsilon_{\ell t},$
instrumenting $\Delta x_{\ell t}$ with $z_{\ell t,FD}=\sum_{n}s_{\ell n,t-1}\Delta g_{nt}$,
where $\Delta$ is the first-differencing operator for both observations
and shocks. This strategy has been employed, for example, by \textcite{AutorDorn2001}
as we discuss in Section \ref{subsec:Application}.\footnote{There is another argument for fixing the shares in a pre-period that
arises when the current shares are affected by lagged shocks in a
way that is correlated with unobservables $\varepsilon_{\ell t}$.
In the labor supply example, suppose local labor markets vary in flexibility,
with stronger reallocation of employment to industries with larger
increases in import tariffs in flexible markets. If import tariffs
are random but persistent, industries with growing tariffs will be
increasingly concentrated in regions with flexible labor markets and
Assumption 1 will be violated if such flexibility is correlated with
$\varepsilon_{\ell t}$. This concern is not specific to panel data,
but may also arise in cross-sections or repeated cross-sections.}

\subsection{Multiple Shocks and Treatments\label{subsec:Main-Multiple-shocks}}

In some shift-share designs one may be interested in leveraging several
shift-share instruments, corresponding to multiple sets of shocks
satisfying Assumptions 3 and 4. For example while \textcite{AutorDorn2001}
construct an instrument from average Chinese import growth across
eight non-U.S. countries, in principle the industry shocks from each
individual country may be each thought to be as-good-as-randomly assigned.
\textcite{Jaeger2017} instrument two treatments\textemdash the current
and lagged immigration rates\textemdash with two shift-share instruments.
\textcite{Bombardini2018} estimate the reduced-form effects of two
shift-share variables: the regional growth of all exports and the
regional growth of exports in pollution-intensive sectors. In Appendix
\ref{subsec:appx-multiple_shocks} we show that our quasi-experimental
framework extends to these settings, in which the exposure shares
used to construct the instruments are the same but the shocks differ.
The key insight is that SSIV regressions with multiple instruments\textemdash with
and without multiple endogenous variables\textemdash again have an
equivalent representation as particular shock-level IV estimators,
although the equivalence result is more complex under overidentification.

\section{\label{sec:InferenceDiagnostics}Shock-Level Inference and Testing}

A shock-level view also brings new insights to SSIV inference and
testing. In this section we first show how a problem with conventional
SSIV inference, first studied by \textcite{Adao}, has a convenient
solution based on our shock-level equivalence result. We then discuss
how other novel shock-level procedures can be used to assess first-stage
relevance and to implement valid falsification tests of shock exogeneity.
Lastly, we summarize a variety of Monte-Carlo simulations illustrating
the finite-sample properties of SSIV.

\subsection{\label{subsec:inference}Exposure-Robust Standard Errors}

As with consistency, SSIV inference is complicated by the fact that
the observed shocks $g_{n}$ and any unobserved shocks $\nu_{n}$
induce dependencies in the instrument $z_{\ell}$ and residual $\varepsilon_{\ell}$
across observations with similar exposure shares. This problem can
be understood as an extension of the standard clustering concern \parencite{Moulton1986},
in which the instrument and residual are correlated across observations
within predetermined clusters, with the additional complication that
in SSIV every pair of observations with overlapping shares may have
correlated $(z_{\ell},\varepsilon_{\ell})$. \textcite{Adao} develop
a novel approach to conducting valid inference in presence of exposure-based
clustering, building on our quasi-experimental framework for identification.

Our equivalence result in Section \ref{subsec:numerical-equiv} motivates
a convenient alternative approach to valid SSIV inference. By estimating
SSIV coefficients with an equivalent shock-level IV regression, one
directly obtains valid (``exposure-robust'') standard errors under
the assumptions in \textcite{Adao} and an additional condition on
the controls that we discuss below.\footnote{This solution generalizes a well-known approach to addressing conventional
group clustering \parencite[p. 313]{Angrist2008}: by estimating a
regression at the level of as-good-as-random variation (here, shocks)
one avoids inferential biases due to clustering (here, by shock exposure).}
\begin{lyxlist}{00.00.0000}
\item [{\textbf{Proposition}}] \noindent \textbf{5 }Consider $s_{n}$-weighted
IV estimation of the second stage equation
\begin{align}
\bar{y}_{n}^{\perp} & =\alpha+\beta\bar{x}_{n}^{\perp}+q_{n}^{\prime}\delta+\bar{\varepsilon}_{n}^{\perp}\label{eq:fullSEss}
\end{align}
where $\tilde{w}_{\ell}=\sum_{n}s_{\ell n}q_{n}$ is included in the
control vector $w_{\ell}$ used to compute $\bar{y}_{n}^{\perp}$
and $\bar{x}_{n}^{\perp}$, and $\bar{x}_{n}^{\perp}$ is instrumented
by $g_{n}$. The IV estimate of $\beta$ is numerically equivalent
to the SSIV estimate $\hat{\beta}$. Furthermore, when Assumptions
B3\textendash B6 in Appendix \ref{sec:appx_Inference} hold and $\sum_{\ell}e_{\ell}x_{\ell}^{\perp}z_{\ell}\xrightarrow{p}\pi$
for $\pi\ne0$, the conventional heteroskedasticity-robust standard
error for $\hat{\beta}$ yields asymptotically-valid confidence intervals
for $\beta$.
\end{lyxlist}
\begin{proof}
\noindent See Appendix \ref{sec:appx_Inference}.
\end{proof}
Equation (\ref{eq:fullSEss}) extends the previous shock-level estimating
equation (\ref{eq:ind_reg}) by including a vector of controls $q_{n}$
which, as in Proposition 4, are included in the SSIV control vector
$w_{\ell}$ as exposure-weighted averages. The first result in Proposition
5 is that the addition of these controls does not alter the coefficient
equivalence established in Proposition 1. The second result states
conditions, which strengthen those of Proposition 4, under which conventional
shock-level standard errors from estimation of (\ref{eq:fullSEss})
yield valid asymptotic inference on $\beta$.\footnote{Appendix \ref{sec:appx_Inference} also establishes two related results
regarding specifications without controls and an alternative inference
procedure to improve finite-sample performance.}

While our previous results do not restrict the structure of the control
vector $w_{\ell}$, Proposition 5 (specifically, Assumption B4) allows
for only two types of controls. All sources of shock-level confounding
have to be captured by controls with a shift-share structure (i.e.
$\sum_{n}s_{\ell n}q_{n}$, as in Proposition 4); the other controls
should not be asymptotically correlated with the instrument although
they may increase the asymptotic efficiency of the estimator. While
valid shift-share inference with general control vectors remains an
open problem, we show in Appendix \ref{sec:appx_Inference} that the
standard errors from Proposition 5 are asymptotically conservative
under a much weaker assumption, which allows for controls of the form
$\sum_{n}s_{\ell n}p_{n}+u_{\ell}$, where $p_{n}$ are unobserved
confounders and $u_{\ell}$ is noise.\footnote{\textcite{Adao} provide asymptotically valid standard errors in a
special case of this weaker assumption: when the average variance
of the noise $u_{\ell}$ is asymptotically small for all controls
that are necessary for identification. Our standard errors remain
asymptotically conservative in this case.}

Our shock-level approach to estimating exposure-robust standard errors
offers three practical features. First, it can be performed with standard
statistical software packages given a simple initial transformation
of the data (i.e. to obtain $\bar{y}_{n}^{\perp}$, $\bar{x}_{n}^{\perp}$,
and $s_{n}$), for which we have released a Stata package \emph{ssaggregate}
(see footnote \ref{fn:ssagg}). Second, it is readily extended to
settings where shocks are clustered or autoregressive, as in Assumptions
5 and 6 respectively. Conventional cluster-robust or heteroskedastic-and-autocorrelation-consistent
(HAC) standard error calculations applied to equation (\ref{eq:fullSEss})
are then valid. Third, the shock-level inference approach works when
$N>L$ or when some exposure shares are collinear.\footnote{This is not possible with the standard error calculation of \textcite{Adao}
because their procedure involves projecting $z_{\ell}^{\perp}$ on
the vector of shares in order to account for the shock-level confounders
underlying the approximate shift-share controls. This issue can be
empirically relevant: for instance, employment shares of some industries
are collinear in the \textcite{AutorDorn2001} setting.}

\subsection{Falsification and Relevance Tests\label{subsec:A1_A2_tests}}

Our Proposition 5 also provides a practical way to perform valid regression-based
tests of shock orthogonality (i.e. falsification tests) and first-stage
relevance. As a falsification test of Assumption 3, one may regress
an observed proxy $r_{\ell}$ for the unobserved residual $\varepsilon_{\ell}$
on the instrument $z_{\ell}$ (controlling for $w_{\ell})$. Examples
of such $r_{\ell}$ include baseline characteristics realized prior
to the shocks or lagged observations of the outcome (yielding a so-called
``pre-trend'' test). To make the magnitude of the placebo coefficient
more interpretable, one may also be regress $r_{\ell}$ on $x_{\ell}$
while instrumenting with $z_{\ell}$. To address the exposure clustering
problem that may arise in these regressions, the shock-level regression
of Proposition 5 can be used, yielding valid inference on these coefficients.
If a researcher instead starts from a shock-level confounder $r_{n}$,
they can construct its observation-level average $r_{\ell}=\sum_{n}s_{\ell n}r_{n}$
and proceed similarly; a simpler test regresses $r_{n}$ on $g_{n}$
directly (weighting by $s_{n}$ and controlling for $q_{n}$).\footnote{While pre-trend and other $\ell$-level balance tests are also useful
in the alternative \textcite{GPSS} framework (albeit with a different
approach to inference), this shock-level test is specific to our approach
to identification. We emphasize that all tests discussed here are
meant to falsify the quasi-random shock assignment assumption made
\emph{a priori}, and not to test the two frameworks against each other.}

Similarly, Proposition 5 yields a convenient way to test first-stage
relevance in the OLS regression of $x_{\ell}$ on $z_{\ell}$ and
$w_{\ell}$. We note that the equivalent shock-level regression is
IV, not OLS (see footnote \ref{fn:reduced-form-equiv}). The first
stage \emph{F}-statistic, which is a common heuristic for relevance,
is then obtained as a squared \emph{t}-statistic.\footnote{We generalize this result to the case of multiple shift-share instruments
in Appendix \ref{subsec:appx-multiple_shocks} by detailing the appropriate
construction of the ``effective'' first-stage \emph{F}-statistic
of \textcite{MontielOleaPflueger2013}, again based on an equivalent
shock-level IV regression.}

\subsection{Monte-Carlo Simulations}

Though the exposure-robust standard errors obtained from estimating
equation (\ref{eq:fullSEss}) are asymptotically valid, it is useful
to verify that they offer appropriate coverage with a finite number
of observations and shocks. In Appendix \ref{subsec:Finite-Sample-Performance}
we provide Monte-Carlo simulations confirming that the finite-sample
performance of the equivalent regression (\ref{eq:fullSEss}) is comparable
to that of more conventional shock-level IV regressions, in which
the outcome and instrument are not aggregated from a common set of
$y_{\ell}$ and $x_{\ell}$. The asymptotic approximation performs
well even with a Herfindahl concentration index $\sum_{n}s_{n}^{2}$
of $1/20$ (which can be compared to a shock-level regression with
20 equal-sized industries); the conventional rule of thumb for detecting
weak instruments based on the appropriate constructed first-state
\emph{F}-statistic applies equally well to SSIV estimators. These
results indicate that a researcher who is comfortable with the finite-sample
performance of a shock-level analysis with some set of $g_{n}$ should
also be comfortable using such shocks in SSIV, provided there is sufficient
variation in exposure shares to yield a strong SSIV first stage.

\section{Shift-Share IV in Practice\label{sec:InPractice}}

We now summarize and illustrate the practical implications of our
econometric framework. We first characterize the kinds of empirical
settings to which the foregoing framework may be applied. We then
apply the framework to the influential setting of \textcite{AutorDorn2001}.

\subsection{A Taxonomy of SSIV Settings\label{subsec:Taxonomy}}

Our framework can be applied to various empirical settings. To characterize
these settings, we distinguish between three cases of SSIVs employed
in the literature. 

In the first case, the shift-share instrument is based on a set of
shocks which can itself be thought of as an instrument. For example,
the $g_{n}$ which enter $z_{\ell}$ might correspond to a set of
observed growth rates that could be plausibly thought of as being
randomly assigned to a large number of industries. Our framework shows
how the shift-share instrument maps these shocks to the level of observed
outcomes and treatments (e.g., geographic regions). A researcher who
is comfortable with the identification conditions and finite-sample
performance of an industry-level analysis based on $g_{n}$ should
generally also be comfortable applying our framework, provided there
is sufficient variation in the exposure shares and treatment to yield
a strong first stage. \textcite{AutorDorn2001} and the corresponding
industry-level analysis conducted by \textcite{AADHP2016} give a
prime example of this case, as we show below.

Empirical settings covered by this first case belong to various fields
in economics, with outcomes and shocks defined at levels ($\ell$
and $n$, respectively) different than regions and industries. In
international trade, \textcite{Hummels2014} estimate the wage effects
of offshoring across Danish importing firms $\ell$. They leverage
a shift-share instrument for offshoring based on shocks to export
supply by type of intermediate inputs and origin country; titanium
hinges from Japan is an example of an $n$. While they translate these
shocks to the firm level by using the lagged composition of firm imports
as the shares, one could imagine an analysis of Danish imports at
the input-by-country level directly that would leverage the same supply
shocks. In finance, \textcite{Xu2018} examines the long-term effects
of financial shocks on exports across countries $\ell$. Her shift-share
instrument is based on a disruption that affected some but not all
London-based banks $n$ in 1866, with country-specific exposure shares
measuring pre-1866 market shares of those banks in each country. In
line with considering bank shocks as-good-as-randomly assigned, she
reports that affected and unaffected banks were balanced on various
observable characteristics. In the immigration literature, \textcite{Peri2016}
estimate the effect of immigrant STEM workers on the labor market
outcomes of natives across U.S. cities $\ell$. They exploit variation
in the supply of STEM workers across migration origin countries $n$
and over time that arises from plausibly exogenous shifts to national
H1-B policy. Similarly, in a literature on innovation, \textcite{Stuen2012}
leverage education policy shocks in foreign countries as a supply
shock to U.S. doctoral programs. 

In the second case, a researcher does not directly observe a large
set of quasi-experimental shocks, but can still conceive of an underlying
set of $g_{n}$ which if observed would be a useful instrument. Constructing
the instrument then requires an initial step where these shocks are
estimated in-sample, potentially introducing mechanical biases. In
the canonical setting of \textcite{Bartik1991} and \textcite{Blanchard1992},
for example, a local labor demand instrument is sought, with the ideal
$g_{n}$ measuring an aggregate change in industry labor demand that
may be assumed orthogonal to local labor supply shocks. Aggregate
demand changes are however not directly observed and must be estimated
from national industry employment growth (often using leave-out corrections,
as in \textcite{autorduggan03} and \textcite{Diamond2016}). We have
discussed how our framework generalizes to this more involved setting
in Section \ref{subsec:Main-estimated-shocks}, showing the additional
assumptions required for the estimation error to be asymptotically
ignorable. While this case differs from the first in terms of the
instrument construction, the underlying logic of our framework still
applies. This case also covers instruments in the immigration literature,
as in \textcite{Card2001} and \textcite{Card2009}, where latent
shocks to out-migration from foreign countries can be thought to be
as-good-as-randomly assigned but are estimated from aggregate in-migration
flows in the U.S.

The third case is conceptually distinct, in that the $g_{n}$ underlying
the (perhaps idealized) instrument cannot be naturally viewed as an
instrument itself. This could either be because it is not plausible
that these shocks are as-good-as-randomly assigned, even conditionally
on shock-level observables, or because there are too few shocks. Identification
in this case may instead follow from exogeneity of the exposure shares,
as suggested by \textcite{GPSS}. 

Share exogeneity may be a more plausible approach in the third case
when the exposure shares are ``tailored'' to the specific economic
question, and to the particular endogenous variable included in the
model. In this case, the scenario considered in Section \ref{subsec:moment-equiv}\textemdash that
there are unobserved shocks $\nu_{n}$ which enter $\varepsilon_{\ell}$
through the shares\textemdash may be less of a concern. \textcite{Mohnen2019},
for example, uses the age profile of older workers in local labor
markets as the exposure shares of a shift-share instrument for the
change in the local elderly employment rate in the following decade.
He argues, based on economic intuition, that these tailored shares
are uncorrelated with unobserved trends in youth employment rates.
This argument notably does not require one to specify the age-specific
shocks $g_{n}$, which only affect power of the instrument (in fact,
the shocks are dispensed with altogether in robustness checks that
directly instrument with the shares). Similarly, \textcite{Algan2017}
use the lagged share of the construction sector in the regional economy
as an instrument for unemployment growth during the Great Recession,
arguing that it does not predict changes in voting outcomes in other
ways. With a single industry the identification assumption reduces
to that of conventional difference-in-differences with continuous
treatment intensity and our framework cannot be applied.

In contrast, our framework may be more appropriate in settings where
shocks are tailored to a specific question while the shares are ``generic,''
in that they could conceivably measure an observation's exposure to
multiple shocks (both observed and unobserved). Both \textcite{AutorDorn2001}
and \textcite{AcemogluRestrepo}, for example, build shift-share instruments
with similar lagged employment shares but different shocks\textemdash rising
trade with China and the adoption of industrial robots, respectively.
According to the \textcite{GPSS} view, these papers use essentially
the same instruments (lagged employment shares) for different endogenous
variables (growth of import competition and growth of robot adoption),
and are therefore mutually inconsistent. Our framework helps reconcile
these identification strategies, provided the variation in each set
of shocks can be described as arising from a natural experiment. In
principle, shares and shocks may simultaneously provide valid identifying
variation, but in practice it would seem unlikely for both sources
of variation to be \emph{a priori }plausible in the same setting.

This discussion highlights that plausibility of our framework, as
with the alternative framework of \textcite{GPSS}, depends on the
details of the SSIV application. We encourage practitioners to use
our framework only after establishing an \emph{a priori }argument
for the plausibility of exogenous shocks. Various diagnostics on the
extent of shock variation and falsification of this assumption may
then be conducted to assess \emph{ex post} the plausibility of exogenous
shocks. We next illustrate this approach in the \textcite{AutorDorn2001}
setting.

\subsection{Application to Autor, Dorn, and Hanson (2013)\label{subsec:Application}}

Our application to \textcite[henceforth ADH]{AutorDorn2001} aims
to illustrate our theoretical framework only, and not to reassess
their substantive findings. In line with this goal, we first describe
how the ADH instrument could be thought to leverage quasi-experimental
shocks and discuss potential threats to this identification strategy.
We then illustrate the tools and lessons that follow from our framework,
demonstrating steps that researchers can emulate in their own SSIV
applications. Specifically, we analyze the distribution of shocks
to assess the plausibility of Assumption 4 (many conditionally uncorrelated
shocks), use balance tests to corroborate the plausibility of Assumption~3
(conditional quasi-random shock assignment), use equivalent shock-level
IV regressions to obtain exposure-robust inference, and analyze the
sensitivity of the results to the inclusion of different shock-level
controls. This analysis shows how our quasi-experimental framework
can help understand the identifying variation in the ADH SSIV design.

\subsubsection{Setting and Intuition for Identification\label{subsec:ApplicationSetting}}

ADH use a shift-share IV to estimate the causal effect of rising import
penetration from China on U.S. local labor markets. They do so with
a repeated cross section of 722 commuting zones $\ell$ and 397 four-digit
SIC manufacturing industries $n$ over two periods $t$, 1990-2000
and 2000-2007. In these years U.S. commuting zones were exposed to
a dramatic rise in import penetration from China, a historic change
in trade patterns commonly referred to as the ``China shock.'' Variation
in exposure to this change across commuting zones results from the
fact that different areas were initially specialized in different
industries which saw different changes in the aggregate U.S. growth
of Chinese imports. ADH combine import changes across industries in
eight comparable developed economies (as shocks) with lagged industry
employment (as exposure shares) to construct their shift-share instrument.

To illustrate our framework in this setting we focus on ADH's primary
outcome of the change in total manufacturing employment as a percentage
of working-age population during period $t$ in location $\ell$,
which we write as $y_{\ell t}$. The treatment variable $x_{\ell t}$
measures local exposure to the growth of imports from China in \$1,000
per worker. The vector of controls $w_{\ell t}$, which comes from
the preferred specification of ADH (Column 6 of their Table 3), contains
start-of-period measures of labor force demographics, period fixed
effects, Census region fixed effects, and the start-of-period total
manufacturing share to which we return below. The shift-share instrument
is $z_{\ell t}=\sum_{n}s_{\ell nt}g_{nt}$, where $s_{\ell nt}$ is
the share of manufacturing industry $n$ in total employment in location
$\ell$ (measured a decade before each period $t$ begins) and $g_{nt}$
is industry $n$'s growth of imports from China in the eight comparable
economies over period $t$ (also expressed in \$1,000 per U.S. worker).\footnote{\label{fn:ADH-details}To be precise, local exposure to the growth
of imports from China is constructed for period $t$ as $x_{\ell t}=\sum_{n}s_{\ell nt}^{\text{current}}g_{nt}^{\text{US}}$.
Here $g_{nt}^{\text{US}}=\frac{\Delta M_{nt}^{\text{US}}}{E_{nt}^{\text{current}}}$
is the growth of U.S. imports from China in thousands of dollars ($\Delta M_{nt}^{\text{US}}$)
divided by the industry employment in the U.S. at the beginning of
the current period ($E_{nt}^{\text{current}}$) and $s_{\ell nt}^{\text{current}}$
are local employment shares, also measured at the beginning of the
period. The instrument, in contrast, is constructed as $z_{\ell t}=\sum_{n}s_{\ell nt}g_{nt}$
with $g_{nt}=\frac{\Delta M_{nt}^{\text{8 countries}}}{E_{nt}}$,
where $\Delta M_{nt}^{\text{8 countries}}$ measures the growth of
imports from China in eight comparable economies (in thousands of
U.S. dollars) and both local employment shares $s_{\ell nt}$ and
U.S. employment $E_{nt}$ are lagged by 10 years. The eight countries
are Australia, Denmark, Finland, Germany, Japan, New Zealand, Spain,
and Switzerland. Note that \textcite{AutorDorn2001} express the same
instrument differently, based on employment shares relative to the
industry total, rather than the regional total. Our way of writing
$z_{\ell t}$ aims to clearly separate the exposure shares from the
industry shocks, highlighting the shift-share structure of the instrument.} Importantly, the sum of lagged manufacturing shares across industries
($S_{\ell t}=\sum_{n}s_{\ell nt}$) is not constant across locations
and periods, placing the ADH instrument in the ``incomplete shares''
class discussed in Section \ref{subsec:Incomplete-shares}. All regressions
are weighted by $e_{\ell t}$, which measures the start-of-period
population of the commuting zone, and all variables are measured in
ten-year equivalents.

To see how the ADH instrument can be viewed as leveraging quasi-experimental
shocks, consider an idealized experiment generating random variation
in the growth of imports from China across industries. One could imagine,
for example, random variation in industry-specific productivities
in China affecting import growth both in the U.S. and in comparable
economies. This would yield a set of observed productivity changes
$g_{nt}$ which would plausibly satisfy our Assumption 1. Assumption~2
would further hold when the productivity shocks are idiosyncratic
across many industries, with small average exposure to each shock
across commuting zones. Weaker versions of this experimental ideal,
in which productivity shocks can be partly predicted by industry observables
and are only weakly dependent across industries, are accommodated
by the extensions in Section \ref{subsec:Conditional-Quasi-Random-Assignm}.
For example, in ADH's repeated cross section one might invoke Assumption
3 in allowing the average shock to vary across periods, in recognition
that the 1990s and 2000s were very different trade environments, as
China joined the World Trade Organization in 2001. Here $q_{nt}$
would indicate periods.

ADH's approach can be seen as approximating this idealized experiment
by using observed changes in trade patterns between China and a group
of developed countries outside the United States. Trade between the
U.S. and China depends on changes in U.S. supply and demand conditions,
which may have direct effects on employment dynamics in U.S. regions.
In contrast, variation in the ADH $g_{nt}$ reflects only Chinese
productivity shocks and the various supply and demand shocks in the
non-U.S. developed countries. In this way, the ADH strategy can be
understood as eliminating bias from shocks that are specific to the
United States.

This discussion gives an \emph{a priori} justification for thinking
of the ADH instrument as leveraging quasi-experimental shocks within
the two periods. Nonetheless, since the ADH shocks are not truly randomized,
one may still worry that they are confounded by other unobserved characteristics.
For example, China's factor endowment may imply that it specializes
in low-skill industries, which could have been on different employment
trends in the U.S. even absent increased trade with China. Similarly,
one can imagine a common component of import growth in the U.S. and
the group of comparable developed economies due to correlated technological
shocks in those countries, which may have a direct effect on U.S.
labor markets. Given these potential concerns, it will be important
to assess the plausibility of Assumption 3 in this setting by conducting
within-period falsification tests of the kind we describe in Section
\ref{subsec:A1_A2_tests}. It will also be important to assess whether
there is sufficient variation in the ADH shocks for Assumption 4 to
hold.

Before applying these tests, it is worth highlighting that the assumption
of exogenous exposure shares, as discussed by \textcite{GPSS}, is
likely to be \emph{a priori} implausible in the ADH setting. As indicated
in Section \ref{subsec:moment-equiv}, any unobserved shocks $\nu_{nt}$
invalidate the share exogeneity assumption if they enter the error
$\varepsilon_{\ell t}$ in a manner which is correlated with the shares.
Because ADH use generic manufacturing employment shares to instrument
for a specific treatment variable, the possibility of other industry
shocks entering $\varepsilon_{\ell t}$ looms large. These unobserved
shocks could take many forms, for example heterogeneous speeds of
automation, secular changes in consumer demand, or changes in factor
prices which differentially affect industries based on their skill
intensity. This is in contrast to our Assumption 3 which allows for
any of these unobserved shocks as long as they are uncorrelated with
$g_{nt}$ across industries, conditionally on observables $q_{nt}$.\footnote{This assumption, which allows one to isolate import competition from
other industry shocks, is standard in similar industry-level analyses
(e.g. \cite{AADHP2016}) and can be tested with falsification tests,
as we do in Section \ref{subsec:appl-balance}.}

With a plausible justification of our framework in hand, we next illustrate
its application.

\subsubsection{\label{subsec:appl-Properties-of-shocks}Properties of Industry Shocks
and Exposure Shares}

Our quasi-experimental view of the ADH research design places particular
emphasis on the variation in Chinese import growth rates $g_{nt}$
and their average exposure $s_{nt}$ across industries and periods.
With few or insufficiently-variable shocks, or highly concentrated
shocks exposure, the large-$N$ asymptotic approximation developed
in Section \ref{sec:QE_assignment} is unlikely to be a useful tool
for characterizing the finite-sample behavior of the SSIV estimator.
We thus first summarize the distribution of $g_{nt}$, as well as
the industry-level weights from our equivalence result, $s_{nt}\propto\sum_{\ell}e_{\ell t}s_{\ell nt}$
(normalized to add up to one in the entire sample).\footnote{Note that $s_{nt}$ would be proportional to lagged industry employment
if the ADH regression weights $e_{\ell t}$ were lagged regional employment.
ADH however use a slightly different $e_{\ell t}$: the start-of-period
commuting zone population.}

In summarizing the industry-level variation it is instructive to recall
that the ADH instrument is constructed with ``incomplete'' manufacturing
shares. Per the discussion in Section \ref{subsec:Incomplete-shares},
this means that absent any regression controls the SSIV estimator
uses variation not only in manufacturing industry shocks but also
implicitly the variation in the 10-year lagged total manufacturing
share $S_{\ell t}$ across commuting zones and periods. In practice,
ADH control for the start-of-period manufacturing share, which is
highly\textemdash though not perfectly\textemdash correlated with
$S_{\ell t}$. We thus summarize the ADH shocks both with and without
the ``missing'' shock $g_{0t}=0$, which here represents the lack
of a ``China shock'' in service (i.e. non-manufacturing) industries.
Given that trade with China was very different in the 1990s and 2000s,
we focus on the within-period variation in manufacturing shocks.

Table 1 reports summary statistics for the ADH shocks $g_{nt}$ computed
with importance weights $s_{nt}$, and characterizes these weights.
Column 1 includes the ``missing'' service industry shock of zero
in each period. It is evident that with this shock the distribution
of $g_{nt}$ is unusual: for example, its interquartile range is zero.
This is because the service industry accounts for a large fraction
of total employment ($s_{0t}$ is 71.9\% of the period total in the
1990s and 79.5\% in the 2000s). As a result we see a high concentration
of industry exposure as measured by the inverse of its Herfindahl
index (HHI), $1/\sum_{n,t}s_{nt}^{2}$, which corresponds to the effective
sample size of our equivalent regression and plays a key role in Assumption
2. With the ``missing'' shock included, the effective sample size
is only 3.5. For an HHI computed at the level of three-digit industry
codes $\sum_{c}s_{c}^{2}$, where $s_{c}$ aggregates exposure across
the two periods and industries within the same 3-digit group $c$,
it is even lower, at 1.7. This suggests even less industry-level variation
is available when shocks are allowed to be serially correlated or
clustered by groups. Furthermore, the mean of manufacturing shocks
is significantly different from the zero shock of the missing service
industry.\footnote{The weighted mean of manufacturing shocks is 7.4, with a standard
error clustered at the 3-digit SIC level (as in our analysis below)
of 1.3.} Together, these analyses suggest that the service industry should
be excluded from the identifying variation, because it is likely to
violate both Assumption 1 ($\expec{g_{nt}\mid\bar{\varepsilon},s}\neq g_{0t}=0$)
and Assumption 2 ($\sum_{n,t}s_{nt}^{2}$ is not close to zero).

Column 2 of Table 1 therefore summarizes the sample with the service
industry excluded. The distribution of shocks is now much more regular,
with an average of 7.4, a standard deviation of 20.9 and an interquartile
range of 6.6. The inverse HHI of the $s_{nt}$ is also relatively
high: 191.6 across industry-by-period cells and 58.4 when exposure
is aggregated by SIC3 group. The largest shock weights in this column
are only 3.4\% across industry-by-periods and 6.5\% across SIC3 groups.
This suggests a sizable degree of variation at the industry level,
consistent with Assumption 2. In general, we recommend that researchers
report the inverse of the HHI of shock-level average exposure as a
simple way of describing their effective sample size. A first-stage
\emph{F}-statistic, which we discuss appropriate computation of in
Section \ref{subsec:A1_A2_tests}, will provide a formal test of the
power of the shock variation.

Finally, column 3 of Table 1 summarizes the distribution of within-period
manufacturing shocks, which would be leveraged by an assumption of
conditional quasi-experimental assignment (Assumption 3). The column
confirms that even conditional on period there is sizable residual
shock variation. The standard deviation and interquartile range of
shock residuals (obtained from regressing shocks on period fixed effects
with $s_{nt}$ weights) are only mildly smaller than in Column 2,
despite the higher mean shock in the later period, at 12.6 versus
3.6.

Besides the condition on the effective sample size, Assumption 2 (and
its clustered version in Assumption 5) requires the shocks to be sufficiently
mutually uncorrelated. To assess the plausibility of this assumption
and choose the appropriate level of clustering for exposure-robust
standard errors, we next analyze the correlation patterns of shocks
across manufacturing industries using available industry classifications
and the time dimension of the pooled cross section. In particular,
we compute intra-class correlation coefficients (ICCs) of shocks within
different industry groups, as one might do to correct for conventional
clustering parametrically (e.g. \textcite[p. 312]{Angrist2008}).\footnote{Note that similar ICC calculations could be implemented in a setting
that directly regresses industry outcomes on industry shocks, such
as \textcite{AADHP2016}. Mutual correlation in the instrument is
a generic concern that is not specific to shift-share designs, although
one that is rarely tested for. Getting the correlation structure in
shocks right is especially important for inference in our framework,
since the outcome and treatment in the industry-level regression ($\bar{y}_{nt}^{\perp}$
and $\bar{x}_{nt}^{\perp}$) are by construction correlated across
industries.} These ICCs come from a random effects model, providing a hierarchical
decomposition of residual within-period shock variation:
\begin{align}
g_{nt} & =\mu_{t}+a_{\text{ten}(n),t}+b_{\text{sic2}(n),t}+c_{\text{sic3}(n),t}+d_{n}+e_{nt},\label{eq:mixed}
\end{align}
where $\mu_{t}$ are period fixed effects; $a_{\text{ten}(n),t}$,
$b_{\text{sic2}(n),t}$, and $c_{\text{sic3}(n),t}$ denote time-varying
(and possibly auto-correlated) random effects generated by the ten
industry groups in \textcite{AADHP2016}, 20 groups identified by
SIC2 codes, and 136 groups corresponding to SIC3 codes, respectively;
and $d_{n}$ is a time-invariant industry random effect (across our
397 four-digit SIC industries). Following convention, we estimate
equation (\ref{eq:mixed}) as a hierarchical linear model by maximum
likelihood, assuming Gaussian residual components.\footnote{In particular we estimate an unweighted mixed-effects regression using
Stata's \emph{mixed }command, imposing an exchangeable variance matrix
for $(a_{\text{ten}(n),1},a_{\text{ten}(n),2})$, $(b_{\text{sic2}(n),1},b_{\text{sic2}(n),2})$,
and $(c_{\text{sic3}(n),1},c_{\text{sic3}(n),2})$.}

Table 2 reports estimated ICCs from equation (\ref{eq:mixed}), summarizing
the share of the overall shock residual variance due to each random
effect. These reveal moderate clustering of shock residuals at the
industry and SIC3 level (with ICCs of 0.169 and 0.073, respectively).
At the same time, there is less evidence for clustering of shocks
at a higher SIC2 level and particularly by ten cluster groups (ICCs
of 0.047 and 0.016, respectively, with standard errors of comparable
magnitude). This supports the assumption that shocks are mean-independent
across SIC3 clusters, so it will be sufficient to cluster standard
errors at the level of SIC3 groups, as \textcite{AADHP2016} do in
their conventional industry-level IV regressions. The inverse HHI
estimates in Table 1 indicate that at this level of shock clustering
there is still an adequate effective sample size.

\subsubsection{Falsification Tests \label{subsec:appl-balance}}

We next implement falsification tests of ADH shock orthogonality,
which provide a way of assessing the plausibility of Assumption 3.
Following Section \ref{subsec:A1_A2_tests}, we do this in two ways,
both different from conventional falsification tests sometimes run
in SSIV settings. First, we regress potential proxies for the unobserved
residual (i.e., any unobserved industry labor demand or labor supply
shock) on the instrument $z_{\ell}$ but use exposure-robust inference
that takes into account the inherent dependencies of the data. Second,
we regress potential industry-level confounders directly on the shocks
(while again clustering by SIC3). While this second type of falsification
tests would be standard in industry-level analyses, such as \textcite{AADHP2016},
it has rarely been used to assess the plausibility of SSIV designs
(with \textcite{Xu2018}, mentioned above, being a rare exception).

Choosing the set of potential confounders for these exercises is a
context-specific issue, which should be justified separately in every
application. To discipline our illustrative exercise, we use the industry-level
production controls in \textcite{AADHP2016} and the regional controls
in ADH. Consistent with our \emph{a priori }view of the quasi-experiment,
we maintain only the period fixed effects as controls when evaluating
balance on these other observables. For the industry-level balance
test this amounts to regressing each potential confounder on the manufacturing
shocks (normalized to have a unit variance) and period fixed effects,
weighting by average industry employment shares. Regional balance
coefficients are obtained by regressing each potential confounder
on the shift-share instrument (normalized to have a unit variance)
and the share-weighted average of period effects (i.e., the period-interacted
sum-of-shares), since ADH is a setting with incomplete shares. To
obtain exposure-robust standard errors, we implement these regressions
at the shock level, as discussed above.

Panel A of Table 3 reports the results of our industry-level balance
tests. The five \textcite{AADHP2016} production controls are an industry's
share of production workers in employment in 1991, the ratio of its
capital to value-added in 1991, its log real wages in 1991, the share
of its investment devoted to computers in 1990, and the share of its
high-tech equipment in total investment in 1990.\footnote{The last two controls are missing for five out of 397 industries.
We impute the missing values by the medians in the SIC3 industry group
or, when not available, in the SIC2 group.} Broadly, these variables reflect the structure of employment and
technology across industries. If the ADH shocks are as-good-as-randomly
assigned to industries within periods, we expect them to not predict
these predetermined variables. Panel A shows that there is indeed
no statistically significant correlation within periods, consistent
with Assumption 3.

Panel B of Table 3 reports the results of our regional balance tests.
The five ADH controls are the fraction of a commuting zone's population
who is college-educated, the fraction of its population who is foreign-born,
the fraction of its workers who are female, its fraction of employment
in routine occupations, and the average offshorability index of its
occupations. Broadly, these variables reflect the composition of a
region's workforce. We again find no statistically significant relationships
between these variables and the shift-share instrument within periods,
except for the foreign-born population fraction. Locations exposed
to a large ADH trade shock tend to have a higher fraction of immigrants,
suggesting that they may be subject to different labor supply dynamics.
We explore the importance of this imbalance for the SSIV coefficient
estimate in sensitivity tests below.

Finally, the last two rows of the same panel conduct a regional ``pre-trends''
analysis. We regress the pre-trend variables from ADH\textemdash manufacturing
employment growth in the 1970s and 1980s\textemdash on the shift-share
instrument, using the same specification as in the previous rows.
We find no relationship between the shift-share instrument and manufacturing
employment growth in the 1980s, but there is a positive statistically
significant relationship with manufacturing employment growth in the
1970s. Both findings are similar to those from ADH's pre-trend analysis.

Overall, we fail to reject imbalance in ten out of the twelve potential
confounders at conventional levels of statistical significance. How
to proceed when some balance tests fail is a general issue in quasi-experimental
analyses and has to be decided in the context of an application. One
might view the balance failures as sufficient evidence against Assumption
3 to seek alternative shocks or more appropriate shock-level controls.
Alternatively, one may argue that the observed imbalances are unlikely
to invalidate the research design. ADH, for example, note that the
positive relationship they find between the shift-share instrument
and manufacturing employment in the 1970s occurs in the distant past,
while the insignificant relationship in the 1980s demonstrates that
the relationship between rising China trade exposure and declining
manufacturing employment was absent in the decade immediately prior
to China\textquoteright s rise. Similarly the imbalance of the foreign-born
share that we observe need not generate a bias in the estimate if
it is not strongly correlated with the second-stage residual. To gauge
this potential for omitted variable bias one can include such variables
as controls in the SSIV specification and check sensitivity of the
coefficient; we report results of this exercise next.

\subsubsection{Main Estimates and Sensitivity Analyses\label{subsec:appl-Main-Estimates}}

We next estimate the effects of import competition on local labor
market outcomes, leveraging within-period exogeneity of the industry
shocks $g_{nt}$. We then check sensitivity of results to inclusion
of the \textcite{AutorDorn2001} regional controls and \textcite{AADHP2016}
industry-level controls.

Table 4 reports SSIV coefficients from regressing regional manufacturing
employment growth in the U.S. on the growth of import competition
from China, instrumented by predicted Chinese import growth.\footnote{Appendix Table C1 reports estimates for other outcomes in ADH: growth
rates of unemployment, labor force non-participation, and average
wages, corresponding to columns 3 and 4 of Table 5 and column 1 of
Table 6 in ADH.} Per the results in Section \ref{subsec:inference}, we estimate these
coefficients with equivalent industry-level regressions in order to
obtain valid exposure-robust standard errors. Consistent with the
above analysis of shock ICCs, we cluster standard errors at the SIC3
level. We also report first-stage \emph{F}-statistics with corresponding
exposure-robust inference. As discussed in Section \ref{subsec:A1_A2_tests},
these come from industry-level IV regressions of the aggregated treatment
and instrument (i.e. $\bar{x}_{nt}^{\perp}$ on $\bar{z}_{nt}^{\perp}$),
instrumented with shocks and weighting by $s_{nt}$. The \emph{F}-statistics
are well above the conventional threshold of ten in all columns of
the table.

Column 1 first replicates column 6 of Table 3 in \textcite{AutorDorn2001}
by including in $w_{\ell t}$ period fixed effects, Census division
fixed effects, start-of-period conditions (\% college educated, \%
foreign-born, \% employment among women, \% employment in routine
occupations, and the average offshorability index), and the start-of-period
manufacturing share. The point estimate is -0.596, with a corrected
standard error of 0.114.\footnote{Appendix Table C2 implements three alternative methods for conducting
inference in Table 4, reporting conventional state-clustered standard
errors as in ADH (which are not exposure-robust), the \textcite{Adao}
standard errors (which are asymptotically equivalent to ours but differ
in finite samples), and null-imposed confidence intervals obtained
from shock-level Lagrange multiplier tests (which may have better
finite-sample properties). Consistent with the theoretical discussion
in Appendix \ref{sec:appx_Inference}, the conventional standard errors
are generally too low, while the \textcite{Adao} standard errors
are slightly larger than those from Table 4 in most columns. Imposing
the null widens the confidence interval more substantially, by 30\textendash 50\%,
although more so on the left end, suggesting that much larger effects
are not rejected by the data. This last finding is consistent with
\textcite{Adao}, except that we use the equivalent industry-level
regression to compute the null-imposed confidence interval.\label{fn:adh-se-comparison}}

As noted, the ADH specification in column 1 does not include the lagged
manufacturing share control $S_{\ell t}$, which is necessary to solve
the incomplete shares issue in Section \ref{subsec:Incomplete-shares},
though it does include a highly correlated control (start-of-period
manufacturing share). In column 2 of Table 4 we isolate within-manufacturing
variation in shocks by replacing the latter sum-of-share control with
the former. The SSIV point estimate remains almost unchanged, at -0.489
(with a standard error of 0.100). Here exposure-robust standard errors
are obtained from an industry-level regression that drops the implicit
service sector shock of $g_{0t}=0$.

Isolating the within-period variation in manufacturing shocks requires
further controls in the incomplete shares case, as discussed in Section
\ref{subsec:Panels}. Specifically, column 3 controls for lagged manufacturing
shares interacted with period indicators, which are the share-weighted
sums of period effects in $q_{nt}$. This is equivalent to the use
of period fixed effects in the industry-level analysis of \textcite{AADHP2016}.
With these controls the SSIV point estimate is -0.267 with an exposure-robust
standard error of 0.099.\footnote{Appendix Figure C1 reports binned scatter plots that illustrate the
first-stage and reduced-form industry-level relationships corresponding
to the column 3 specification. This estimate can be interpreted as
a weighted average of two period-specific shift-share IV coefficients.
Column 1 of Appendix Table C3 shows the underlying estimates, from
a just-identified IV regression where both treatment and the instrument
are interacted with period indicators (as well as the manufacturing
share control, as in column 3), with exposure-robust standard errors
obtained by the equivalent industry-level regression discussed in
Section \ref{sec:InferenceDiagnostics}. The estimated effect of increased
Chinese import competition is negative in both periods (\textendash 0.491
and \textendash 0.225). Other columns repeat the analysis for other
outcomes.} While the coefficient remains statistically and economically significant,
it is smaller in magnitude than the estimates in columns 1 and 2.
The difference stems from the fact that 2000\textendash 07 saw both
a faster growth in imports from China (e.g., due to its entry to the
WTO) and a faster decline in U.S. manufacturing. The earlier columns
attribute the faster manufacturing decline to increased trade with
China, while the specification in Column 3 controls for any unobserved
shocks specific to the manufacturing sector overall in the 2000s (e.g.,
any demand or supply shock affecting the manufacturing sector, which
could include automation, innovation, falling consumer demand due
to income effects, etc.). Conventional industry-level IV regressions
control for such unobserved shocks with period fixed effects, as in
Table 3, column 1 of \textcite{AADHP2016}.\footnote{In principle, China could have affected the path of the U.S. manufacturing
sector as a whole, and thus the variation in the average China shock
across periods may be informative about the effects of interest. However,
because of the multiplicity of shocks that may affect the manufacturing
sector as a whole in a given period, this variation cannot be viewed
as a quasi-experimental source of variation for the impact of trade
with China on employment and other outcomes. This is why industry-level
studies of the China shock use period fixed effects, possibly reducing
power but substantially improving robustness of the estimates. In
the ADH application the estimation power is not actually reduced,
as the Table 4 column 3 standard error is even slightly smaller than
that in the previous columns.} The translation of their industry specification into the regional
setup of ADH requires interacting the lagged manufacturing shares
with period indicators, a simple but important insight of our framework.

Column 4 implements a simple sensitivity test to assess the stability
of the results when the controls from ADH are omitted. This test is
motivated by the result of the balance test in panel B of Table 3,
which indicated that the shift-share instrument was correlated with
the share of foreign born population. It is therefore instructive
to see whether the headline regression coefficient is sensitive to
the inclusion of this and other controls. In fact, we find that the
results remain very similar without controls, with a point estimate
of -0.314 and an exposure-robust standard error of 0.134. We proceed
by keeping the ADH controls for the remainder of the analysis.

Further columns of Table 4 parallel the specifications of \textcite[Table 3]{AADHP2016}
that include further industry-level controls. This illustrates how
our framework makes it straightforward to introduce more detailed
industry-level controls in SSIV, which are commonly used in industry-level
studies of the China shock. The validity of these estimates relies
on weaker versions of conditional random assignment (Assumption 3),
and robustness of the coefficients is therefore reassuring. Specifically,
\textcite{AADHP2016} control for fixed effects of ten broad industry
groups (one-digit manufacturing sectors) in column 2 of their Table
3. By Proposition 4, we can exploit shock variation within these industry
groups in the SSIV design by controlling for the lagged shares of
exposure to these industry groups (and including fixed effects of
these groups in the equivalent industry regressions for correct inference,
per Section \ref{subsec:inference}). The resulting point estimate
in column 5 of Table 4 remains very similar to that of column 3, at
-0.310 with a standard error of 0.134.

Column 6 instead parallels the specification of \textcite{AADHP2016}
that includes production controls, which we used for the balance tests
in Panel A of Table 3. This is done by controlling for the regional
share-weighted sums of those controls. The results remain virtually
unchanged, with a regression coefficient of -0.293 and an exposure-robust
standard error of 0.125.

Finally, column 7 instead introduces industry fixed effects, again
following \textcite{AADHP2016}. This specification is more ambitious
because it isolates changes in trade with China within each four-digit
SIC industry, across the two periods. To translate the industry fixed
effects to the location-level setup, we control for the lagged location-specific
share of exposure to each industry.\footnote{If the shares used to construct the instrument were time-invariant,
a more conventional and intuitive way to exploit over-time variation
in the shocks would be by including the regional fixed effects in
the regression, as Section \ref{subsec:Panels} explained. In the
ADH setting where the shares vary over time, they need to be controlled
for directly.} The magnitude of the regression coefficient increases, to -0.432,
with an exposure-robust standard error of 0.205. Broadly, these results
demonstrate the stability of the SSIV regression coefficient under
alternative sets of controls, corresponding to different assumptions
of conditional quasi-random shock assignment.

The appendix reports estimates from additional specifications. Appendix
Table C4 includes additional controls corresponding to other specifications
of \textcite{AADHP2016}, Table 3: for example, controlling for observed
changes in employment in the pre-periods or combining multiple sets
of controls. The regression coefficients remain stable across all
specifications. Appendix Table C5 instead shows robustness of the
coefficients to using overidentified SSIV procedures (leveraging variation
in eight country-specific Chinese import growth, instead of the ADH
total), illustrating the theoretical results of Section \ref{subsec:Main-Multiple-shocks}.
The table also reports a \emph{p}-value for the shock-level overidentification
test of 0.142, providing further support to the identification assumptions.

\subsubsection{Discussion}

Taken together, the sensitivity, falsification, and overidentification
exercises suggest that the ADH approach can be reasonably viewed as
leveraging exogenous shock variation via our framework. This is notably
in contrast to the analysis of \textcite{GPSS}, who find the ADH
exposure shares to be implausible instruments via different balance
and overidentification tests. This contrast should perhaps come as
no surprise. As mentioned, the exogeneity of industry employment shares
is an \emph{ex ante} implausible research design, because it is invalidated
by any unobserved labor demand or supply shocks across industries
(which we view as an inherent feature of the economy).

In contrast, our approach relies on the exogeneity of the specific
ADH trade shocks, allowing for endogenous exposure shares. With this
view, the potential confounders are a more specific set of unobserved
industry shocks (namely, unobserved shocks that would correlate with
the ADH shocks), rather than any unobserved shocks. In principle,
the conditions for shock orthogonality could still fail because of
these specific unobserved shocks. In practice, our balance tests indicate
that there is little evidence to suggest that the ADH shocks are confounded.

Our ADH application therefore illustrates two points. First, the assumptions
of our framework are plausible, both \emph{ex ante} and \emph{ex post},
in an influential empirical setting, where an alternative SSIV framework
is inapplicable. Second, our framework helps researchers translate
shock-level identifying assumptions to appropriate SSIV regression
controls, falsify those assumptions with appropriate balance tests,
and perform correct inference.

\section{Conclusion}

Shift-share instruments combine a common set of observed shocks with
variation in shock exposure. In this paper, we provide a quasi-experimental
framework for the validity of such instruments based on identifying
variation in the shocks, allowing the exposure shares to be endogenous.
Our framework revolves around novel equivalence results: the orthogonality
between a shift-share instrument and an unobserved residual can be
represented as the orthogonality between the underlying shocks and
a shock-level unobservable, and SSIV regression coefficients can be
obtained from a transformed shock-level regression with shocks directly
used as an instrument. Shift-share instruments are therefore valid
when shocks are idiosyncratic with respect to an exposure-weighted
average of the unobserved factors determining the outcome variable,
and yield consistent IV estimates when the number of shocks is large
and they are sufficiently dispersed in terms of their average exposure.

Through various extensions and illustrations, we show how our quasi-experimental
SSIV framework can guide empirical work in practice. By controlling
for exposure-weighted averages of shock-level confounders, researchers
can isolate more plausibly exogenous variation in shocks, such as
over time or within narrow industry groups. By estimating SSIV coefficients,
placebo regressions, and first stage \emph{F}-statistics at the level
of shocks, researchers can conveniently perform exposure-robust inference
that accounts for the inherent non-standard clustering of observations
with common shock exposure. Our shock-level analysis also raises new
concerns: SSIV designs with few or insufficiently dispersed shocks
may have effectively small samples, despite there being many underlying
observations, and instruments constructed from exposure shares that
do not add up to a constant require appropriate controls in order
to isolate quasi-random shock variation. We illustrate these practical
implications in an application to the influential study of \textcite{AutorDorn2001}.

In sum, our analysis formalizes the claim that SSIV identification
and consistency may arise from the exogeneity of shocks, while providing
new guidance for SSIV estimation and inference that may be applied
across a number of economic fields, including international trade,
labor economics, urban economics, macroeconomics, and public finance.
Our shock-level assumptions connect SSIV in these settings to conventional
shock-level IV estimation, bringing shift-share instruments to more
familiar econometric territory and facilitating the assessment of
SSIV credibility in practice.

\newpage{}

\section*{Figures and Tables}

\vspace*{\fill}\addcontentsline{toc}{section}{Figures and Tables}
\begin{center}
Table 1: Shock Summary Statistics in the \textcite{AutorDorn2001}
Setting\vspace{2.5mm}
\par\end{center}

\begin{center}
{\small{}}%
\begin{tabular}{lccc}
\toprule 
 & {\small{}(1)} & {\small{}(2)} & {\small{}(3)}\tabularnewline
\midrule
\midrule 
{\small{}Mean} & {\small{}1.79} & {\small{}7.37} & {\small{}0}\tabularnewline
{\small{}Standard deviation} & {\small{}10.79} & {\small{}20.92} & {\small{}20.44}\tabularnewline
{\small{}Interquartile range} & {\small{}0} & {\small{}6.61} & {\small{}6.11}\tabularnewline\addlinespace[0.3cm]
{\small{}\uline{Specification}} &  &  & \tabularnewline
{\small{}Excluding service industries} &  & {\small{}$\checked$} & {\small{}$\checked$}\tabularnewline
{\small{}Residualizing on period FE} &  &  & {\small{}$\checked$}\tabularnewline\addlinespace[0.3cm]
\multicolumn{4}{l}{{\small{}\uline{Effective sample size (\mbox{$1/HHI$} of \mbox{$s_{nt}$}
weights)}}}\tabularnewline
{\small{}Across industries and periods} & {\small{}3.5} & {\small{}191.6} & {\small{}191.6}\tabularnewline
{\small{}Across SIC3 groups} & {\small{}1.7} & {\small{}58.4} & {\small{}58.4}\tabularnewline\addlinespace[0.3cm]
{\small{}\uline{Largest \mbox{$s_{nt}$} weight}} &  &  & \tabularnewline
{\small{}Across industries and periods} & {\small{}0.398} & {\small{}0.035} & {\small{}0.035}\tabularnewline
{\small{}Across SIC3 groups} & {\small{}0.757} & {\small{}0.066} & {\small{}0.066}\tabularnewline\addlinespace[0.3cm]
{\small{}\uline{Observation counts}} &  &  & \tabularnewline
{\small{}\# of industry-period shocks} & {\small{}796} & {\small{}794} & {\small{}794}\tabularnewline
{\small{}\# of industries} & {\small{}398} & {\small{}397} & {\small{}397}\tabularnewline
{\small{}\# of SIC3 groups} & {\small{}137} & {\small{}136} & {\small{}136}\tabularnewline
\bottomrule
\end{tabular}{\small\par}
\par\end{center}

\begin{singlespace}
\vspace*{-0.2cm}

{\footnotesize{}\noindent Notes: This table summarizes the distribution
of China import shocks $g_{nt}$ across industries $n$ and periods
$t$ in the \textcite{AutorDorn2001} application. Shocks are measured
as the total flow of imports from China in eight developed economics
outside of the United States. All statistics are weighted by the average
industry exposure shares $s_{nt}$; shares are measured from lagged
manufacturing employment, as described in Section \ref{subsec:ApplicationSetting}.
Column 1 includes the non-manufacturing industry aggregate in each
period with a shock of 0, while columns 2 and 3 restrict the sample
to manufacturing industries. Column 3 residualizes manufacturing shocks
on period indicators. We report the effective sample size (the inverse
renormalized Herfindahl index of the $s_{nt}$ weights, as described
in Section \ref{subsec:appl-Properties-of-shocks}) with and without
the non-manufacturing industry, at the industry-by-period level and
at the level of SIC3 groups (aggregated across periods), along with
the largest $s_{nt}$.}{\footnotesize\par}
\end{singlespace}

\vspace*{\fill}

\newpage{}
\begin{center}
\vspace*{\fill}
\par\end{center}

\begin{singlespace}
\begin{center}
Table 2: Shock Intra-Class Correlations in the \textcite{AutorDorn2001}
Setting\vspace{2.5mm}
\par\end{center}
\end{singlespace}

\begin{center}
{\small{}}%
\begin{tabular}{lcc}
\toprule 
 & {\small{}Estimate} & {\small{}SE}\tabularnewline
 & {\small{}(1)} & {\small{}(2)}\tabularnewline
\midrule
\midrule 
\multicolumn{3}{l}{{\small{}\uline{Shock ICCs}}}\tabularnewline
{\small{}10 sectors} & {\small{}0.016} & {\small{}(0.022)}\tabularnewline
{\small{}SIC2} & {\small{}0.047} & {\small{}(0.052)}\tabularnewline
{\small{}SIC3} & {\small{}0.073} & {\small{}(0.057)}\tabularnewline
{\small{}Industry} & {\small{}0.169} & {\small{}(0.047)}\tabularnewline\addlinespace[0.3cm]
\multicolumn{3}{l}{{\small{}\uline{Period means}}}\tabularnewline
{\small{}1990s} & {\small{}4.65} & {\small{}(1.38)}\tabularnewline
{\small{}2000s} & {\small{}16.87} & {\small{}(3.34)}\tabularnewline\addlinespace[0.3cm]
{\small{}\# of industry-periods} & \multicolumn{2}{c}{{\small{}794}}\tabularnewline
\bottomrule
\end{tabular}{\small\par}
\par\end{center}

\begin{singlespace}
\vspace*{-0.2cm}

{\footnotesize{}\noindent Notes: This table reports intra-class correlation
coefficients for the \textcite{AutorDorn2001} manufacturing shocks,
estimated from the hierarchical model described in Section \ref{subsec:appl-Properties-of-shocks}.
Estimates come from a maximum likelihood procedure with an exchangeable
covariance structure for each industry and sector random effect and
with period fixed effects. Robust standard errors are reported in
parentheses.}{\footnotesize\par}
\end{singlespace}

\vspace*{\fill}

\newpage{}
\begin{center}
\vspace*{\fill}
\par\end{center}

\begin{center}
Table 3: Shock Balance Tests in the \textcite{AutorDorn2001} Setting
\par\end{center}

\begin{center}
Panel A: Industry-Level Balance
\par\end{center}

\begin{center}
{\small{}}%
\begin{tabular}{llcc}
\toprule 
{\small{}Balance variable} &  & Coef. & {\small{}SE}\tabularnewline
\midrule
\midrule 
{\small{}Production workers\textquoteright{} share of employment,
1991} &  & {\small{}-0.011} & {\small{}(0.012)}\tabularnewline
{\small{}Ratio of capital to value-added, 1991} &  & {\small{}-0.007} & {\small{}(0.019)}\tabularnewline
{\small{}Log real wage (2007 USD), 1991} &  & {\small{}-0.005} & {\small{}(0.022)}\tabularnewline
{\small{}Computer investment as share of total, 1990} &  & {\small{}0.750} & {\small{}(0.465)}\tabularnewline
{\small{}High-tech equipment as share of total investment, 1990} &  & {\small{}0.532} & {\small{}(0.296)}\tabularnewline\addlinespace[0.3cm]
{\small{}\# of industry-periods} &  & \multicolumn{2}{c}{{\small{}794}}\tabularnewline
\bottomrule
\end{tabular}{\small\par}
\par\end{center}

\begin{center}
\vspace{1cm}
Panel B: Regional Balance
\par\end{center}

\begin{center}
{\small{}}%
\begin{tabular}{llcc}
\toprule 
{\small{}Balance variable} &  & {\small{}Coef.} & {\small{}SE}\tabularnewline
\midrule
\midrule 
{\small{}Start-of-period \% of college-educated population} &  & {\small{}0.915} & {\small{}(1.196)}\tabularnewline
{\small{}Start-of-period \% of foreign-born population} &  & {\small{}2.920} & {\small{}(0.952)}\tabularnewline
{\small{}Start-of-period \% of employment among women} &  & {\small{}-0.159} & {\small{}(0.521)}\tabularnewline
{\small{}Start-of-period \% of employment in routine occupations} &  & {\small{}-0.302} & {\small{}(0.272)}\tabularnewline
{\small{}Start-of-period average offshorability index of occupations} &  & {\small{}0.087} & {\small{}(0.075)}\tabularnewline\addlinespace[0.3cm]
{\small{}Manufacturing employment growth, 1970s} &  & {\small{}0.543} & {\small{}(0.227)}\tabularnewline
{\small{}Manufacturing employment growth, 1980s} &  & {\small{}0.055} & {\small{}(0.187)}\tabularnewline\addlinespace[0.3cm]
{\small{}\# of region-periods} &  & \multicolumn{2}{c}{{\small{}1,444}}\tabularnewline
\bottomrule
\end{tabular}{\small\par}
\par\end{center}

\vspace*{-0.5cm}

\begin{singlespace}
{\footnotesize{}\noindent Notes: Panel A of this table reports coefficients
from regressions of the industry-level covariates in \textcite{AADHP2016}
on the \textcite{AutorDorn2001} shocks, controlling for period indicators
and weighting by average industry exposure shares. Standard errors
are reported in parentheses and allow for clustering at the level
of three-digit SIC codes. Panel B reports coefficients from regressions
of commuting zone-level covariates and pre-trends from \textcite{AutorDorn2001}
on the shift-share instrument, controlling for period indicators interacted
with the lagged manufacturing share. Balance variables (the first
five rows of this panel) vary across the two periods, while pre-trends
(the last two rows) do not. SIC3-clustered exposure-robust standard
errors are reported in parentheses and obtained from equivalent industry-level
IV regressions as described in Section \ref{subsec:appl-balance}.
Independent variables in both panels are normalized to have a variance
of one in the sample.}{\footnotesize\par}
\end{singlespace}
\begin{center}
\vspace*{\fill}
\par\end{center}

\newpage{}
\begin{center}
\vspace*{\fill}
\par\end{center}

\begin{center}
Table 4: Shift-Share IV Estimates of the Effect of Chinese Imports
on Manufacturing Employment
\par\end{center}

\begin{center}
{\small{}}%
\begin{tabular}{lccccccc}
\toprule 
 & {\small{}(1)} & {\small{}(2)} & {\small{}(3)} & {\small{}(4)} & {\small{}(5)} & {\small{}(6)} & {\small{}(7)}\tabularnewline
\midrule
\midrule 
{\small{}Coefficient} & {\small{}-0.596} & {\small{}-0.489} & {\small{}-0.267} & {\small{}-0.314} & {\small{}-0.310} & {\small{}-0.290} & {\small{}-0.432}\tabularnewline
 & {\small{}(0.114)} & {\small{}(0.100)} & {\small{}(0.099)} & {\small{}(0.107)} & {\small{}(0.134)} & {\small{}(0.129)} & {\small{}(0.205)}\tabularnewline\addlinespace[0.3cm]
{\small{}\uline{Regional controls}} &  &  &  &  &  &  & \tabularnewline
{\small{}\textcite{AutorDorn2001} controls} & {\small{}$\checked$} & {\small{}$\checked$} & {\small{}$\checked$} &  & {\small{}$\checked$} & {\small{}$\checked$} & {\small{}$\checked$}\tabularnewline
{\small{}Start-of-period mfg. share} & {\small{}$\checked$} &  &  &  &  &  & \tabularnewline
{\small{}Lagged mfg. share} &  & {\small{}$\checked$} & {\small{}$\checked$} & {\small{}$\checked$} & {\small{}$\checked$} & {\small{}$\checked$} & {\small{}$\checked$}\tabularnewline
{\small{}Period-specific lagged mfg. share} &  &  & {\small{}$\checked$} & {\small{}$\checked$} & {\small{}$\checked$} & {\small{}$\checked$} & {\small{}$\checked$}\tabularnewline
{\small{}Lagged 10-sector shares} &  &  &  &  & {\small{}$\checked$} &  & {\small{}$\checked$}\tabularnewline
{\small{}Local \textcite{AADHP2016} controls} &  &  &  &  &  & {\small{}$\checked$} & \tabularnewline
{\small{}Lagged industry shares} &  &  &  &  &  &  & {\small{}$\checked$}\tabularnewline\addlinespace[0.3cm]
{\small{}SSIV first stage }\emph{\small{}F}{\small{}-stat.} & {\small{}185.6} & {\small{}166.7} & {\small{}123.6} & {\small{}272.4} & {\small{}64.6} & {\small{}63.3} & {\small{}27.6}\tabularnewline
{\small{}\# of region-periods} & {\small{}1,444} & {\small{}1,444} & {\small{}1,444} & {\small{}1,444} & {\small{}1,444} & {\small{}1,444} & {\small{}1,444}\tabularnewline
{\small{}\# of industry-periods} & {\small{}796} & {\small{}794} & {\small{}794} & {\small{}794} & {\small{}794} & {\small{}794} & {\small{}794}\tabularnewline
\bottomrule
\end{tabular}{\small\par}
\par\end{center}

\begin{singlespace}
\vspace*{-0.2cm}

{\footnotesize{}\noindent Notes: This table reports shift-share IV
coefficients from regressions of regional manufacturing employment
growth in the U.S. on the growth of import competition from China,
instrumented with predicted China import growth as described in Section
\ref{subsec:ApplicationSetting}. Column 1 replicates column 6 of
Table 3 in \textcite{AutorDorn2001} by controlling for period fixed
effects, Census division fixed effects, start-of-period conditions
(\% college educated, \% foreign-born, \% employment among women,
\% employment in routine occupations, and the average offshorability
index), and the start-of-period manufacturing share. Column 2 replaces
the start-of-period manufacturing shares control with the lagged manufacturing
shares underlying the instrument, while column 3 interacts this control
with period indicators. Column 4 removes the Census division fixed
effects and start-of-period covariates. Columns 5\textendash 7 instead
add exposure-weighted sums of industry controls from \textcite{AADHP2016}:
indicators of 10 industry sectors (column 5), production controls
(column 6), and indicators of 397 industries (column 7). Production
controls are: employment share of production workers, ratio of capital
to value-added, log real wage (all measured in 1991); and computer
investment as share of total and high-tech equipment as share of total
employment (both measured in 1990). Exposure-robust standard errors
(reported in parentheses) and first-stage }\emph{\footnotesize{}F}{\footnotesize{}-statistics
are obtained from equivalent industry-level IV regressions, as described
in the text, allowing for clustering of shocks at the level of three-digit
SIC codes. For commuting zone controls which have a shift-share structure
(all controls starting with the lagged manufacturing share), we include
the corresponding $q_{nt}$ controls in the industry-level IV regression.
The sample in columns 2\textendash 7 includes 722 locations (commuting
zones) and 397 industries, each observed in two periods; the estimate
in column 1 implicitly includes an additional two observations for
the non-manufacturing industry with a shock of zero in each period.}{\footnotesize\par}
\end{singlespace}
\begin{center}
\vspace*{\fill}
\par\end{center}

\begin{center}
\newpage{}
\par\end{center}

\begin{spacing}{1.1}
\printbibliography
\newpage{}
\end{spacing}

\appendix
\begin{center}
\textbf{\Large{}\uline{Appendix to ``Quasi-Experimental Shift-Share
Research Designs''}}\\
\par\end{center}

\begin{center}
{\large{}Kirill Borusyak, UCL}\\
{\large{}Peter Hull, University of Chicago}\\
{\large{}Xavier Jaravel, London School of Economics}\\
\emph{\large{}December 2020}{\large\par}
\par\end{center}

\tableofcontents{}

\pagebreak\addtocontents{toc}{\protect\setcounter{tocdepth}{2}}\newrefsection

\section{Appendix Results}

\subsection{Heterogeneous Treatment Effects\label{sec:hetFX}}

In this appendix we consider what a linear SSIV identifies when the
structural relationship between $y_{\ell}$ and $x_{\ell}$ is nonlinear.
We show that under a first-stage monotonicity condition the large-sample
SSIV coefficient estimates a convexly weighted average of heterogeneous
treatment effects. This holds even when the instrument has different
effects on the outcome depending on the underlying realization of
shocks, for example when $y_{\ell}=\sum_{n}s_{\ell n}\tilde{\beta}_{\ell n}x_{\ell n}+\varepsilon_{\ell}$
with $\tilde{\beta}_{\ell n}$ capturing the effects of (possibly
unobserved) observation- and shock-specific treatments $x_{\ell n}$
making up the observed $x_{\ell}=\sum_{n}s_{\ell n}x_{\ell n}$.

Consider a general structural outcome model of
\begin{equation}
y_{\ell}=y(x_{\ell1},\dots,x_{\ell R},\varepsilon_{\ell}),\label{eq:nonlinear_outcome}
\end{equation}
where the $R$ treatments are given by $x_{\ell r}=x_{r}(g,\eta_{\ell r})$
with $g$ collecting the vector of shocks $g_{n}$ and with $\eta_{\ell}=(\eta_{\ell1},\dots,\eta_{\ell R})$
capturing first-stage heterogeneity. We consider an IV regression
of $y_{\ell}$ on some aggregated treatment $x_{\ell}=\sum_{r}\alpha_{\ell r}x_{\ell r}$
with $\alpha_{\ell r}\ge0$. Note that this nests the case of a single
aggregate treatment ($R=1$ and $\alpha_{\ell1}=1$) with arbitrary
effect heterogeneity, as well as the special case above $(R=N$ and
$\alpha_{\ell r}=s_{\ell n}$). We abstract away from controls $w_{\ell}$
and assume each shock is as-good-as-randomly assigned (mean-zero and
mutually independent) conditional on the vector of second-stage unobservables
$\varepsilon_{\ell}$ and the matrices of first-stage unobservables
$\eta_{\ell r}$, exposure shares $s_{\ell n}$, importance weights
$e_{\ell}$, and aggregation weights $\alpha_{\ell r}$, collected
in $\mathcal{I}=\left\{ \varepsilon_{\ell},e_{\ell},\left\{ \eta_{\ell r},\alpha_{\ell r}\right\} _{r},\left\{ s_{\ell n}\right\} _{n}\right\} _{\ell}$.
This assumption is stronger than Assumption 3 but generally necessary
in a non-linear setting while still allowing for the endogeneity of
exposure shares. For further notational simplicity we assume that
$y(\cdot,\varepsilon_{\ell})$ and each $x_{r}(\cdot,\eta_{\ell r})$
are almost surely continuously differentiable, such that $\beta_{\ell r}(\cdot)=\frac{\partial}{\partial x_{r}}y(\cdot,\varepsilon_{\ell})$
captures the effect, for observation $\ell$, of marginally increasing
treatment $r$ on the outcome and $\pi_{\ell nr}(\cdot)=\frac{\partial}{\partial g_{n}}x_{r}(\cdot,\eta_{\ell r})$
captures the effect of marginally increasing the $n$th shock on the
$r$th treatment at $\ell$.

Under an appropriate law of large numbers, the shift-share IV estimator
approximates the IV estimand:
\begin{align}
\hat{\beta} & =\frac{\expec{\sum_{\ell}e_{\ell}z_{\ell}y_{\ell}}}{\expec{\sum_{\ell}e_{\ell}z_{\ell}x_{\ell}}}+o_{p}(1)=\frac{\sum_{\ell}\sum_{n}\expec{s_{\ell n}e_{\ell}g_{n}y_{\ell}}}{\sum_{\ell}\sum_{n}\sum_{r}\expec{s_{\ell n}e_{\ell}g_{n}\alpha_{\ell r}x_{\ell r}}}+o_{p}(1).\label{eq:hetfx_SSIV}
\end{align}
Given this, we have the following result:
\begin{lyxlist}{00.00.0000}
\item [{\textbf{Proposition}}] \textbf{A1} When $\pi_{\ell nr}([\breve{g};g_{-n}])\ge0$
almost surely for all $\breve{g}\in\mathbb{R}$, equation (\ref{eq:hetfx_SSIV})
can be written
\begin{align}
\hat{\beta} & =\frac{\sum_{\ell}\sum_{n}\sum_{r}\expec{\int_{-\infty}^{\infty}\tilde{\beta}_{\ell nr}(\breve{g})\omega_{\ell nr}(\breve{g})}d\gamma}{\sum_{\ell}\sum_{n}\sum_{r}\expec{{\it \int_{-\infty}^{\infty}}\omega_{\ell nr}(\breve{g})}d\gamma}+o_{p}(1),\label{eq:het_fx1}
\end{align}
where $\omega_{\ell nr}(\breve{g})\ge0$ almost surely and 
\begin{align}
\tilde{\beta}_{\ell nr}(\breve{g}) & =\frac{\beta_{\ell r}(x_{1}([\breve{g};g_{-n}],\eta_{\ell1}),\dots x_{R}([\breve{g};g_{-n}],\eta_{\ell R}))}{\alpha_{\ell r}}
\end{align}
is a rescaled treatment effect, evaluated at $(x_{1}([\breve{g};g_{-n}],\eta_{\ell1}),\dots x_{R}([\breve{g};g_{-n}],\eta_{\ell R})$
for $[\breve{g};g_{-n}]=(g_{1},\dots,g_{n-1},\breve{g},g_{n+1},\dots g_{N})^{\prime}$.
\end{lyxlist}
\begin{proof}
\noindent See Appendix \ref{sec:A1_pf}.
\end{proof}
This shows that in large samples $\hat{\beta}$ estimates a convex
average of rescaled treatment effects, $\tilde{\beta}_{\ell nr}(\breve{g})$,
when the first stage is monotone in each shock. Appendix \ref{sec:A1_pf}
shows that the weights $\omega_{\ell nr}(\breve{g})$ are proportional
to the first-stage effects $\pi_{\ell nr}([\breve{g};g_{-n}])$, exposure
shares $s_{\ell n}$, regression weights $e_{\ell}$, treatment aggregation
weights $\alpha_{\ell r}$, and a function of the shock distribution.
In the case without aggregation, i.e. $R=\alpha_{\ell r}=1$, there
is no rescaling in the $\tilde{\beta}_{\ell nr}(\breve{g})$. Equation
(\ref{eq:het_fx1}) then can be seen as generalizing the result of
\textcite{AGI2000}, on the identification of heterogeneous effects
of continuous treatments, to the continuous shift-share instrument
case. Intuition for the $\omega_{\ell nr}(\breve{g})$ weights follows
similarly from this connection. With aggregation\textemdash that is,
when the realization of shocks may have heterogeneous effects on $y_{\ell}$
holding the aggregated $x_{\ell}$ fixed\textemdash equation (\ref{eq:het_fx1})
shows that SSIV captures a convex average of treatment effects per
aggregated unit. Thus in the leading example of $y_{\ell}=\sum_{n}s_{\ell n}\tilde{\beta}_{\ell n}x_{\ell n}+\varepsilon_{\ell}$
and $x_{\ell}=\sum_{n}s_{\ell n}x_{\ell n}$, this result establishes
identification of a convex average of the $\tilde{\beta}_{\ell n}$.
In this way the result generalizes \textcite{Adao}, who establish
the identification of convex averages of rescaled treatment effects
in reduced form shift-share regressions.

\subsection{\label{subsec:appx-no-Unobserved-Shocks}Unobserved $n$-level Shocks
Violate Share Exogeneity}

In this appendix, we show that the assumption of SSIV share exogeneity
from \textcite{GPSS} is violated when there are unobserved shocks
$\nu_{n}$ that affect outcomes via the exposure shares $s_{\ell n}$,
i.e. when the residual has the structure

\begin{equation}
\varepsilon_{\ell}=\sum_{n}s_{\ell n}\nu_{n}+\check{\varepsilon}_{\ell}.\label{eq:shift-share-resid}
\end{equation}
We consider large-sample violations share exogeneity in terms of the
asymptotic non-ignorability of the $\bar{\varepsilon}_{n}$ terms
in the equivalent moment condition (\ref{eq:id_rewrite}). It is intuitive
that the cross-sectional dependence between $s_{\ell n}$ and $\varepsilon_{\ell}$
will not asymptotically vanish when $N$ is fixed (as in \textcite{GPSS})
and each $\nu_{n}$ shock contributes significantly to the residual,
causing $\bar{\varepsilon}_{n}\not\xrightarrow{p}0$ for some or all
$n$. We next prove this result and show that it generalizes to the
case of increasing $N$, where the contribution of each $\nu_{n}$
to the variation in $\varepsilon_{\ell}$ becomes small. The intuition
here is that the SSIV relevance condition generally requires individual
observations to be sufficiently concentrated in a small number of
shocks (see Section \ref{subsec:A1andA2}), and under this condition
the share exogeneity violations remain asymptotically non-ignorable
even as $N\rightarrow\infty$.

We define share endogeneity as non-vanishing $\var{\bar{\varepsilon}_{n}}$
at least for some $n$. This will tend to make the SSIV estimator
inconsistent, unless shocks are as-good-as-randomly assigned (Assumption
1), even if the importance weights of individual shocks, $s_{n}$,
converge to zero (Assumption 2). Here we treat $e_{\ell}$ and $s_{\ell n}$
as non-stochastic to show this result with simple notation.
\begin{lyxlist}{00.00.0000}
\item [{\textbf{Proposition}}] \textbf{A2 }Suppose condition (\ref{eq:shift-share-resid})
holds with the $\nu_{n}$ mean-zero and uncorrelated with the $\check{\varepsilon}_{\ell}$
and with each other, and with $\var{\nu_{n}}=\sigma_{n}^{2}\ge\sigma_{\nu}^{2}$
for a fixed $\sigma_{\nu}^{2}>0$. Also assume $H_{L}=\sum_{\ell}e_{\ell}\sum_{n}s_{\ell n}^{2}\stackrel{}{\to}\bar{H}>0$
such that first-stage relevance can be satisfied. Then there exists
a constant $\delta>0$ such that $\max_{n}\var{\bar{\varepsilon}_{n}}>\delta$
for sufficiently large $L$.
\end{lyxlist}
\begin{proof}
\noindent See Appendix \ref{sec:A2_pf}.
\end{proof}

\subsection{Comparing SSIV and Native Shock-Level Regression Estimands\label{subsec:abstract_model}}

In this appendix we illustrate economic differences between the estimands
of two regressions that researchers may consider: SSIV using outcome
and treatment observations $y_{\ell}$ and $x_{\ell}$ (which we show
in Proposition 1 are equivalent to certain shock-level IV regressions),
and more conventional shock-level IV regressions using ``native''
$y_{n}$ and $x_{n}$. These outcomes and treatments capture the same
economic concepts as the original $y_{\ell}$ and $x_{\ell}$, in
contrast to the constructed $\bar{y}_{n}$ and $\bar{x}_{n}$ discussed
in Section \ref{subsec:numerical-equiv}. In line with the labor supply
and other key SSIV examples, we will for concreteness refer to the
$\ell$ and $n$ as indexing regions and industries, respectively.
We consider the case where both the outcome and treatment can be naturally
defined at the level of region-by-industry cells (henceforth, cells)\textemdash $y_{\ell n}$
and $x_{\ell n}$, respectively\textemdash and thus suitable for aggregation
across either dimension with some weights $E_{\ell n}$ (e.g., cell
employment growth rates aggregated with lagged cell employment weights):
$y_{\ell}=\sum_{n}s_{\ell n}y_{\ell n}$ for $s_{\ell n}=\frac{E_{\ell n}}{\sum_{n'}E_{\ell n'}}$
and $y_{n}=\sum_{\ell}\omega_{\ell n}y_{\ell n}$ for $\omega_{\ell n}=\frac{E_{\ell n}}{\sum_{\ell'}E_{\ell'n}}$,
with analogous expressions for $x_{\ell}$ and $x_{n}$. We further
define $E_{\ell}=\sum_{n}E_{\ell n}$ and $E_{n}=\sum_{\ell}E_{\ell n}$.\footnote{This formulation nests reduced-form shift-share regressions when $x_{\ell n}=g_{n}$
for each $\ell$. The labor supply example of Section \ref{subsec:setup}
fits only partially in this formal setup because the industry or regional
wage growth $y_{n}$ is not equal to a weighted average of wage growth
across cells: reallocation of employment affects the average wage
growth even in the absence of wage changes in any given cell.}

We consider the estimands of two regression specifications: $\beta$
from the regional level model (\ref{eq:causalmodel}), instrumented
by $z_{\ell}$ and weighted by $e_{\ell}=E_{\ell}/E$ for $E=\sum_{\ell}E_{\ell}$,
and $\beta_{\text{ind}}$ from a simpler industry-level IV regression
of
\begin{equation}
y_{n}=\beta_{\text{ind}}x_{n}+\varepsilon_{n},\label{eq:reg_indlevel}
\end{equation}
instrumented by the industry shock $g_{n}$ and weighted by $s_{n}=E_{n}/E$.
For simplicity we do not include any controls in either specification
and implicitly condition on $\left\{ E_{\ell n}\right\} _{\ell,n}$
(and some other variables as described below), viewing them as non-stochastic.\footnote{Note that we thereby condition on the shares $s_{\ell n}$ and importance
weights $e_{\ell}$. Yet we still allow for share endogeneity by not
restricting $\expec{\varepsilon_{\ell n}}$ to be zero.}

We show that $\beta$ and $\beta_{\text{ind}}$ generally differ when
there are within-region spillover effects or when treatment effects
are heterogeneous. We study these cases in turn, maintaining several
assumptions: (i) a first stage relationship analogous to the one considered
in Section \ref{subsec:A1andA2}:
\begin{equation}
x_{\ell n}=\pi_{\ell n}g_{n}+\eta_{\ell n},
\end{equation}
for non-stochastic $\pi_{\ell n}\ge\bar{\pi}>0$, (ii) a stronger
version of our Assumption 1 that imposes $\expec{g_{n}}=\expec{g_{n}\varepsilon_{\ell n^{\prime}}}=\expec{g_{n}\eta_{\ell n^{\prime}}}=0$
for all $\ell$, $n$, and $n^{\prime}$, with $\varepsilon_{\ell n'}$
denoting the unobserved cell-level residual of each model, (iii) the
assumption that $g_{n}$ is uncorrelated with $g_{n^{\prime}}$ for
all $n$ and $n^{\prime}$, and (iv) that all appropriate laws of
large numbers hold.

\paragraph*{Within-Region Spillover Effects}

Suppose the structural model at the cell level is given by
\begin{equation}
y_{\ell n}=\beta_{0}x_{\ell n}-\beta_{1}\sum_{n^{\prime}}s_{\ell n^{\prime}}x_{\ell n^{\prime}}+\varepsilon_{\ell n}.\label{eq:spillover}
\end{equation}
Here $\beta_{0}$ captures the direct effect of the shock on the cell
outcome, and $\beta_{1}$ captures a within-region spillover effect.
The local employment effects of industry demand shocks from the model
in Appendix \ref{subsec:appx-labor-demand-model} fit in this framework,
see equation (\ref{eq:g_ln_model}).\footnote{In the labor supply example from the main text $y_{\ell n}$ is the
cell wage, which is equalized within the region, and $x_{\ell n}$
is cell employment. Equation (\ref{eq:spillover}) therefore holds
for $\beta_{0}=0$ and $-\beta_{1}$ being the inverse labor supply
elasticity.} The following proposition shows that the SSIV estimand $\beta$ captures
the effect of treatment net of spillovers (i.e. $\beta_{0}-\beta_{1}$),
whereas $\beta_{\text{ind}}$ subtracts the spillover only partially;
this is intuitive since the spillover effect is fully contained within
regions but not within industries.
\begin{lyxlist}{00.00.0000}
\item [{\textbf{Proposition}}] \textbf{A3} Suppose equation (\ref{eq:spillover})
holds and the average local concentration index $H_{L}=\sum_{\ell,n}e_{\ell}s_{\ell n}^{2}$
is bounded from below by a constant $\bar{H}_{L}>0$. Further assume
$\pi_{\ell n}=\bar{\pi}$ and $\var{g_{n}}=\sigma_{g}^{2}$ for all
$\ell$ and $n$. Then the SSIV estimator satisfies
\begin{equation}
\hat{\beta}=\beta_{0}-\beta_{1}+o_{p}(1)\label{eq:spillovers_SSIV}
\end{equation}
while the native industry-level IV estimator satisfies
\begin{equation}
\hat{\beta}_{\text{ind}}=\beta_{0}-\beta_{1}H_{L}+o_{p}(1),\label{eq:spillovers_IV}
\end{equation}
If $\beta_{1}\ne0$ (i.e. in presence of within-region spillovers),
$\hat{\beta}$ and $\hat{\beta}_{\text{ind}}$ asymptotically coincide
if and only if $H_{L}\stackrel{p}{\to}1$, which corresponds to the
case where the average region is asymptotically concentrated in one
industry.
\end{lyxlist}
\begin{proof}
\noindent See Appendix \ref{sec:A3_pf}.
\end{proof}

\paragraph*{Treatment Effect Heterogeneity}

Now consider a different structural model which allows for heterogeneity
in treatment effects:
\begin{equation}
y_{\ell n}=\beta_{\ell n}x_{\ell n}+\varepsilon_{\ell n}.\label{eq:het_model}
\end{equation}
We also allow the first-stage coefficients $\pi_{\ell n}$ and shock
variance $\sigma_{n}^{2}$ to vary. The following proposition shows
that $\beta$ and $\beta_{\text{ind}}$ differ in how they average
effect $\beta_{\ell n}$ (here treated as non-stochastic) across the
$(\ell,n)$ cells. The weights corresponding to the SSIV estimand
$\beta$ are relatively higher for cells that represent a larger fraction
of the regional economy. This follows because in the regional regression
$s_{\ell n}$ determines the cell's weight in both the outcome and
the shift-share instrument, while in the industry regression only
the former argument applies. Heterogeneity in the $\pi_{\ell n}$
and $\sigma_{n}^{2}$, in contrast, has equivalent effects on the
weighting scheme of both estimands.
\begin{lyxlist}{00.00.0000}
\item [{\textbf{Proposition}}] \textbf{A4 }In the casual model (\ref{eq:het_model}),
\begin{equation}
\hat{\beta}=\frac{\sum_{\ell,n}E_{\ell n}s_{\ell n}\pi_{\ell n}\sigma_{n}^{2}\cdot\beta_{\ell n}}{\sum_{\ell,n}E_{\ell n}s_{\ell n}\pi_{\ell n}\sigma_{n}^{2}}+o_{p}(1)
\end{equation}
and
\begin{equation}
\hat{\beta}_{\text{ind}}=\frac{\sum_{\ell,n}E_{\ell n}\pi_{\ell n}\sigma_{n}^{2}\cdot\beta_{\ell n}}{\sum_{\ell,n}E_{\ell n}\pi_{\ell n}\sigma_{n}^{2}}+o_{p}(1),
\end{equation}
\end{lyxlist}
\begin{proof}
\noindent See Appendix \ref{sec:A4_pf}.
\end{proof}

\subsection{\label{sec:appx-Rotemberg}Connection to Rotemberg Weights}

In this appendix we rewrite the decomposition of the SSIV coefficient
$\hat{\beta}$ from \textcite{GPSS} that gives rise to their ``Rotemberg
weight'' interpretation, and show that these weights measure the
leverage of shocks in our equivalent shock-level IV regression. We
then show that, in our framework, skewed Rotemberg weights do not
measure sensitivity to misspecification (of share exogeneity) and
do not pose a problem for SSIV consistency. We finally discuss the
implications of high-leverage observations for SSIV inference.

Proposition 1 implies the following decomposition:
\begin{align}
\hat{\beta} & =\frac{\sum_{n}s_{n}g_{n}\bar{y}_{n}^{\perp}}{\sum_{n}s_{n}g_{n}\bar{x}_{n}^{\perp}}=\sum_{n}\alpha_{n}\hat{\beta}_{n},\label{eq:Rot-decomp}
\end{align}
where 
\begin{equation}
\hat{\beta}_{n}=\frac{\bar{y}_{n}^{\perp}}{\bar{x}_{n}^{\perp}}=\frac{\sum_{\ell}e_{\ell}s_{\ell n}y_{\ell}^{\perp}}{\sum_{\ell}e_{\ell}s_{\ell n}x_{\ell}^{\perp}}
\end{equation}
and
\begin{equation}
\alpha_{n}=\frac{s_{n}g_{n}\bar{x}_{n}^{\perp}}{\sum_{n'}s_{n'}g_{n'}\bar{x}_{n'}^{\perp}}.
\end{equation}
This is a shock-level version of the decomposition discussed in \textcite{GPSS}:
$\hat{\beta}_{n}$ is the IV estimate of $\beta$ that uses share
$s_{\ell n}$ as the instrument, and $\alpha_{n}$ is the so-called
Rotemberg weight.

To see the connection with leverage (defined, typically in the context
of OLS, as the derivative of each observation's fitted value with
respect to its outcome) in our equivalent IV regression, note that
\begin{equation}
\frac{\partial\left(\bar{x}_{n}^{\perp}\hat{\beta}\right)}{\partial\bar{y}_{n}^{\perp}}=\bar{x}_{n}^{\perp}\frac{s_{n}g_{n}}{\sum_{n'}s_{n'}g_{n'}\bar{x}_{n'}^{\perp}}=\alpha_{n}.
\end{equation}
In this way, $\alpha_{n}$ measures the sensitivity of $\hat{\beta}$
to $\hat{\beta}_{n}$.

In the preferred interpretation of \textcite{GPSS}, exposure to each
shock is a valid instrument such that $\hat{\beta}_{n}\stackrel{p}{\to}\beta$
for each $n$. However, in our framework deviations of $\hat{\beta}_{n}$
from $\beta$ reflect nonzero $\bar{\varepsilon}_{n}$ in large samples,
and such share endogeneity is not ruled out; thus $\alpha_{n}$ does
not have the same sensitivity-to-misspecification interpretation.
Moreover, a high leverage of certain shocks (``skewed Rotemberg weights,''
in the language of \textcite{GPSS}) is not a problem for consistency
in our framework, provided it results from a heavy-tailed and high-variance
distribution of shocks (that still satisfies our regularity conditions,
such as finite shock variance), and each $s_{n}$ is small as required
by Assumption 2.

Nevertheless, skewed $\alpha_{n}$ may cause issues with SSIV inference,
as would high leverage observations in any regression. In general,
the estimated residuals $\hat{\bar{\varepsilon}}_{n}^{\perp}$ of
high-leverage observations will tend to be biased toward zero, which
may lead to underestimation of the residual variance and too small
standard errors \parencite[e.g.,][]{ColinCameron2015}. This issue
can be addressed, for instance, by computing confidence intervals
with the null imposed, as \textcite{Adao} recommend and as we discuss
in Section \ref{subsec:inference}. In practice our Monte-Carlo simulations
in Appendix \ref{subsec:Finite-Sample-Performance} find that the
coverage of conventional exposure-robust confidence intervals to be
satisfactory even with Rotemberg weights as skewed as those reported
in the applications of \textcite{GPSS} analysis.

\subsection{Consistency of Control Coefficients\label{sec:controlconsistency}}

This appendix shows how the control coefficient $\gamma$, defined
in footnote \ref{fn:model_derivation}, can be consistently estimated
as required in Proposition 3 (Assumption B2). We discuss conditions
for $\sum_{\ell}e_{\ell}w_{\ell}\varepsilon_{\ell}\xrightarrow{p}\expec{\sum_{\ell}e_{\ell}w_{\ell}\varepsilon_{\ell}}$,
where by definition $\expec{\sum_{\ell}e_{\ell}w_{\ell}\varepsilon_{\ell}}=0$.
Consistency of the estimator $\hat{\gamma}=\gamma+\left(\sum_{\ell}e_{\ell}w_{\ell}w_{\ell}^{\prime}\right){}^{-1}\sum_{\ell}e_{\ell}w_{\ell}\varepsilon_{\ell}$
follows, provided the elements of $\left(\sum_{\ell}e_{\ell}w_{\ell}w_{\ell}^{\prime}\right)^{-1}$
are stochastically bounded (i.e., $O_{p}(1)$). For simplicity we
consider control vectors $w_{\ell}$ of fixed length.

The argument for convergence of $\sum_{\ell}e_{\ell}w_{\ell}\varepsilon_{\ell}$
depends on the source of randomness in $w_{\ell}$ and $\varepsilon_{\ell}$.
We consider two characteristic cases. In the first case, $(e_{\ell},w_{\ell}^{\prime},\varepsilon_{\ell})^{\prime}$
can be viewed as \emph{iid} or clustered in a conventional way. For
example, $w_{\ell}$ and $\varepsilon_{\ell}$ may contain observed
and unobserved local labor supply shocks which are uncorrelated across
markets, clusters of markets (e.g. states), or beyond a given distance
threshold. In this case conventional laws of large numbers can be
used to establish $\sum_{\ell}e_{\ell}w_{\ell}\varepsilon_{\ell}\xrightarrow{p}0$.
For instance if $(e_{\ell},w_{\ell}^{\prime},\varepsilon_{\ell})^{\prime}$
is \emph{iid }then $\sum_{\ell}e_{\ell}w_{\ell}\varepsilon_{\ell}$
gives a vector of sample averages of mutually uncorrelated mean-zero
random variables, which weakly converge to zero when the $e_{\ell}$
weights are asymptotically dispersed ($\expec{\sum_{\ell}e_{\ell}^{2}}\rightarrow0$)
and when $\text{\ensuremath{\expec{w_{\ell}^{2}\varepsilon_{\ell}^{2}\mid e}}}$
is uniformly bounded.

In the second case, either $w_{\ell}$ or $\varepsilon_{\ell}$ has
a shift-share structure like $z_{\ell}$: i.e. $w_{\ell}=\sum_{n}s_{\ell n}q_{n}$
for an observed $q_{n}$ (in line with our Proposition 3) or $\varepsilon_{\ell}=\sum_{n}s_{\ell n}\nu_{n}$
for an unobserved $\nu_{n}$ (capturing, for example, a set of unobserved
industry-level factors averaged with the employment weights $s_{\ell n}$).
In this case convergence of $\sum_{\ell}e_{\ell}w_{\ell}\varepsilon_{\ell}$
can be shown to follow similarly to the convergence of the sample
analog of the instrument moment condition (\ref{eq:identification}).
If, for instance, $\varepsilon_{\ell}=\sum_{n}s_{\ell n}\nu_{n}$
with $\expec{\nu_{n}\mid s,w}=0$ and $\cov{\nu_{n},\nu_{m}\mid s,w}=0$
for $w=\left\{ w_{n}\right\} _{n}$, then for each control $\sum_{\ell}e_{\ell}w_{\ell m}\varepsilon_{\ell}=\sum_{n}s_{n}\nu_{n}\bar{w}_{nm}$
weakly converges when the $s_{n}$ weights are dispersed ($\expec{\sum_{n}s_{n}^{2}}\rightarrow0$)
and both $\var{\nu_{n}\mid s,w}$ and $\expec{\bar{w}_{nm}^{2}\mid s}$
are uniformly bounded. This argument can be extended to the case where
either $w_{\ell}$ or $\varepsilon_{\ell}$ is formed from different
exposure shares $\tilde{s}_{\ell k}$, perhaps defined over a different
range of $K$ observed $q_{k}$ or unobserved $\nu_{k}$, and when
the $q_{k}$ or $\nu_{k}$ are clustered or otherwise weakly mutually
correlated.

More generally, the two cases can be combined to settings where $w_{\ell}=\sum_{k}\tilde{s}_{\ell k}q_{k}+\check{w}_{\ell}$
and $\varepsilon_{\ell}=\sum_{k^{\prime}}\tilde{s}_{\ell k^{\prime}}\nu_{k^{\prime}}+\check{\varepsilon}_{\ell}$
where $(e_{\ell},\check{w}_{\ell}^{\prime},\check{\varepsilon}_{\ell})^{\prime}$
is \emph{iid }or conventionally clustered and where $q_{k}$ and $\nu_{k^{\prime}}$
are many weakly correlated random shocks or, even more generally,
allowing for multiple shift-share terms with different exposure shares.

\subsection{Estimated Shocks\label{subsec:appx-Estimated_shocks}}

This appendix establishes the formal conditions for the SSIV estimator,
with or without a leave-one-out correction, to be consistent when
shocks $g_{n}$ are noisy estimates of some latent $g_{n}^{\ast}$
satisfying Assumptions 1 and 2. We also propose a heuristic measure
that indicates whether the leave-one-out correction is likely to be
important and compute it for the \textcite{Bartik1991} setting. Straightforward
extensions to other split-sample estimators follow.

Suppose a researcher estimates shocks via a weighted average of variables
$g_{\ell n}$. That is, given weights $\omega_{\ell n}\ge0$ such
that $\sum_{\ell}\omega_{\ell n}=1$ for all $n$, she computes
\begin{equation}
g_{n}=\sum_{\ell}\omega_{\ell n}g_{\ell n}.\label{eq:g_hat-1-1}
\end{equation}
A leave-one-out (LOO) version of the shock estimator is instead
\begin{equation}
g_{n,-\ell}=\frac{\sum_{\ell'\ne\ell}\omega_{\ell'n}g_{\ell'n}}{\sum_{\ell'\ne\ell}\omega_{\ell'n}}.
\end{equation}
We assume that each $g_{\ell n}$ is a noisy version of the same latent
shock $g_{n}^{\ast}$:
\begin{equation}
g_{\ell n}=g_{n}^{\ast}+\psi_{\ell n},\label{eq:g_gstar}
\end{equation}
where $g_{n}^{\ast}$ satisfies Assumptions 1 and 2 and $\psi_{\ell n}$
is estimation error (in Section \ref{subsec:Main-estimated-shocks}
we considered the special case of $\psi_{\ell n}\propto\varepsilon_{\ell}$).
This implies a feasible shift-share instrument of $z_{\ell}=z_{\ell}^{\ast}+\psi_{\ell}$
and its LOO version $z_{\ell}^{LOO}=z_{\ell}^{\ast}+\psi_{\ell}^{LOO}$,
where $z_{\ell}^{\ast}=\sum_{n}s_{\ell n}g_{n}^{\ast}$, $\psi_{\ell}=\sum_{n}s_{\ell n}\sum_{\ell'}\omega_{\ell'n}\psi_{\ell'n}$,
and $\psi_{\ell}^{LOO}=\sum_{n}s_{\ell n}\frac{\sum_{\ell'\ne\ell}\omega_{\ell'n}\psi_{\ell'n}}{\sum_{\ell'\ne\ell}\omega_{\ell'n}}$.
Consistency with these instruments, given a first stage, requires
that $\sum_{\ell}e_{\ell}\varepsilon_{\ell}\psi_{\ell}\stackrel{p}{\to}0$
and $\sum_{\ell}e_{\ell}\varepsilon_{\ell}\psi_{\ell}^{LOO}\stackrel{p}{\to}0$
respectively.

We now present three sets of results. First, we establish a simple
sufficient condition under which the LOO instrument satisfies $\sum_{\ell}e_{\ell}\varepsilon_{\ell}\psi_{\ell}^{LOO}\stackrel{p}{\to}0$.
We also propose stronger conditions that guarantee consistency of
LOO-SSIV. Second, we explore the conditions under which the covariance
between $\varepsilon_{\ell}$ and $\psi_{\ell n}$ is ignorable, i.e.
asymptotically does not lead to a ``mechanical'' bias of the conventional
non-leave-one-out estimator. We propose a heuristic measure that is
large when the bias is likely to be small. Lastly, we apply these
ideas to the setting of \textcite{Bartik1991} using the data from
\textcite{GPSS}. In line with previous appendices, we condition on
$s_{\ell n}$, $\omega_{\ell n}$, and $e_{\ell}$ and treat them
as non-stochastic for notational convenience. We also assume the SSIV
regressions are estimated without controls $w_{\ell}$.

\paragraph{LOO Identification and Consistency}

The following proposition establishes three results. The first is
the most important one, providing the condition for orthogonality
to hold . The second strengthens this condition so that the estimator
converges, which naturally requires that most shocks are estimated
with sufficient amount of data. A tractable case of complete specialization
is considered in last part, where there should be many more observations
than shocks.
\begin{lyxlist}{00.00.0000}
\item [{\textbf{Proposition}}] \noindent \textbf{A5}
\end{lyxlist}
\begin{enumerate}
\item If $\expec{\varepsilon_{\ell}\psi_{\ell'n}}=0$ for all $\ell\ne\ell'$
and $n$, then $\expec{\sum_{\ell}e_{\ell}\varepsilon_{\ell}\psi_{\ell,LOO}}=0$.
\item If $\expec{\left(\varepsilon_{\ell},\psi_{\ell n}\right)\mid\left\{ \left(\varepsilon_{\ell^{\prime}},\psi_{\ell^{\prime}n^{\prime}}\right)\right\} _{\ell^{\prime}\ne\ell,n^{\prime}}}=0$
for all $\ell$ and $n$, then the LOO estimator is consistent, provided
it has a first stage and two regularity conditions hold: $\expec{\left|\varepsilon_{\ell_{1}}\varepsilon_{\ell_{2}}\psi_{\ell_{1}^{\prime}n_{1}}\psi_{\ell_{2}^{\prime}n_{2}}\right|}\le B$
for a constant $B$ and all $\left(\ell_{1},\ell_{2},\ell_{1}^{\prime},\ell_{2}^{\prime},n_{1},n_{2}\right)$
and
\begin{equation}
\sum_{\substack{\left(\ell_{1},\ell_{2},\ell_{1}^{\prime},\ell_{2}^{\prime}\right)\in\mathcal{J},\\
n_{1},n_{2}
}
}e_{\ell_{1}}e_{\ell_{2}}s_{\ell_{1}n_{1}}s_{\ell_{2}n_{2}}\frac{\omega_{\ell_{1}^{\prime}n_{1}}}{\sum_{\ell\ne\ell_{1}}\omega_{\ell n_{1}}}\frac{\omega_{\ell_{2}^{\prime}n_{2}}}{\sum_{\ell\ne\ell_{2}}\omega_{\ell n_{2}}}\to0,\label{eq:LOO_sum}
\end{equation}
with $\mathcal{J}$ denoting the set of tuples $\left(\ell_{1},\ell_{2},\ell_{1}^{\prime},\ell_{2}^{\prime}\right)$
for which one of the two conditions hold: (i) $\ell_{1}=\ell_{2}$
and $\ell_{1}^{\prime}=\ell_{2}^{\prime}\ne\ell_{1}$, (ii) $\ell_{1}=\ell_{2}^{\prime}$
and $\ell_{2}=\ell_{1}^{\prime}\ne\ell_{1}$.
\item Condition (\ref{eq:LOO_sum}) is satisfied if $\frac{N}{L}\to0$ in
the special case where each region is specialized in one industry,
i.e. $s_{\ell n}=\mathbf{1}\left[n=n(\ell)\right]$ for some $n(\cdot)$,
there are no importance weights ($e_{\ell}=\frac{1}{L}$), and shocks
estimated by simple LOO averaging among observations exposed to a
given shock ($\omega_{\ell n}=\frac{1}{L_{n}}$ for $L_{n}=\sum_{\ell}\mathbf{1}\left[n(\ell)=n\right]$),
assuming further that $L_{n}\ge2$ for each $n$ so that the LOO estimator
is well-defined.
\end{enumerate}
\begin{proof}
\noindent See Appendix \ref{sec:A5_pf}.
\end{proof}
The condition in the first part of Proposition A5 would be quite innocuous
in random samples of $\ell$ \textendash{} the environment in which
leave-one-out adjustments are often considered (e.g. \textcite{JIVE})
\textendash{} but is strong without random sampling. It requires $\varepsilon_{\ell}$
and $\psi_{\ell'n}$ to be uncorrelated for $\ell^{\prime}\ne\ell$,
which may easily be violated when both $\ell$ and $\ell^{\prime}$
are exposed to the same shocks\textemdash a situation in which excluding
own observation is not sufficient. Moreover, since we have conditioned
on the exposure shares throughout, $\expec{\varepsilon_{\ell}\psi_{\ell'n}}=0$
generally requires either $\varepsilon_{\ell}$ or $\psi_{\ell'n}$
to have a zero \emph{conditional} mean\textemdash the share exogeneity
assumption applied to either the residuals or the estimation error.
At the same time, this condition does not require $\expec{\varepsilon_{\ell}\psi_{\ell'n}}=0$
for $\ell=\ell^{\prime}$, which reflects the benefit of LOO: eliminating
the mechanical bias from the residual directly entering shock estimates.

\paragraph{Heuristic for Importance of LOO Correction}

We now return to the non-LOO SSIV estimator. As in Proposition A5,
we assume that $\expec{\varepsilon_{\ell}\psi_{\ell^{\prime}n}}=0$
for $\ell^{\prime}\ne\ell$ and all $n$, so the LOO estimator is
consistent under the additional regularity conditions. We also assume,
without loss of generality, that $z_{\ell}$ is mean-zero. Then the
``mechanical bias'' mentioned in Section \ref{subsec:Main-estimated-shocks}
is the only potential problem: under appropriate regularity conditions
(similar to those in part 2 of Proposition A5),
\begin{align}
\hat{\beta}-\beta & =\frac{\expec{\sum_{\ell}e_{\ell}\varepsilon_{\ell}\psi_{\ell}}}{\expec{\sum_{\ell}e_{\ell}z_{\ell}x_{\ell}}}+o_{p}(1)\nonumber \\
 & =\frac{\sum_{\ell,n}e_{\ell}s_{\ell n}\omega_{\ell n}\expec{\varepsilon_{\ell}\psi_{\ell n}}}{\expec{\sum_{\ell}e_{\ell}z_{\ell}x_{\ell}}}+o_{p}(1).\label{eq:non-loo-bias}
\end{align}
With $\left|\expec{\varepsilon_{\ell}\psi_{\ell n}}\right|$ bounded
by some $B_{1}>0$ for all $\ell$ and $n$, the numerator of (\ref{eq:non-loo-bias})
is bounded by $H_{N}B_{1}$, for an observable composite of the relevant
shares $H_{N}=\sum_{\ell,n}e_{\ell}s_{\ell n}\omega_{\ell n}$. The
structure of the shares also influences the strength of the first
stage in the denominator. Imposing our standard model of the first
stage from Section \ref{subsec:A1andA2} (but specified based on the
latent shock $g_{n}^{\ast}$), i.e. $x_{\ell}=\sum_{n}s_{\ell n}x_{\ell n}$
for $x_{\ell n}=\pi_{\ell n}g_{n}^{\ast}+\eta_{\ell n}$, $\eta_{\ell n}$
mean-zero and uncorrelated with $g_{n^{\prime}}^{\ast}$ for all $\ell,n,n^{\prime}$,
$\var{g_{n}^{\ast}}\ge\bar{\sigma}_{g}^{2}>0$ and $\pi_{\ell n}\ge\bar{\pi}>0$,
yields:
\begin{align}
\expec{\sum_{\ell}e_{\ell}z_{\ell}x_{\ell}} & =\sum_{\ell}e_{\ell}\expec{\left(\sum_{n}s_{\ell n}\left(g_{n}^{\ast}+\psi_{\ell n}\right)\right)\left(\sum_{n^{\prime}}s_{\ell n^{\prime}}\left(\pi_{\ell n}g_{n^{\prime}}^{\ast}+\eta_{\ell n^{\prime}}\right)\right)}\nonumber \\
 & =\sum_{\ell,n}e_{\ell}s_{\ell n}^{2}\cdot\pi_{\ell n}\var{g_{n}^{\ast}}+\sum_{\ell}e_{\ell}\sum_{n,n^{\prime}}s_{\ell n}s_{\ell n^{\prime}}\expec{\psi_{\ell n}\left(\pi_{\ell n}g_{n^{\prime}}^{\ast}+\eta_{\ell n^{\prime}}\right)}.\label{eq:estimated_shock_FS}
\end{align}
Excepting knife-edge cases where the two terms in (\ref{eq:estimated_shock_FS})
cancel out, $\expec{\sum_{\ell}e_{\ell}z_{\ell}^{\perp}x_{\ell}}\not\to0$
provided $H_{L}=\sum_{\ell,n}e_{\ell}s_{\ell n}^{2}\ge\bar{H}$ for
some fixed $\bar{H}>0$.

We thus define the following heuristic:
\begin{equation}
H=\frac{H_{L}}{H_{N}}=\frac{\sum_{\ell,n}e_{\ell}s_{\ell n}^{2}}{\sum_{\ell,n}e_{\ell}s_{\ell n}\omega_{\ell n}}.\label{eq:heuristic}
\end{equation}
When $H$ is large, we expect the non-LOO SSIV estimator to be relatively
insensitive to the mechanical bias generated by the average covariance
between $\psi_{\ell n}$ and $\varepsilon_{\ell}$, and thus similar
to the LOO estimator.

We note an important special case. Suppose all weights are derived
from variable $E_{\ell n}$ (e.g. lagged employment level in region
$\ell$ and industry $n$) as $s_{\ell n}=\frac{E_{\ell n}}{E_{\ell}}$,
$\omega_{\ell n}=\frac{E_{\ell n}}{E_{n}}$, and $e_{\ell}=\frac{E_{\ell}}{E}$,
for $E_{\ell}=\sum_{n}E_{\ell n}$, $E_{n}=\sum_{\ell}E_{\ell n}$,
and $E=\sum_{\ell}E_{\ell}$. Then
\begin{align}
H_{N} & =\sum_{\ell,n}\frac{E_{\ell}}{E}\frac{E_{\ell n}}{E_{\ell}}\frac{E_{\ell n}}{E_{n}}=\sum_{\ell,n}\frac{E_{n}}{E}\left(\frac{E_{\ell n}}{E_{n}}\right)^{2}=\sum_{n}s_{n}\sum_{\ell}\omega_{\ell n}^{2},
\end{align}
where $s_{n}=\frac{E_{n}}{E}$ is the weight in our equivalent shock-level
regression. Therefore, $H_{N}$ is the weighted average across $n$
of $n$-specific Herfindahl concentration indices, while $H_{L}$
is the weighted average across $\ell$ of $\ell$-specific Herfindahl
indices. With $E_{\ell n}$ denoting lagged employment, $H$ is high
(and thus we expect the LOO correction to be unnecessary) when employment
is much more concentrated across industries in a typical region than
it is concentrated across regions for a typical industry.

The formula simplifies further with $E_{\ell n}=\one\left[n=n(\ell)\right]$
for all $\ell,n$, corresponding to the case of complete specialization
of observations in shocks with no regression or shock estimation weights,
as in part 3 of Proposition A5. In that case,
\begin{align}
H & =\frac{1}{\sum_{\ell}\frac{1}{L}\frac{1}{L_{n(\ell)}}}=\frac{1}{\frac{1}{L}\sum_{n}\sum_{\ell\colon n(\ell)=n}\frac{1}{L_{n}}}=\frac{L}{N}.
\end{align}
Our heuristic is therefore large when there are many observations
per estimated shock.\footnote{Here $1/H=N/L$ is proportional to the ``bias'' of the non-LOO estimator,
which is similar to how the finite-sample bias of conventional 2SLS
is proportional to the number of instruments over the sample size
(\cite{nagar1959}).}

\paragraph{Application to \textcite{Bartik1991}}

We finally apply our insights to the \textcite{Bartik1991} setting,
using the \textcite{GPSS} replication code and data. Table C6 reports
the results. Column 1 shows the estimates of the inverse local labor
supply elasticity using SSIV estimators with and without the LOO correction
and using population weights, replicating Table 3, column 2, of \textcite{GPSS}
except with employment on the left-hand side and wages on the right-hand
side.\footnote{\textcite{GPSS} estimate the inverse labor supply elasticity. By
properties of IV estimation, our coefficient is the inverse of theirs.} Column 2 repeats the analysis without the population weights.\footnote{Industry growth shocks in this column are the same as in Column 1,
again estimated with employment weights.} We find all estimates to range between 1.2 and 1.3, showing that
in practice for \textcite{Bartik1991} the LOO correction does not
play a substantial role.

This is however especially true without weights, where the LOO and
conventional SSIV estimators are 1.30 and 1.29, respectively. Our
heuristic provides an explanation: $H$ is almost 8 times bigger when
computed without weights. The intuition is that large commuting zones,
such as Los Angeles and New York, may constitute a substantial fraction
of employment in industries of their comparative advantage. This generates
a potential for the mechanical bias: labor supply shocks in those
regions affect shock estimates; this bias is avoided by LOO estimators.
However, the role of the largest commuting zones is only significant
in weighted regressions (by employment or, as in \textcite{GPSS},
population).

\subsection{Equilibrium Industry Growth in a Model of Local Labor Markets\label{subsec:appx-labor-demand-model}}

This appendix develops a simple model of regional labor supply and
demand, similar to the model in \textcite{Adao2018a}. Our goal is
to show how the national growth rate of industry employment can be
viewed as a noisy version of the national industry-specific labor
demand shocks, and how regional labor supply shocks (along with some
other terms) generate the ``estimation error.''

Consider an economy that consists of a set of $L$ regions. In each
region $\ell$ there is a prevailing wage $W_{\ell}$, and labor supply
has constant elasticity $\phi$:
\begin{equation}
E_{\ell}=M_{\ell}W_{\ell}^{\phi},\label{eq:labor-supply}
\end{equation}
where $E_{\ell}$ is total regional employment and $M_{\ell}$ is
the supply shifter that depends on the working-age population, the
outside option, and other factors. Labor demand in each industry $n$
is given by a constant-elasticity function
\begin{align}
E_{\ell n} & =A_{n}\xi_{\ell n}W_{\ell}^{-\sigma},\label{eq:labor-demand}
\end{align}
where $E_{\ell n}$ is employment, $A_{n}$ is the national industry
demand shifter, $\xi_{\ell n}$ is its idiosyncratic component, and
$\sigma$ is the elasticity of labor demand. The equilibrium is given
by
\begin{equation}
\sum_{n}E_{\ell n}=E_{\ell}.\label{eq:local-eqm}
\end{equation}

Now consider small changes in fundamentals $A_{n}$, $\xi_{\ell n}$
and $M_{\ell}$. We use log-linearization around the observed equilibrium
and employ the \textcite{Jones1965} hat algebra notation, with $\hat{v}$
denoting the relative change in $v$ between the equilibria. We then
establish:
\begin{lyxlist}{00.00.0000}
\item [{\textbf{Proposition}}] \textbf{A6 }After a set of small changes
to fundamentals, the national industry employment growth is characterized
by
\begin{equation}
g_{n}=\sum_{\ell}\omega_{\ell n}g_{\ell n},\label{eq:natl-empl-growth}
\end{equation}
for $\omega_{\ell n}=E_{\ell n}/\sum_{\ell'}E_{\ell'n}$ denoting
the share of region $\ell$ in industry employment, and the change
in region-by-industry employment $g_{\ell n}$ is characterized by
\begin{equation}
g_{\ell n}=g_{n}^{\ast}+\frac{\sigma}{\sigma+\phi}\varepsilon_{\ell}+\hat{\xi}_{\ell n}-\frac{\sigma}{\sigma+\phi}\sum_{n}s_{\ell n}\left(g_{n}^{\ast}+\hat{\xi}_{\ell n}\right),\label{eq:g_ln_model}
\end{equation}
where $g_{n}^{\ast}=\hat{A}_{n}$ is the national industry labor demand
shock, $\varepsilon_{\ell}=\hat{M}_{\ell}$ is the regional labor
supply shock, and $s_{\ell n}=E_{\ell n}/\sum_{n'}E_{\ell n'}$.
\end{lyxlist}
\begin{proof}
\noindent See Appendix \ref{sec:A6_pf}.
\end{proof}
The first term in (\ref{eq:g_ln_model}) justifies our interpretation
of the observed industry employment growth as a noisy estimate of
the latent labor demand shock $g_{n}^{\ast}$. The other terms constitute
the ``estimation error.'' The first of them is proportional to the
residual of the labor supply equation, $\varepsilon_{\ell}$; we have
previously established the conditions under which it may or may not
confound SSIV estimation. The other terms, that we abstracted away
from in Section \ref{subsec:Main-estimated-shocks}, include the idiosyncratic
demand shock $\hat{\xi}_{\ell n}$ and shift-share averages of both
national and idiosyncratic demand shocks. If the model is correct,
all of these are uncorrelated with $\varepsilon_{\ell}$, thus not
affecting Assumption 1.

\subsection{SSIV Consistency in Short Panels\label{sec:paneldata}}

This appendix shows how alternative shock exogeneity assumptions imply
the consistency of panel SSIV regressions with many fixed effect coefficients.
We consider the incidental parameters problem in ``short'' panels,
with fixed $T$ and $L\rightarrow\infty$ and with unit fixed effects,
in which case the control coefficient $\gamma$ cannot be consistently
estimated with the fixed effects included in $w_{\ell}$. We show
how an analog of Assumption 3 can be instead applied to a demeaned
shock-level unobservable that partials out the fixed effect nuisance
coefficients. A similar argument applies to period fixed effects in
the fixed $L$ and $T\rightarrow\infty$ asymptotic.

Suppose for the linear causal model $y_{\ell t}=\beta x_{\ell t}+\epsilon_{\ell t}$
and control vector $w_{\ell t}$ (which includes unit FEs), we define
$\gamma=\expec{\sum_{\ell}e_{\ell t}w_{\ell t}^{\Delta}w_{\ell t}^{\Delta\prime}}^{-1}\expec{\sum_{\ell}e_{\ell t}w_{\ell t}^{\Delta}\epsilon_{\ell t}^{\Delta}}$
where $v_{\ell t}^{\Delta}$ is a subvector of the (weighted) unit-demeaned
observation of variable $v_{\ell t}$, $v_{\ell t}-\frac{\sum_{\tau}e_{\ell\tau}v_{\ell\tau}}{\sum_{\tau}e_{\ell\tau}}$,
that drops any elements that are identically zero (e.g. those corresponding
to the unit FEs in $w_{\ell t}$). Note we have assumed no perfect
multicollinearity in the remaining elements such that $\expec{\sum_{\ell}e_{\ell t}w_{\ell t}^{\Delta}w_{\ell t}^{\Delta\prime}}$
is invertible. We can then write $y_{\ell t}^{\Delta}=\beta x_{\ell t}^{\Delta}+w{}_{\ell t}^{\Delta\prime}\gamma+\varepsilon_{\ell t}^{\Delta}$.
Suppose also that $\sum_{\ell}e_{\ell t}z_{\ell t}x_{\ell t}^{\perp}\xrightarrow{p}\pi$
for some $\pi\ne0$ and the analog of Assumption B2 for unit-demeaned
controls holds. Then, following the proof to Proposition 3, $\hat{\beta}$
is consistent if and only if\emph{
\begin{align}
\sum_{n,t}s_{nt}g_{nt}\bar{\varepsilon}_{nt}^{\Delta} & \xrightarrow{p}0,\label{eq:panel_orth}
\end{align}
}

\noindent where $s_{nt}=\sum_{\ell}e_{\ell t}s_{\ell nt}$ and $\bar{\varepsilon}_{nt}^{\Delta}=\frac{\sum_{\ell}e_{\ell t}s_{\ell nt}\varepsilon_{\ell t}^{\Delta}}{\sum_{\ell}e_{\ell t}s_{\ell nt}}$.
This condition is satisfied when analogs of Assumptions 1,2, and B1
hold, or under the various extensions discussed in Section \ref{sec:QE_assignment}.
In particular when $w_{\ell t}$ contains $t$-specific FE the key
assumption of quasi-experimental shock assignment is $\expec{g_{nt}\mid\bar{\varepsilon}^{\Delta},s}=\mu_{t}$,
for all $n$ and $t$, allowing endogenous period-specific shock means
$\mu_{t}$ via Proposition 4. This assumption avoids the incidental
parameters problem by considering shocks as-good-as-randomly assigned
given the set of unobserved $\bar{\varepsilon}_{nt}^{\Delta}$, each
of which is a function of the time-varying $\varepsilon_{\ell p}$
across all periods $p$.

An intuitive special case is when the exposure shares and importance
weights are time-invariant: $s_{\ell nt}=s_{\ell n0}$ and $e_{\ell t}=e_{\ell0}$.
Then the weights in (\ref{eq:panel_orth}) are also time-invariant,
$s_{nt}=s_{n0}$, and 
\begin{align}
\bar{\varepsilon}_{nt}^{\Delta} & =\frac{\sum_{\ell}e_{\ell0}s_{\ell n0}\varepsilon_{\ell t}^{\Delta}}{\sum_{\ell}e_{\ell}s_{\ell n0}}\nonumber \\
 & =\frac{\sum_{\ell}e_{\ell0}s_{\ell n0}\left(\varepsilon_{\ell t}-\frac{1}{T}\sum_{\tau}\varepsilon_{\ell\tau}\right)}{\sum_{\ell}e_{\ell0}s_{\ell n0}}\nonumber \\
 & =\bar{\varepsilon}_{nt}-\frac{1}{T}\sum_{\tau}\bar{\varepsilon}_{n\tau},
\end{align}
where $\bar{\varepsilon}_{nt}=\frac{\sum_{\ell}e_{\ell0}s_{\ell n0}\varepsilon_{\ell t}}{\sum_{\ell}e_{\ell0}s_{\ell n0}}$
is an aggregate of period-specific unobservables $\varepsilon_{\ell t}$.
It is then straightforward to extend Propositions 3 and 4 under a
shock-level assumption of strong exogeneity, i.e. that $\expec{g_{nt}\mid\bar{\varepsilon},s}=\mu_{n}+\zeta_{t}$
for all $n$ and $t$. Here endogenous $n$-specific shock means are
permitted by the observation in Section \ref{subsec:Panels}, that
share-weighted $n$-specific FEs at the shock level are subsumed by
$\ell$-specific FEs in the SSIV regression when shares and weights
are time-invariant.

\subsection{\label{subsec:SSIV-Relevance-panels}SSIV Relevance with Panel Data}

This appendix shows that holding the exposure shares fixed in a pre-period
is likely to weaken the SSIV first-stage in panel regressions. Consider
a panel extension of the first stage model used in Section \ref{subsec:A1andA2},
where $x_{\ell t}=\sum_{n}s_{\ell nt}x_{\ell nt}$ with $x_{\ell nt}=\pi_{\ell nt}g_{nt}+\eta_{\ell nt}$,
$\pi_{\ell n}\ge\bar{\pi}$ for some fixed $\bar{\pi}>0$, and the
$g_{nt}$ are mutually independent and mean-zero with variance $\sigma_{nt}^{2}\ge\bar{\sigma}_{g}^{2}$
for fixed $\sigma_{g}^{2}>0$, independently of $\left\{ \eta_{\ell nt}\right\} _{\ell,n,t}$.
As in other appendices, we here treat $s_{\ell nt}$, $e_{\ell t}$,
and $\pi_{\ell nt}$ as non-stochastic. Then an SSIV regression with
$z_{\ell t}=\sum_{n}s_{\ell nt}^{*}g_{nt}$ as an instrument, where
$s_{\ell nt}^{*}$ is either $s_{\ell nt}$ (updated shares) or $s_{\ell n0}$
(fixed shares), yields a first-stage of 
\begin{align}
\expec{\sum_{\ell}\sum_{t}e_{\ell t}z_{\ell t}x_{\ell t}} & \ge\bar{\sigma}_{g}^{2}\bar{\pi}\sum_{\ell}\sum_{t}e_{\ell t}\sum_{n}s_{\ell nt}^{*}s_{\ell tn}.
\end{align}
For panel SSIV relevance we require the $e_{\ell t}$-weighted average
of $\sum_{n}s_{\ell nt}^{*}s_{\ell nt}$ to not vanish asymptotically.
With updated shares this is satisfied when the Herfindahl index of
an average observation-period (across shocks) is non-vanishing, while
in the fixed shares case the overlap of shares in periods $0$ and
$t$, $\sum_{n}s_{\ell n0}s_{\ell nt}$, may become weak or even vanish
as $T\rightarrow\infty$, on average across observations.

\subsection{SSIV with Multiple Endogenous Variables or Instruments\label{subsec:appx-multiple_shocks}}

This appendix first generalizes our equivalence result to SSIV regressions
with multiple endogenous variables and instruments, and discusses
corresponding extensions of our quasi-experimental framework via the
setting of \textcite{Jaeger2017}. We also describe how to construct
the effective first-stage \emph{F}-statistic of \textcite{MontielOleaPflueger2013}
for SSIV with one endogenous variable but multiple instruments. We
then consider new shock-level IV procedures in this framework, which
can be used for efficient estimation and specification testing. Finally,
we illustrate these new procedures in the \textcite{AutorDorn2001}
setting.

\paragraph{Generalized Equivalence and SSIV Consistency}

We consider a class of SSIV estimators of an outcome model with multiple
treatment channels,
\begin{align}
y_{\ell} & =\beta^{\prime}x_{\ell}+\gamma^{\prime}w_{\ell}+\varepsilon_{\ell},\label{eq:two_endog}
\end{align}
where $x_{\ell}=(x_{1\ell},\dots,x_{K\ell})^{\prime}$ is instrumented
by $z_{\ell}=(z_{1\ell},\dots,z_{J\ell})^{\prime}$, for $z_{j\ell}=\sum_{n}s_{\ell n}g_{jn}$
and $J\ge K$, and observations are weighted by $e_{\ell}$. Members
of this class are parameterized by a (possibly stochastic) full-rank
$K\times J$ matrix $\boldsymbol{c}$, which is used to combine the
instruments into a vector of length $J$, $\boldsymbol{c}z_{\ell}$.
For example the two-stage least squares (2SLS) estimator sets $\boldsymbol{c}=\boldsymbol{x}^{\perp\prime}\boldsymbol{e}\boldsymbol{z}(\boldsymbol{z}^{\perp\prime}\boldsymbol{e}\boldsymbol{z}^{\perp})^{-1}$,
where $\boldsymbol{z}^{\perp}$ stacks observations of the residualized
$z_{\ell}^{\perp\prime}$. IV estimates using a given combination
are written as
\begin{align}
\hat{\beta} & =(\boldsymbol{c}\boldsymbol{z}^{\prime}\boldsymbol{e}\boldsymbol{x}^{\perp})^{-1}\boldsymbol{c}\boldsymbol{z}^{\prime}\boldsymbol{e}\boldsymbol{y}^{\perp},
\end{align}
where $\boldsymbol{y}^{\perp}$ and\textbf{ $\boldsymbol{x}^{\perp}$
}stack observations of the residualized $y_{\ell}^{\perp}$ and $x_{\ell}^{\perp\prime}$,
$\boldsymbol{z}$ stacks observations of $z_{\ell}^{\prime}$, and
$\boldsymbol{e}$ is an $L\times L$ diagonal matrix of $e_{\ell}$
weights. In just-identified IV models (i.e. $J=K$) the two $\boldsymbol{c}$'s
cancel in this expression and all IV estimators are equivalent. Note
that while the shocks $g_{jn}$ are different across the multiple
instruments, we assume here that the exposure shares $s_{\ell n}$
are all the same.

As in Proposition 1, $\hat{\beta}$ can be equivalently obtained by
a particular shock-level IV regression. Intuitively, when the shares
are the same, $\boldsymbol{c}z_{\ell}$ also has a shift-share structure
based on a linear combination of shocks $\boldsymbol{c}g_{n}$, and
thus Proposition 1 extends. Formally, write $\boldsymbol{z}=\boldsymbol{sg}$
where $\boldsymbol{s}$ is an $L\times N$ matrix of exposure shares
and $\boldsymbol{g}$ stacks observations of the shock vector $g_{n}^{\prime}$;
then,
\begin{align}
\hat{\beta} & =(\boldsymbol{c}\boldsymbol{g}^{\prime}\boldsymbol{s}^{\prime}\boldsymbol{e}\boldsymbol{x}^{\perp})^{-1}\boldsymbol{c}\boldsymbol{g}^{\prime}\boldsymbol{s}^{\prime}\boldsymbol{e}\boldsymbol{y}^{\perp}\nonumber \\
 & =(\boldsymbol{c}\boldsymbol{g}^{\prime}\boldsymbol{S}\bar{\boldsymbol{x}}^{\perp})^{-1}(\boldsymbol{c}\boldsymbol{g}^{\prime}\boldsymbol{S}\bar{\boldsymbol{y}}^{\perp}),\label{eq:two-EV}
\end{align}
where $\boldsymbol{S}$ is an $N\times N$ diagonal matrix with elements
$s_{n}$, $\bar{\boldsymbol{x}}^{\perp}$ is an $N\times K$ matrix
with elements $\bar{x}_{kn}^{\perp}$, and $\bar{\boldsymbol{y}}^{\perp}$
is an $N\times1$ vector of $\bar{y}_{n}^{\perp}$. This is the formula
for an $s_{n}$-weighted IV regression of $\bar{y}_{n}^{\perp}$ on
$\bar{x}_{1n}^{\perp},\dots,\bar{x}_{Kn}^{\perp}$ with shocks as
instruments, no constant, and the same $\boldsymbol{c}$ matrix. Furthermore,
as in Proposition 1, 
\begin{align}
\iota^{\prime}\boldsymbol{S}\bar{\boldsymbol{y}}^{\perp} & =\sum_{n}s_{n}\bar{y}_{n}^{\perp}=\sum_{\ell}e_{\ell}\left(\sum_{n}s_{\ell n}\right)y_{\ell}^{\perp}=\sum_{\ell}e_{\ell}y_{\ell}^{\perp}=0,
\end{align}
and similarly for $\iota^{\prime}\boldsymbol{S}\bar{\boldsymbol{x}}^{\prime}$,
where $\iota$ is a $N\times1$ vector of ones. Therefore, the same
estimate is obtained by including a constant in this IV procedure
(and the same result holds including a shock-level control vector
$q_{n}$ provided $\sum_{n}s_{\ell n}$ has been included in $w_{\ell}$,
as in Proposition 5). The \textbf{$\boldsymbol{c}$ }matrix is again
redundant in the just-identified case.

A natural generalization of the quasi-experimental framework of Section
\ref{sec:QE_assignment} follows. Rather than rederiving all of these
results, we discuss them intuitively in the setting of \textcite{Jaeger2017}.
Here $y_{\ell}$ denotes the growth rate of wages in region $\ell$
in a given period (residualized on Mincerian controls), $x_{1\ell}$
is the immigrant inflow rate in that period, and $x_{2\ell}$ is the
previous period's immigration rate. The residual $\varepsilon_{\ell}$
captures changes to local productivity and other regional unobservables.
\textcite[Table 5]{Jaeger2017} estimate this model with two ``past
settlement'' instruments $z_{1\ell}=\sum_{n}s_{\ell n}g_{1n}$ and
$z_{2\ell}=\sum_{n}s_{\ell n}g_{2n}$, where $s_{\ell n}$ is the
share of immigrants from country of origin $n$ in location $\ell$
at a previous reference date and $\boldsymbol{g}_{n}=(g_{1n},g_{2n})^{\prime}$
gives the current and previous period's national immigration rate
from $n$. When this path of immigration shocks is as-good-as-randomly
assigned with respect to the aggregated productivity shocks $\bar{\varepsilon}_{n}$
(satisfying a generalized Assumption 1), the\textbf{ $\boldsymbol{g}_{n}$}
are uncorrelated across countries and $\expec{\sum_{n}s_{n}^{2}}\rightarrow0$
(satisfying a generalized Assumption 2), and appropriately generalized
regularity conditions hold, the multiple-treatment shock orthogonality
condition is satisfied: $\sum_{n}s_{n}g_{kn}\bar{\varepsilon}_{n}\xrightarrow{p}0$
for each $k$. Then under the relevance condition from Proposition
2, again appropriately generalized, the SSIV estimates are consistent:
$\hat{\beta}\xrightarrow{p}\beta$.

\paragraph{Effective First-Stage \emph{F}-statistics}

With one endogenous variable and multiple instruments, the \textcite{MontielOleaPflueger2013}
effective first-stage \emph{F}-statistic provides a state-of-art heuristic
for detecting a weak first-stage. Here we describe a correction to
it for SSIV that generalizes the \emph{F}-statistic in the single
instrument case discussed in Section \ref{subsec:A1_A2_tests}. The
Stata command \emph{weakssivtest}, provided with our replication archive,
implements this correction.\footnote{Our package extends the \emph{weakivtest} command developed by \textcite{Pflueger2015}.}

Consider a structural first stage with multiple instruments and one
endogenous variable:
\begin{equation}
x_{\ell}=\pi^{\prime}z_{\ell}+\rho w_{\ell}+\eta_{\ell}.
\end{equation}
Suppose each of the shocks satisfies Assumption 3, i.e. $\expec{g_{jn}\mid\bar{\varepsilon},q,s}=\mu_{j}^{\prime}q_{n}$,
where $\sum_{n}s_{\ell n}q_{n}$ is included in $w_{\ell}$, and the
residual shocks $g_{jn}^{\ast}=g_{jn}-\mu_{j}^{\prime}q_{n}$ are
independent from $\left\{ \eta_{\ell}\right\} _{\ell}$. The \textcite{MontielOleaPflueger2013}
effective \emph{F}-statistic for the 2SLS regression of $y_{\ell}$
on $x_{\ell}$, instrumenting with $z_{1\ell},\dots,z_{J\ell}$, controlling
for $w_{\ell}$, and weighting by $e_{\ell}$, is given by
\begin{equation}
F_{\text{eff}}=\frac{\left(\sum_{\ell}e_{\ell}x_{\ell}^{\perp}z_{\ell}^{\perp}\right)^{\prime}\left(\sum_{\ell}e_{\ell}x_{\ell}^{\perp}z_{\ell}^{\perp}\right)}{\tr\left(\hat{V}\right)},\label{eq:Feff}
\end{equation}
where $\hat{V}$ estimates $V=\var{\sum_{\ell}e_{\ell}z_{\ell}^{\perp}\eta_{\ell}}$.
Note that, as before, the first-stage covariance of the original SSIV
regression equals that of the equivalent shock-level one from Proposition
5:
\begin{align}
\sum_{\ell}e_{\ell}x_{\ell}^{\perp}z_{\ell}^{\perp} & =\sum_{\ell}e_{\ell}x_{\ell}^{\perp}z_{\ell}=\sum_{n}s_{n}g_{n}\bar{x}_{n}^{\perp}=\sum_{n}s_{n}g_{n\perp}\bar{x}_{n}^{\perp},
\end{align}
where $g_{n\perp}$ is the residuals from an $s_{n}$-weighted projection
of $g_{n}$ on $q_{n}$, which consistently estimates $g_{n}^{\ast}$.
A natural extension of Proposition 5 to many mutually-uncorrelated
shocks further implies that $V$ is well-approximated by
\begin{equation}
\hat{V}=\sum_{n}s_{n}^{2}g_{n\perp}g_{n\perp}^{\prime}\bar{\hat{\eta}}_{n}^{2},\label{eq:MOP_variance}
\end{equation}
where, per the discussion in Section \ref{subsec:A1_A2_tests}, $\bar{\hat{\eta}}_{n}$
denotes the residuals from an IV regression of $\bar{x}_{n}^{\perp}$
on $\bar{z}_{1n}^{\perp},\dots,\bar{z}_{Jn}^{\perp}$, instrumented
with $g_{1n},\dots,g_{Jn}$, weighted by $s_{n}$ and controlling
for $q_{n}$. Plugging this $\hat{V}$ into (\ref{eq:Feff}) yields
the corrected effective first-stage \emph{F}-statistic.

\paragraph{Efficient Shift-Share GMM}

In overidentified settings ($J>K$), it is natural to consider which
estimators are most efficient; for quasi-experimental SSIV, this can
be answered by combining the asymptotic results of \textcite{Adao}
with the classic generalized methods of moments (GMM) theory of \textcite{hansen82}.
Here we show how standard shock-level IV procedures (such as 2SLS)
may yield efficient coefficient estimates $\hat{\beta}^{*}$, depending
on the variance structure of multiple quasi-randomly assigned shocks.

We first note that the equivalence result (\ref{eq:two-EV}) applies
to SSIV-GMM estimators as well:
\begin{align}
\hat{\beta} & =\arg\min_{b}\left(\boldsymbol{y}^{\perp}-\boldsymbol{x}^{\perp}b\right)^{\prime}\boldsymbol{e}\boldsymbol{z}\boldsymbol{W}\boldsymbol{z}^{\prime}\boldsymbol{e}\left(\boldsymbol{y}^{\perp}-\boldsymbol{x}^{\perp}b\right)\nonumber \\
 & =\arg\min_{b}\left(\bar{\boldsymbol{y}}^{\perp}-\bar{\boldsymbol{x}}^{\perp}b\right)^{\prime}\boldsymbol{S}\boldsymbol{g}\boldsymbol{W}\boldsymbol{g}^{\prime}\boldsymbol{S}\left(\bar{\boldsymbol{y}}^{\perp}-\bar{\boldsymbol{x}}^{\perp}b\right),\label{eq:ssiv_gmm}
\end{align}
where $\boldsymbol{W}$ is an $J\times J$ moment-weighting matrix.
This leads to an IV estimator with $\boldsymbol{c}=\bar{\boldsymbol{x}}^{\perp\prime}\boldsymbol{S}\boldsymbol{g}\boldsymbol{W}$.
For 2SLS estimation, for example, $\boldsymbol{W}=(\boldsymbol{z}^{\perp\prime}\boldsymbol{e}\boldsymbol{z}^{\perp})^{-1}$.
Under appropriate regularity conditions, the efficient choice of $\boldsymbol{W}^{*}$
consistently estimates the inverse asymptotic variance of $\boldsymbol{z}^{\prime}\boldsymbol{e}\left(\boldsymbol{y}^{\perp}-\boldsymbol{x}^{\perp}\beta\right)=\boldsymbol{g}^{\prime}S\bar{\varepsilon}+o_{p}(1)$.
Generalizations of results in \textcite{Adao} can then be used to
characterize this $\boldsymbol{W}^{*}$ when shocks are as-good-as-randomly
assigned with respect to $\bar{\varepsilon}$. Given an estimate $\hat{\boldsymbol{W}^{*}}$,
an efficient coefficient estimate $\hat{\beta}^{*}$ is given by shock-level
IV regressions (\ref{eq:two-EV}) that set $\boldsymbol{c}^{*}=\bar{\boldsymbol{x}}^{\perp\prime}\boldsymbol{S}\boldsymbol{g}\boldsymbol{\hat{W}}^{*}$.
A $\chi_{J-K}^{2}$ test statistic based on the minimized objective
in (\ref{eq:ssiv_gmm}) can be used for specification testing.

As an example, suppose shocks are conditionally homoskedastic with
the same variance-covariance matrix across $n$, $\var{\boldsymbol{g}_{n}\mid\bar{\varepsilon},s}=\boldsymbol{G}$
for a constant $J\times J$ matrix $\boldsymbol{G}$. Then the optimal
$\hat{\beta}^{*}$ is obtained by a shock-level 2SLS regression of
$\bar{y}_{n}^{\perp}$ on all $\bar{x}_{kn}^{\perp}$ (instrumented
by $g_{jn}$ and weighted by $s_{n}$). We show this in the case of
no controls (and mean-zero shocks) for notational simplicity. Then,
\begin{align}
\var{\boldsymbol{g}^{\prime}S\left(\bar{\boldsymbol{y}}^{\perp}-\bar{\boldsymbol{x}}^{\perp}\beta\right)} & =\expec{\bar{\varepsilon}^{\prime}\boldsymbol{S}\boldsymbol{g}\boldsymbol{g}^{\prime}\boldsymbol{S}\bar{\varepsilon}}\nonumber \\
 & =\tr\left(\expec{\bar{\varepsilon}^{\prime}\boldsymbol{S}\boldsymbol{G}\boldsymbol{S}\bar{\varepsilon}}\right)\nonumber \\
 & =k\boldsymbol{G}
\end{align}
for $k=tr\left(\expec{\boldsymbol{S}\bar{\varepsilon}\bar{\varepsilon}^{\prime}\boldsymbol{S}}\right)$.
The optimal weighting matrix thus should consistently estimate $\boldsymbol{G}$,
which is satisfied by $\hat{\boldsymbol{G}}=\boldsymbol{g}^{\prime}\boldsymbol{S}\boldsymbol{g}$.
Under appropriate regularity conditions, a feasible optimal GMM estimate
is thus given by 
\begin{align}
\hat{\beta}^{\ast} & =(\bar{\boldsymbol{x}}^{\perp\prime}\boldsymbol{S}\boldsymbol{g}\hat{\boldsymbol{G}}^{-1}\boldsymbol{g}^{\prime}\boldsymbol{S}\bar{\boldsymbol{x}}^{\perp})^{-1}(\bar{\boldsymbol{x}}^{\perp\prime}\boldsymbol{S}\boldsymbol{g}\hat{\boldsymbol{G}}^{-1}\boldsymbol{g}^{\prime}\boldsymbol{S}\bar{\boldsymbol{y}}^{\perp})\nonumber \\
 & =\left(\left(P_{\boldsymbol{g}}\bar{\boldsymbol{x}}^{\perp}\right)^{\prime}\boldsymbol{S}\bar{\boldsymbol{x}}^{\perp}\right)^{-1}\left(P_{\boldsymbol{g}}\bar{\boldsymbol{x}}^{\perp}\right)^{\prime}\boldsymbol{S}\bar{y}^{\perp},\label{eq:2sls}
\end{align}
where $P_{\boldsymbol{g}}=\boldsymbol{g}(\boldsymbol{g}^{\prime}\boldsymbol{S}\boldsymbol{g})^{-1}\boldsymbol{g}^{\prime}\boldsymbol{S}$
is an $s_{n}$-weighted shock projection matrix. This is the formula
for an $s_{n}$-weighted IV regression of $\bar{y}_{n}^{\perp}$ on
the fitted values from projecting the $\bar{x}_{kn}^{\perp}$ on the
shocks, corresponding to the 2SLS regression above. Straightforward
extensions of this equivalence between optimally-weighted estimates
of $\beta$ and shock-level overidentified IV procedures follow in
the case of heteroskedastic or clustered shocks, in which case the
2SLS estimator (\ref{eq:2sls}) is replaced by the estimator of \textcite{white82}.
We emphasize that these shock-level estimators are generally different
than 2SLS or \textcite{white82} estimators at the level of original
observations, which are optimal under conditional homoskedasticity
and independence\emph{ }assumptions placed on the residual $\varepsilon_{\ell}$
(assumptions which are generally violated in our quasi-experimental
framework).

\paragraph{Many Shocks in \textcite{AutorDorn2001}}

Appendix Table C5 illustrates different shock-level overidentified
IV estimators in the setting of \textcite{AutorDorn2001}, introduced
in Section \ref{subsec:ApplicationSetting}. ADH construct their shift-share
instrument based on the growth of Chinese imports in eight economies
comparable to the U.S., together. We separate them to produce eight
sets of industry shocks $g_{jn}$, $j=1,\dots,8$, each reflecting
the growth of Chinese imports in one of those countries. As in Section
\ref{subsec:Application}, the outcome of interest is a commuting
zone's growth in total manufacturing employment with the single treatment
variable measuring a commuting zone's local exposure to the growth
of imports from China (see footnote \ref{fn:ADH-details} for precise
variable definitions). The vector of controls coincides with that
of column 3 of Table 4, isolating within-period variation in manufacturing
shocks. Per Section \ref{subsec:inference}, exposure-robust standard
errors are obtained by controlling for period main effects in the
shock-level IV procedures, and we report corrected first stage \emph{F}-statistics
constructed as detailed above.

Column 1 reports estimates of the ADH coefficient $\beta$ using the
industry-level two-stage least squares procedure (\ref{eq:2sls}).
At -0.238, this estimate it is very similar to the just-identified
estimate in column 3 of Table 4. Column 2 shows that we also obtain
a very similar coefficient of -0.247 with an industry-level limited
information maximum likelihood (LIML) estimator. Finally, in column
3 we report a two-step optimal IV estimate of $\beta$ using an industry-level
implementation of the \textcite{white82} estimator. Both the coefficient
and standard error fall somewhat, with the latter consistent with
the theoretical improvement in efficiency relative to columns 1 and
2. From this efficient estimate we obtain an omnibus overidentification
test statistic of 10.92, distributed as chi-squared with seven degrees
of freedom under the null of correct specification. This yields a
\emph{p}-value for the test of joint orthogonality of all eight ADH
shocks of 0.142. Table C5 also reports the corrected effective first-stage
\emph{F}-statistic which measures the strength of the relationship
between the endogenous variable and the eight shift-share instruments
across regions. At 15.10 it is substantially lower than with one instrument
in column 3 of Table 4 but still above the conventional heuristic
threshold of 10.

\subsection{\label{subsec:Finite-Sample-Performance}Finite-Sample Performance
of SSIV: Monte-Carlo Evidence}

In this appendix we study the finite-sample performance of the SSIV
estimator via Monte-Carlo simulation. We base this simulation on the
data of \textcite{AutorDorn2001}, as described in Section \ref{subsec:Application}.
For comparison, we also simulate more conventional shock-level IV
estimators, similar to those used in \textcite{AADHP2016}, which
also estimate the effects of import competition with China on U.S.
employment. We begin by describing the design of these simulations
and the benchmark Monte-Carlo results. We then explore how the simulation
results change with various deviations from the benchmark: with different
levels of industry concentration, different numbers of industries
and regions, and with many shock instruments. Besides showing the
general robustness of our framework, these extensions allow us to
see how informative some conventional rules of thumb are on the finite-sample
performance of shift-share estimators.\footnote{Naturally, these simulation results may be specific to the data-generating
process we consider here, modeled after the ``China shock'' setting
of \textcite{AutorDorn2001}. In practice, we recommend that researchers
perform similar simulations based on their data if they are concerned
with the quality of asymptotic approximation\textemdash a suggestion
that of course applies to conventional shock-level IV analyses as
well.}

\paragraph*{Simulation design}

We base our benchmark data-generating process for SSIV on the specification
in column 3 of Table 4. The outcome variable $y_{\ell t}$ corresponds
to the change in manufacturing employment as a fraction of working-age
population of region $\ell$ in period $t$, treatment $x_{\ell t}$
is a measure of regional import competition with China, and the shift-share
instrument is constructed by combining the industry-level growth of
China imports in eight developed economies, $g_{nt}$, with lagged
regional employment weights of different industries $s_{\ell nt}$.
We also include pre-treatment controls $w_{\ell t}$ as in column
3 of Table 4 and and estimate regressions with regional employment
weights $e_{\ell t}$; see Section \ref{subsec:ApplicationSetting}
for more detail on the \textcite{AutorDorn2001} setting.

In a first step we obtain an estimated SSIV second and first stage
of
\begin{align}
y_{\ell t} & =\hat{\beta}x_{\ell t}+\hat{\gamma}^{\prime}w_{\ell t}+\hat{\varepsilon}_{\ell t},\label{eq:sim-2nd}\\
x_{\ell t} & =\hat{\pi}z_{\ell t}+\hat{\rho}^{\prime}w_{\ell t}+\hat{u}_{\ell t}.\label{eq:sim-1st}
\end{align}
We then generate 10,000 simulated samples by drawing shocks $g_{nt}^{\ast}$,
as detailed below, and constructing the simulated shift-share instrument
$z_{\ell t}^{\ast}=\sum_{n}s_{\ell nt}g_{nt}^{\ast}$ and treatment
$x_{\ell t}^{\ast}=\hat{\pi}z_{\ell t}^{\ast}+\hat{u}_{\ell t}$.
Imposing a true causal effect of $\beta^{\ast}=0$, we use the same
$y_{\ell t}^{\ast}\equiv\hat{\varepsilon}_{\ell t}$ as the outcome
in each simulation (note that it is immaterial whether we include
$\hat{\pi}'w_{\ell t}$ and $\hat{\rho}'w_{\ell t}$, since all our
specifications control for $w_{\ell t}$). By keeping $\hat{\varepsilon}_{\ell t}$
and $\hat{u}_{\ell t}$ fixed, we study the finite sample properties
of the estimator that arises from the randomness of shocks, which
is the basis of the inferential framework of \textcite{Adao}; we
also avoid having to take a stand on the joint data generating process
of $\left(\varepsilon_{\ell t},u_{\ell t}\right)$, which this inference
framework does not restrict.

We estimate SSIV specifications that parallel (\ref{eq:sim-2nd})-(\ref{eq:sim-1st})
from the simulated data
\begin{align}
y_{\ell t}^{\ast} & =\beta^{\ast}x_{\ell t}^{\ast}+\gamma^{\ast\prime}w_{\ell t}+\varepsilon_{\ell t}^{\ast},\\
x_{\ell t}^{\ast} & =\pi^{\ast}z_{\ell t}^{\ast}+\rho^{\ast\prime}w_{\ell t}+u_{\ell t}^{\ast}.
\end{align}
using the original weights $e_{\ell t}$ and controls $w_{\ell t}$.
We then test the (true) hypothesis $\beta^{\ast}=0$ using either
the heteroskedasticity-robust standard errors from the equivalent
industry-level regression or their version with the null imposed,
as in Section \ref{subsec:inference}.\footnote{Note that there is no need for clustering since we generate the shocks
independently across industries in all simulations. We have verified,
however, that allowing for correlation in shocks within industry groups
and using clustered standard errors yields similar results.} As in column 3 of Table 4, we control for period indicators as $q_{nt}$
in the industry-level regression.

Our comparison estimator is a conventional industry-level IV inspired
by \textcite{AADHP2016}. However, we try to keep the IV regression
as similar to the SSIV as possible, thus diverging from \textcite{AADHP2016}
in some details. Specifically, the outcome $y_{nt}$ is the industry
employment growth as measured by these authors. It is defined for
392 out of the 397 industries in \textcite{AutorDorn2001}, so we
drop the remaining five industries in each period. The endogenous
regressor $x_{nt}\equiv g_{nt}^{US}$ (growth of U.S. imports from
China per worker) and the instrument $g_{nt}$ (growth of China imports
into eight developed economies) are those from which we built the
shift-share endogenous regressor and treatment, respectively (see
footnote \ref{fn:ADH-details}). Construction of those variables differ
from \textcite{AADHP2016} who measure imports relative to domestic
absorption rather than employment. We also follow our SSIV analysis
in using period indicators as the only industry-level control variables
$q_{nt}$ and taking identical regression importance weights $s_{nt}$.

The Monte-Carlo strategy for the conventional shock-level IV parallels
the one for SSIV; we obtain an estimated industry-level second and
first stage of
\begin{align}
y_{nt} & =\hat{\beta}_{\text{ind}}x_{nt}+\hat{\gamma}^{\prime}q_{nt}+\hat{\varepsilon}_{nt},\\
x_{nt} & =\hat{\pi}_{\text{ind}}g_{nt}+\hat{\rho}^{\prime}q_{nt}+\hat{u}_{nt}.
\end{align}
using the $s_{nt}$ importance weights. We then perform 10,000 simulations
where we regenerate shocks $g_{nt}^{\ast}$ and regress $y_{nt}^{\ast}=\hat{\varepsilon}_{nt}$
(consistent with a true causal effect of $\beta^{\text{ind}}=0$,
given that we control for $q_{nt}$) on $x_{nt}^{*}=\hat{\pi}_{\text{ind}}g_{nt}^{\ast}+\hat{u}_{nt}$,
instrumenting by $g_{nt}^{\ast}$, controlling for $q_{nt}$, and
weighting by $s_{nt}$. We test $\beta_{\text{ind}}=0$ by using robust
standard errors in this IV regression or the version with the null
imposed, which corresponds to a standard Lagrange Multiplier test
for this true null hypothesis.

In both simulations we report the rejection rate of nominal 5\% level
tests for $\beta=0$ and $\beta_{\text{ind}}=0$ to gauge the quality
of each asymptotic approximation. We do not report the bias of the
estimators because they are all approximately unbiased (more precisely,
the simulated median bias is at most 1\% of the estimator's standard
deviation). However we return to the question of bias at the end of
the section, where we extend the analysis to having many instruments
with a weak first stage.

\paragraph*{Main results}

Table C7 reports the rejection rates for shift-share IV (columns 1
and 2) and conventional industry-level IV (columns 3 and 4) in various
simulations. Specifically, column 1 corresponds to using exposure-robust
standard errors from the equivalent industry-level IV, and column
2 implements the version with the null hypothesis imposed. Columns
3 and 4 parallel columns 1 and 2 when applied to conventional IV:
the former uses heteroskedasticity-robust standard errors and the
latter tests $\beta_{\text{ind}}=0$ with the null imposed, which
amounts to using the Lagrange multiplier test.

The simulations in Panel A vary the data-generating process of the
shocks. Following \textcite{Adao} in row (a) we draw the shocks \emph{iid
}from a normal distribution with the variance matched to the sample
variance of the shocks in the data after de-meaning by year. The rejection
rate is close to the nominal rate of 5\% for both SSIV and conventional
IV (7.6\% and 6.8\%, respectively), and in both cases it becomes even
closer when the null is imposed (5.2\% and 5.0\%).

This simulation may not approximate the data-generating process well
because of heteroskedasticity: smaller industries have more volatile
shocks.\footnote{This is established by unreported regressions of $\left|g_{nt}\right|$
on $s_{nt}$, for year-demeaned $g_{nt}$ from ADH, with or without
weights. The negative relationship is significant at conventional
levels.} To match unrestricted heteroskedasticity, in row (b) we use wild
bootstrap, generating $g_{nt}^{\ast}=g_{nt}\nu_{nt}^{\ast}$ by multiplying
the year-demeaned observed shocks $g_{nt}$ by $\nu_{nt}^{\ast}\stackrel{iid}{\sim}\mathcal{N}(0,1)$
\parencite{Liu1988}. This approach also provides a better approximation
for the marginal distribution of shocks than the normality assumption.
Here the relative performance of SSIV is even better: the rejection
rate is 8.0\% vs. 14.2\% for conventional IV.

We now depart from the row (b) simulation in several directions, as
a case study for the sensitivity of the asymptotic approximation to
different features of the SSIV setup. Specifically, we study the role
of the Herfindahl concentration index across industries, the number
of regions and industries, and the many weak instrument bias. We uniformly
find that the performance of the SSIV estimator is similar to that
of industry-level IV. Our results also suggest that the Herfindahl
index is a useful statistic for measuring the effective number of
industries in SSIV, and the first-stage \emph{F}-statistic is informative
about the weak instrument bias, as usual.

\paragraph*{The Role of Industry Concentration}

Since Assumption 2 requires small concentration of industry importance
weights, measured using the Herfindahl index $\sum_{n,t}s_{nt}^{2}/\left(\sum_{n,t}s_{nt}\right)^{2}$,
Panel B of Table C7 studies how increasing the skewness of $s_{nt}$
towards the bigger industries affects coverage of the tests.\footnote{Note that in ADH $\sum_{n}s_{\ell nt}$ equals the lagged share of
regional manufacturing employment, which is below one. We thus renormalize
the shares when computing the Herfindahl.} For conventional IV this simply amounts to reweighting the regression.
Specifically, for a parameter $\alpha>1$, we use weights
\[
\tilde{s}_{nt}=s_{nt}^{\alpha}\cdot\frac{\sum_{n^{\prime},t^{\prime}}s_{n^{\prime}t^{\prime}}}{\sum_{n^{\prime},t^{\prime}}s_{n^{\prime}t^{\prime}}^{\alpha}}.
\]
We choose the unique $\alpha$ to match the target level of $\widetilde{HHI}$
by solving, numerically, 
\begin{equation}
\frac{\sum_{n,t}\left(\tilde{s}_{nt}\right)^{2}}{\left(\sum_{n,t}\tilde{s}_{nt}\right)^{2}}=\widetilde{HHI}.\label{eq:match-HHI}
\end{equation}

Matching the Herfindahl index in SSIV is more complicated since we
need to choose how exactly to amend shares $\tilde{s}_{\ell nt}$
and regional weights $\tilde{e}_{\ell t}$ that would yield $\tilde{s}_{nt}$
from (\ref{eq:match-HHI}). We proceed as follows: we consider the
lagged level of manufacturing employment by industry $E_{\ell nt}=e_{\ell t}s_{\ell nt}$
and the total regional non-manufacturing employment $E_{\ell0t}=e_{\ell t}\left(1-\sum_{n}s_{\ell nt}\right)$.\footnote{The interpretation of $E_{\ell nt}$ as the lagged level is approximate
since $e_{\ell t}$ is measured at the beginning of period in ADH,
while $s_{\ell nt}$ is lagged.} We then define $\tilde{E}_{\ell nt}=E_{\ell nt}\cdot\frac{\tilde{s}_{nt}}{s_{nt}}$
for manufacturing industries (and leave non-manufacturing employment
unchanged, $\tilde{E}_{\ell0t}=E_{\ell0t}$). This increases employment
in large manufacturing industries proportionately in all regions,
while reducing it in smaller ones. We then recompute shares $\tilde{s}_{\ell nt}$
and weights $\tilde{e}_{\ell t}$ accordingly:
\begin{align*}
\tilde{e}_{\ell t} & =\sum_{n=0}^{N}\sum_{t}\tilde{E}_{\ell nt},\\
\tilde{s}_{\ell nt} & =\frac{\tilde{E}_{\ell nt}}{\tilde{e}_{\ell t}}.
\end{align*}

Rows (c)\textendash (e) of Table C7 Panel B implement this procedure
for target Herfindahl levels of $1/50$, $1/20$, and $1/10$, respectively.
For comparison, the Herfindahl in the actual ADH data is $1/191.6$
(Table 1, column 2). The table finds that even with the Herfindahl
index of $1/20$ (corresponding to the ``effective'' number of shocks
of 20 in both periods total) the rejection rate is still around 7\%,
a level that may be considered satisfactory. It also shows that the
rejection rate grows when the Herfindahl is even higher, at $1/10$,
suggesting that the Herfindahl can be used as an indicative rule of
thumb. More importantly, the rejection rates are similar for SSIV
and conventional industry-level IV, as before.

\paragraph*{Varying the Number of Industries and Regions}

The asymptotic sequence we consider in Section \ref{subsec:A1andA2}
relies on both $N$ and $L$ growing. Here we study how the quality
of the asymptotic approximation depends on these parameters.

First, to consider the case of small $N$, we aggregate industries
in a natural way: from 397 four-digit manufacturing SIC industries
into 136 three-digit ones and further into 20 two-digit ones and reconstruct
the endogenous right-hand side variable and the instrument using aggregated
data.\footnote{Specifically, we aggregate imports from China to the U.S. and either
developed economies as well as the number of U.S. workers by manufacturing
industry to construct the new $g_{nt}$ and $g_{nt}^{\text{US}}$.
We then aggregate the shares $s_{\ell nt}$ and $s_{\ell nt}^{\text{current}}$
to construct $x_{\ell t}$ and $z_{\ell t}$ (see footnote \ref{fn:ADH-details}
for formulas). We do not change the regional outcome, controls, or
importance weights. For conventional IV, we additionally reconstruct
the outcome (industry employment growth) by aggregating employment
levels by year in the \textcite{AADHP2016} data and measuring growth
according to their formulas.} Rows (f) and (g) of Table C7 Panel C report simulation results based
on the aggregated data. They show that rejection rates are similar
to the case of detailed industries, and between SSIV and conventional
IV. This does not mean that disaggregated data are not useful: the
dispersion of the simulated distribution (not reported) increases
with industry aggregation, reducing test power. However, standard
errors correctly reflect this variability, resulting in largely unchanged
test coverage rates.

Second, to study the implications of having fewer regions $L$, we
select a random subset of them in each simulation. The results are
presented in Rows (h) and (i) of Panel C for $L=100$ and $25$, compared
to the original $L=722$, respectively.\footnote{When we select regions, we always keep observations from both periods
for each selected region. We keep the second- and first-stage coefficients
from the full sample to focus on the noise that arises from shock
randomness.} They show once again that rejection rates are not significantly affected
(even though unreported standard errors expectedly increase).

\paragraph*{Many Weak Instruments}

In this final simulation we return to the question of SSIV bias. Since
our previous simulations confirm that just-identified SSIV is median-unbiased,
we turn to the case of multiple instruments. We show that the problem
of many weak instruments is similar between SSIV and conventional
IV, and that first-stage \emph{F}-statistics, when properly constructed,
can serve as useful heuristics.

For clarity, we begin by describing the procedure for the conventional
shock-level IV that is a small departure from Column 3 of Table C7.
For a given number of instruments $J\ge1$, in each simulation we
generate $g_{jnt}^{\ast}$, $j=1,\dots,J$, independently across $j$
using wild bootstrap (as in Table C7 Row (b)).\footnote{For computational reasons we perform only $15,000/J$ simulations
when $J>1$ (but 10,000 for $J=1$ as before).} We make only the first instrument relevant by setting $x_{nt}^{\ast}=\hat{\pi}_{\text{ind}}g_{1nt}^{\ast}+\sum_{j=2}^{J}0\cdot g_{jnt}^{\ast}+\hat{u}_{nt}$.
We then estimate the IV regression of $y_{nt}^{\ast}\equiv\hat{\varepsilon}_{nt}$
on $x_{nt}^{\ast}$, instrumenting with $g_{1nt}^{\ast},\dots,g_{Jnt}^{\ast}$,
controlling for $q_{nt}$, and weighting by $s_{nt}$. We use robust
standard errors and compute the effective first-stage $F$-statistic
using the \textcite{MontielOleaPflueger2013} method.

The procedure for SSIV is more complex but as usual parallels the
one for the conventional shock-level IV as much as possible. Given
simulated shocks $g_{jnt}^{\ast}$, we construct shift-share instruments
$z_{j\ell t}^{\ast}=\sum_{\ell}s_{\ell nt}g_{jnt}^{\ast}$ and make
only the first of them relevant, $x_{\ell t}^{\ast}=\hat{\pi}z_{1\ell t}^{\ast}+\sum_{j=2}^{J}0\cdot z_{jnt}^{\ast}+\hat{u}_{\ell t}$.
Since the equivalence result from Section \ref{subsec:numerical-equiv}
need not hold for overidentified SSIV, we rely on the results in Appendix
\ref{subsec:appx-multiple_shocks}: we estimate $\beta^{\ast}$ from
the industry-level regression of $\bar{y}_{nt}^{\ast\perp}$ (based
on $y_{\ell t}^{\ast}=\hat{\varepsilon}_{\ell t}$ as before) on $\bar{x}_{nt}^{\ast\perp}$
by 2SLS, instrumenting by $g_{1nt}^{\ast},\dots,g_{Jnt}^{\ast}$,
controlling for $q_{nt}$ and weighting by $s_{nt}$. We compute robust
standard errors from this regression to test $\beta^{\ast}=0$. For
effective first-stage $F$-statistics, we follow the procedure described
in Appendix \ref{subsec:appx-multiple_shocks} and implemented via
our \emph{weakssivtest} command in Stata.

Table C8 reports the result for $J=1,5,10,25,$ and $50$, presenting
the rejection rate corresponding to the 5\% nominal, the median bias
as a percentage of the simulated standard deviation, and the median
first-stage \emph{F}-statistic. Panel A corresponds to SSIV and Panel
B to the conventional shock-level IV. For higher comparability, we
adjust the first-stage coefficient $\hat{\pi}_{\text{ind}}$ in the
latter in order to make the \emph{F}-statistics approximately match
between the two panels. We find that the median bias is now non-trivial
and grows with $J$, at the same time as the \emph{F}-statistic declines.
However, the level of bias is similar for the two estimators. The
rejection rates tend to be higher for conventional IV than SSIV, although
they converge as $J$ grows.

\newpage{}

\section{Appendix Proofs}

\subsection{Proposition 4 and Extensions\label{sec:consistency}}

This section proves Proposition 4 and extensions that allow for certain
forms of mutual shock dependence (Assumptions 5 and 6). Proposition
3 is obtained as a special case, where $q_{n}=1$. In addition to
Assumptions 3 and 4 and the relevance condition of $\sum_{\ell}e_{\ell}z_{\ell}x_{\ell}^{\perp}\xrightarrow{p}\pi$
with $\pi\neq0$, the proof of Proposition 4 uses two regularity conditions:
\begin{lyxlist}{00.00.0000}
\item [{\textbf{Assumption}}] \noindent \textbf{B1}:\textbf{\emph{ }}$\expec{\tilde{g}_{n}^{2}\mid\bar{\varepsilon},q,s}$
and $\expec{\bar{\varepsilon}_{n}^{2}\mid s}$ are uniformly bounded
by some fixed $B_{g}$ and $B_{\varepsilon}$.
\item [{\textbf{Assumption}}] \noindent \textbf{B2}:\textbf{\emph{ }}$\left\Vert \sum_{\ell}e_{\ell}w_{\ell}\varepsilon_{\ell}\right\Vert _{1}=o_{p}(1)$,
$\max\left|\left(\sum_{\ell}e_{\ell}w_{\ell}w_{\ell}^{\prime}\right){}^{-1}\right|=O_{p}(1)$,
and $\max\left|\sum_{\ell}e_{\ell}w_{\ell}z_{\ell}\right|=O_{p}(1)$.
\end{lyxlist}
The first of these is a weak condition on the second moments of shocks
and shock-level unobservables which we show below permits a shock-level
law of large numbers. The second condition ensures the consistency
of the IV estimate of the control coefficient, $\hat{\gamma}=\left(\sum_{\ell}e_{\ell}w_{\ell}w_{\ell}^{\prime}\right){}^{-1}\sum_{\ell}e_{\ell}w_{\ell}\epsilon_{\ell}=\gamma+\left(\sum_{\ell}e_{\ell}w_{\ell}w_{\ell}^{\prime}\right){}^{-1}\sum_{\ell}e_{\ell}w_{\ell}\varepsilon_{\ell}$
(see footnote \ref{fn:model_derivation}), and stochastic boundedness
of the weighted average $\sum_{\ell}e_{\ell}w_{\ell m}z_{\ell}$,
while generally allowing the length of the control vector to increase
with $L$. We discuss low-level conditions for the consistency of
$\hat{\gamma}$ in Appendix \ref{sec:controlconsistency}.

To prove Proposition 4, we first note that under Assumption B2,
\begin{align}
\sum_{n}s_{n}g_{n}\bar{\varepsilon}_{n}-\sum_{n}s_{n}g_{n}\bar{\varepsilon}_{n}^{\perp} & =\sum_{\ell}e_{\ell}z_{\ell}\left(\varepsilon_{\ell}-\varepsilon_{\ell}^{\perp}\right)\nonumber \\
 & =\left(\sum_{\ell}e_{\ell}z_{\ell}w_{\ell}^{\prime}\right)\left(\hat{\gamma}-\gamma\right)\nonumber \\
 & =\left(\sum_{\ell}e_{\ell}z_{\ell}w_{\ell}^{\prime}\right)\left(\sum_{\ell}e_{\ell}w_{\ell}w_{\ell}^{\prime}\right){}^{-1}\sum_{\ell}e_{\ell}w_{\ell}\varepsilon_{\ell}\xrightarrow{p}0,
\end{align}
so that, when the relevance condition holds,
\begin{align}
\hat{\beta}-\beta & =\frac{\sum_{n}s_{n}g_{n}\bar{\varepsilon}_{n}^{\perp}}{\sum_{n}s_{n}g_{n}\bar{x}_{n}^{\perp}}\nonumber \\
 & =\pi^{-1}\sum_{n}s_{n}g_{n}\bar{\varepsilon}_{n}\left(1+o_{p}(1)\right).
\end{align}
Furthermore, since $\sum_{n}s_{\ell n}=1$, we also have under Assumption
B2 that
\begin{align}
\sum_{n}s_{n}q_{n}^{\prime}\mu\bar{\varepsilon}_{n} & =\left(\sum_{\ell}e_{\ell}\tilde{w}_{\ell}\varepsilon_{\ell}\right)^{\prime}\mu\xrightarrow{p}0.
\end{align}
Thus
\begin{align}
\sum_{n}s_{n}g_{n}\bar{\varepsilon}_{n} & =\sum_{n}s_{n}\tilde{g}_{n}\bar{\varepsilon}_{n}+o_{p}(1),
\end{align}
with
\begin{align}
\expec{\sum_{n}s_{n}\tilde{g}_{n}\bar{\varepsilon}_{n}} & =0
\end{align}
under Assumption 3.

To prove consistency of $\hat{\beta}$, it remains to show that $\var{\sum_{n}s_{n}\tilde{g}_{n}\bar{\varepsilon}_{n}}\rightarrow0$
such that $\sum_{n}s_{n}g_{n}\bar{\varepsilon}_{n}\xrightarrow{p}0$.
Since 
\begin{align}
\expec{\tilde{g}_{n}\tilde{g}_{n^{\prime}}\mid\bar{\varepsilon},q,s} & =\cov{\tilde{g}_{n},\tilde{g}_{n^{\prime}}\mid\bar{\varepsilon},q,s}=0
\end{align}
 under Assumptions 3 and 4, 
\begin{align}
\var{\sum_{n}s_{n}\tilde{g}_{n}\bar{\varepsilon}_{n}} & =\expec{\left(\sum_{n}s_{n}\tilde{g}_{n}\bar{\varepsilon}_{n}\right)^{2}}\nonumber \\
 & =\sum_{n}\sum_{n^{\prime}}\expec{s_{n}s_{n^{\prime}}\tilde{g}_{n}\tilde{g}_{n^{\prime}}\bar{\varepsilon}_{n}\bar{\varepsilon}_{n^{\prime}}}\nonumber \\
 & =\sum_{n}\expec{s_{n}^{2}\expec{\expec{\tilde{g}_{n}^{2}\mid\bar{\varepsilon},q,s}\bar{\varepsilon}_{n}^{2}\mid s}}.
\end{align}
Then, by Assumption B1 and the Cauchy-Schwartz inequality:
\begin{align}
\var{\sum_{n}s_{n}\tilde{g}_{n}\bar{\varepsilon}_{n}} & \le B_{g}B_{\varepsilon}\expec{\sum_{n}s_{n}^{2}}\to0.\label{eq:orthogonality_var}
\end{align}

\paragraph{Extensions}

Similar steps establish equation (\ref{eq:orthogonality_var}) when
Assumption 4 is replaced by either Assumption 5 or 6. Under Assumption
5 we have, for $N\left(c\right)=\left\{ n\colon c(n)=c\right\} $,
\begin{align}
\var{\sum_{n}s_{n}\tilde{g}_{n}\bar{\varepsilon}_{n}} & =\expec{\left(\sum_{c}\sum_{n\in N(c)}s_{n}\tilde{g}_{n}\bar{\varepsilon}_{n}\right)^{2}}\nonumber \\
 & =\expec{\sum_{c}s_{c}^{2}\expec{\left(\sum_{n\in N(c)}\frac{s_{n}}{s_{c}}\tilde{g}_{n}\bar{\varepsilon}_{n}\right)^{2}\mid s}}\nonumber \\
 & =\expec{\sum_{c}s_{c}^{2}\sum_{n,n^{\prime}\in N(c)}\frac{s_{n}}{s_{c}}\frac{s_{n^{\prime}}}{s_{c}}\expec{\tilde{g}_{n}\tilde{g}_{n^{\prime}}\bar{\varepsilon}_{n}\bar{\varepsilon}_{n^{\prime}}\mid s}}\nonumber \\
 & \le B_{g}B_{\varepsilon}\expec{\sum_{c}s_{c}^{2}}\rightarrow0.
\end{align}
Here the last line used Assumption B1 and the Cauchy-Schwartz inequality
twice: to establish, for $n,n'\in N(c)$,
\begin{align}
\expec{\tilde{g}_{n}\tilde{g}_{n^{\prime}}\mid\bar{\varepsilon},q,s} & \le\sqrt{\expec{\tilde{g}_{n}\mid\bar{\varepsilon},q,s}\expec{\tilde{g}_{n^{\prime}}\mid\bar{\varepsilon},q,s}}\nonumber \\
 & \le B_{g}
\end{align}
and
\begin{align}
\expec{\left|\bar{\varepsilon}_{n}\right|\left|\bar{\varepsilon}_{n^{\prime}}\right|\mid s_{c}} & \le\sqrt{\expec{\bar{\varepsilon}_{n}^{2}\mid s}\expec{\bar{\varepsilon}_{n^{\prime}}^{2}\mid s}}\nonumber \\
 & \le B_{\varepsilon}.
\end{align}
If we instead replace Assumption 4 with Assumption 6, we have
\begin{align}
\var{\sum_{n}s_{n}\tilde{g}_{n}\bar{\varepsilon}_{n}} & =\expec{\left(\sum_{n}s_{n}\tilde{g}_{n}\bar{\varepsilon}_{n}\right)^{2}}\nonumber \\
 & =\sum_{n}\sum_{n^{\prime}}\expec{s_{n}s_{n^{\prime}}\expec{\tilde{g}_{n}\tilde{g}_{n^{\prime}}\mid\bar{\varepsilon},q,s}\bar{\varepsilon}_{n}\bar{\varepsilon}_{n^{\prime}}}\nonumber \\
 & \le B_{L}\sum_{n}\sum_{n^{\prime}}f\left(\left|n^{\prime}-n\right|\right)\expec{\left|s_{n}\bar{\varepsilon}_{n}s_{n^{\prime}}\bar{\varepsilon}_{n^{\prime}}\right|}\nonumber \\
 & =B_{L}\left(\sum_{n}\expec{(s_{n}\bar{\varepsilon}_{n})^{2}}f(0)+2\sum_{r=1}^{N-1}\sum_{n=1}^{N-r}\expec{\left|s_{n+r}\bar{\varepsilon}_{n+r}\right|\cdot\left|s_{n}\bar{\varepsilon}_{n}\right|}f(r)\right)\nonumber \\
 & \le\left(B_{L}\sum_{n}\expec{s_{n}^{2}\expec{\bar{\varepsilon}_{n}^{2}\mid s}}\right)\left(f(0)+2\sum_{r=1}^{N-1}f(r)\right)\nonumber \\
 & \le B_{\varepsilon}\left(f(0)+2\sum_{r=1}^{N-1}f(r)\right)\left(B_{L}\expec{\sum_{n}s_{n}^{2}}\right)\rightarrow0,
\end{align}
using $\expec{\bar{\varepsilon}_{n}^{2}\mid s_{n}}<B_{\varepsilon}$
in the last line. Here the second-to-last line follows because for
any sequence of numbers $a_{1},\dots,a_{N}$ and any $r>0$,
\begin{align}
\sum_{n}a_{n}^{2} & \ge\frac{1}{2}\left(\sum_{n=1}^{N-r}a_{n}^{2}+\sum_{n=1}^{N-r}a_{n+r}^{2}\right)\nonumber \\
 & =\frac{1}{2}\sum_{n=1}^{N-r}\left(a_{n}-a_{n+r}\right)^{2}+\sum_{n=1}^{N-r}a_{n}a_{n+r}\nonumber \\
 & \ge\sum_{n=1}^{N-r}a_{n}a_{n+r},
\end{align}
and the same is true in expectation if $a_{n}=\left|s_{n}\bar{\varepsilon}_{n}\right|$
are random variables. We note that allowing $B_{L}$ to grow in the
asymptotic sequence imposes much weaker conditions on the correlation
structure of shocks. For example, with shock importance weights $s_{n}$
approximately equal, i.e. $\sum_{n}s_{n}^{2}=O_{p}\left(1/N\right)$,
it is enough to have $\left|\cov{\tilde{g}_{n},\tilde{g}_{n^{\prime}}\mid\bar{\varepsilon},q,s}\right|\le B_{1}/N^{\alpha}$
for any $\alpha>0$: in this case one can satisfy Assumption 6 by
setting $B_{L}=B_{1}N^{1-\alpha/2}$ and $f(r)=r^{-1-\alpha/2}$.

\subsection{Proposition 5 and Related Results\label{sec:appx_Inference}}

This section proves Proposition 5 and then establishes several additional
results mentioned in Section \ref{subsec:inference}. First, we show
the heteroskedasticity-robust standard error from estimating equation
(\ref{eq:fullSEss}) is numerically equivalent to the baseline IV
standard error of \textcite{Adao} when $w_{\ell}$ contains only
a constant. Second, we show that when Assumption B4 on the structure
of controls is relaxed, the standard errors from Proposition 5 are
conservative. We also discuss the likely difference between our standard
error estimates and those of \textcite{Adao} when Assumption B4 holds.
Finally, we show how the alternative null-imposed inference procedure
of \textcite{Adao} is also conveniently obtained from our equivalent
shock-level regression.

We prove Proposition 5 under additional assumptions that largely follow
\textcite{Adao}:
\begin{lyxlist}{00.00.0000}
\item [{\textbf{Assumption}}] \noindent \textbf{B3}: The first stage satisfies
$x_{\ell}=\sum_{n}s_{\ell n}\pi_{\ell n}g_{n}+\eta_{\ell}$, for all
$\ell$.
\item [{\textbf{Assumption}}] \noindent \textbf{B4}:\emph{ }The control
vector can be partitioned as $w_{\ell}=[\tilde{w}_{\ell}^{\prime},u_{\ell}^{\prime}]^{\prime}$,
for $\tilde{w}_{\ell}=\sum_{n}s_{\ell n}q_{n}$. The vector $q_{n}$
captures all sources of shock confounding: $\expec{g_{n}\mid\mathcal{I}_{L}}=q_{n}^{\prime}\mu$,
for all $n$ and $\mathcal{I}_{L}=\left\{ \left\{ q_{n}\right\} _{n},\left\{ u_{\ell},\epsilon_{\ell},\eta_{\ell},\left\{ s_{\ell n},\pi_{\ell n}\right\} _{n},e_{\ell}\right\} _{\ell}\right\} $.
\item [{\textbf{Assumption}}] \textbf{B5}: The $g_{n}$ are mutually independent
given $\mathcal{I}_{L}$, $\max_{n}s_{n}\rightarrow0$, and $\max_{n}\frac{s_{n}^{2}}{\sum_{n^{\prime}}s_{n^{\prime}}^{2}}\rightarrow0$.
\end{lyxlist}
\textbf{Assumption B6}: $\expec{|g_{n}|^{4+v}\mid\mathcal{I}_{L}}$
is uniformly bounded for some $v>0$ and $\sum_{\ell}e_{\ell}\sum_{n}s_{\ell n}^{2}\var{g_{n}\mid\mathcal{I}_{L}}\pi_{\ell n}\neq0$
almost surely. The support of $\pi_{\ell n}$ is bounded, the fourth
moments of $\epsilon_{\ell}$, $\eta_{\ell}$, $u_{\ell}$, $q_{n}$,
and $\tilde{g}_{n}$ exist and are uniformly bounded, $\sum_{\ell}e_{\ell}w_{\ell}w_{\ell}^{\prime}\xrightarrow{p}\Omega_{ww}$
for positive definite $\Omega_{ww}$, and $\sum s_{n}q_{n}q_{n}^{\prime}\xrightarrow{p}\Omega_{qq}$
for positive definite $\Omega_{qq}$. The control vector $\gamma$
is consistently estimated by $\hat{\gamma}=\left(\sum_{\ell}e_{\ell}w_{\ell}w_{\ell}^{\prime}\right)^{-1}\sum_{\ell}e_{\ell}w_{\ell}\epsilon_{\ell}$.

We note that Assumption B5 both strengthens our baseline Herfindahl
index condition in Assumption 4 and implicitly treats the set of $s_{n}$
as non-stochastic, following Assumption 2 of \textcite{Adao}. The
regularity condition B6 includes the relevant conditions from Assumptions
4 and A.3 of \textcite{Adao}. These assumptions strengthen those
of Proposition 4: Assumptions B3\textendash B6 imply our Assumptions
3, 4, B1, and B2. Relative to \textcite{Adao}, we do not impose that
$L>N$ or that the shares are non-collinear.

To establish the equivalence of IV coefficients in Proposition 5,
note that when $\sum_{n}s_{\ell n}q_{n}$ is included in $w_{\ell}$
\begin{align}
\sum_{n}s_{n}q_{n}\bar{y}_{n}^{\perp} & =\sum_{\ell}e_{\ell}y_{\ell}^{\perp}\left(\sum_{n}s_{\ell n}q_{n}\right)=0
\end{align}
and similarly for $\sum_{n}s_{n}q_{n}\bar{x}_{n}^{\perp}$. The $s_{n}$-weighted
regression of $\bar{y}_{n}^{\perp}$ and $\bar{x}_{n}^{\perp}$ on
$q_{n}$ thus produces a coefficient vector that is numerically zero,
implying the $s_{n}$-weighted and $g_{n}$-instrumented regression
of $\bar{y}_{n}^{\perp}$ on $\bar{x}_{n}^{\perp}$ is unchanged with
the addition of $q_{n}$ controls. Proposition 1 shows that the IV
coefficient from this regression is equivalent to the SSIV estimate
$\hat{\beta}$. 

To establish validity of the standard errors, note that the conventional
heteroskedasticity-robust standard error from for the $s_{n}$-weighted
shock-level IV regression of $\bar{y}_{n}^{\perp}$ on $\bar{x}_{n}^{\perp}$
and $q_{n}$, instrumented by $g_{n}$, is given by
\begin{equation}
\widehat{se}_{\text{equiv}}=\frac{\sqrt{\sum_{n}s_{n}^{2}\hat{\varepsilon}_{n}^{2}\hat{g}_{n}^{2}}}{\left|\sum_{n}s_{n}\bar{x}_{n}^{\perp}g_{n}\right|},\label{eq:se_robust}
\end{equation}
where $\hat{\varepsilon}_{n}=\bar{y}_{n}^{\perp}-\hat{\beta}\bar{x}_{n}^{\perp}$
is the estimated shock-level regression residual (where we used the
fact that the estimated coefficients on $q_{n}$ in that regression
are numerically zero) and $\hat{g}_{n}=g_{n}-\hat{\mu}q_{n}$, where
$\hat{\mu}=\left(\sum_{n}s_{n}q_{n}q_{n}^{\prime}\right)^{-1}\sum_{n}s_{n}q_{n}g_{n}$,
is the residual from a projection of the instrument in equation (\ref{eq:fullSEss})
on the control vector $q_{n}$. By Proposition 1, $\hat{\varepsilon}_{n}$
coincides with the share-weighted aggregate of the SSIV estimated
residuals $\hat{\varepsilon}_{\ell}=y_{\ell}^{\perp}-\hat{\beta}x_{\ell}^{\perp}$:
\begin{align}
\hat{\varepsilon}_{n} & =\frac{\sum_{\ell}e_{\ell}s_{\ell n}y_{\ell}^{\perp}}{\sum_{\ell}e_{\ell}s_{\ell n}}-\hat{\beta}\cdot\frac{\sum_{\ell}e_{\ell}s_{\ell n}x_{\ell}^{\perp}}{\sum_{\ell}e_{\ell}s_{\ell n}}=\frac{\sum_{\ell}e_{\ell}s_{\ell n}\hat{\varepsilon}_{\ell}}{\sum_{\ell}e_{\ell}s_{\ell n}}.
\end{align}
The squared numerator of (\ref{eq:se_robust}) can thus be rewritten
\begin{align}
\sum_{n}s_{n}^{2}\hat{\varepsilon}_{n}^{2}\hat{g}_{n}^{2} & =\sum_{n}\left(\sum_{\ell}e_{\ell}s_{\ell n}\hat{\varepsilon}_{\ell}\right)^{2}\hat{g}_{n}^{2}.
\end{align}
The expression in the denominator of (\ref{eq:se_robust}) estimates
the magnitude of the shock-level first-stage covariance, which matches
the $e_{\ell}$-weighted sample covariance of $x_{\ell}$ and $z_{\ell}$:
\begin{align}
\sum_{n}s_{n}\bar{x}_{n}^{\perp}g_{n} & =\sum_{n}\left(\sum_{\ell}e_{\ell}s_{\ell n}x_{\ell}^{\perp}\right)g_{n}=\sum_{\ell}e_{\ell}x_{\ell}^{\perp}z_{\ell}.
\end{align}
Thus
\begin{equation}
\widehat{se}_{\text{equiv}}=\frac{\sqrt{\sum_{n}\left(\sum_{\ell}e_{\ell}s_{\ell n}\hat{\varepsilon}_{\ell}\right)^{2}\hat{g}_{n}^{2}}}{\left|\sum_{\ell}e_{\ell}x_{\ell}^{\perp}z_{\ell}\right|}.\label{eq:se_equiv}
\end{equation}
We now compare this expression to the standard error formula from
\textcite{Adao}, incorporating the $e_{\ell}$ importance weights.
Equation (39) in their paper yields
\begin{equation}
\widehat{se}_{\text{AKM}}=\frac{\sqrt{\sum_{n}\left(\sum_{\ell}e_{\ell}s_{\ell n}\hat{\varepsilon}_{\ell}\right)^{2}\ddot{g}_{n}^{2}}}{\left|\sum_{\ell}e_{\ell}x_{\ell}^{\perp}z_{\ell}\right|},\label{eq:akmSEs}
\end{equation}
where $\ddot{g}_{n}$ denotes the coefficients from regressing the
residualized instrument $z_{\ell}^{\perp}$ on all shares $s_{\ell n}$,
without a constant; note that to compute this requires $L>N$ and
that the matrix of exposure shares $s_{\ell n}$ is full rank. The
formulas for $\widehat{se}_{\text{equiv}}$ and $\widehat{se}_{\text{AKM}}$
therefore differ only in the construction of shock residuals, $\hat{g}_{n}$
versus $\ddot{g}_{n}$.

We establish the general asymptotic equivalence of $\widehat{se}_{\text{equiv}}^{2}$
and $\widehat{se}_{\text{AKM}}^{2}$, and thus the asymptotic validity
of $\widehat{se}_{\text{equiv}}$, by showing that both capture the
conditional asymptotic variance of $\hat{\beta}$ given $\mathcal{I}_{L}$
under Assumptions B3-B6. Both of the resulting confidence intervals
are then asymptotically valid unconditionally, since if $Pr(\beta\in\widehat{CI}\mid\mathcal{I}_{L})=\alpha$
then $Pr(\beta\in\widehat{CI})=\expec{\expec{\mathbf{1}[\beta\in\widehat{CI}\mid\mathcal{I}_{L}}}=\alpha$
by the law of iterated expectations. Under Assumptions B3-B6, Proposition
A.1 of \textcite{Adao} applies and shows that 
\begin{align}
\sqrt{r_{L}}(\hat{\beta}-\beta) & \stackrel{d}{\to}\mathcal{N}\left(0,\frac{\mathcal{V}}{\pi^{2}}\right)
\end{align}
for $\mathcal{V}=\plim_{L\to\infty}r_{L}\mathcal{V}_{L}$, where $r_{L}=1/\left(\sum_{n}s_{n}^{2}\right)$
and $\mathcal{V}_{L}=\sum_{n}\left(\sum_{\ell}e_{\ell}s_{\ell n}\varepsilon_{\ell}\right)^{2}\var{g_{n}\mid\mathcal{I}_{L}}$,
provided such a limit exists. To establish the asymptotic validity
of $\widehat{se}_{\text{AKM}}$, i.e. that $r_{L}\left(\sum_{n}\left(\sum_{\ell}e_{\ell}s_{\ell n}\hat{\varepsilon}_{\ell}\right)^{2}\ddot{g}_{n}^{2}-\mathcal{V}_{L}\right)\xrightarrow{p}0$,
\textcite{Adao} further assume that $L\ge N$, the matrix of $s_{\ell n}$
is always full rank, and additional regularity conditions (see their
Proposition 5). We establish $r_{L}\left(\sum_{n}\left(\sum_{\ell}e_{\ell}s_{\ell n}\hat{\varepsilon}_{\ell}\right)^{2}\hat{g}_{n}^{2}-\mathcal{V}_{L}\right)\xrightarrow{p}0$,
and thus $r_{L}\sum_{n}\left(\sum_{\ell}e_{\ell}s_{\ell n}\hat{\varepsilon}_{\ell}\right)^{2}\hat{g}_{n}^{2}\xrightarrow{p}\mathcal{V}$,
without imposing those assumptions.

To start, we write $\tilde{g}_{n}=g_{n}-q_{n}^{\prime}\mu$ and decompose
\begin{align}
r_{L}\left(\sum_{n}\left(\sum_{\ell}e_{\ell}s_{\ell n}\hat{\varepsilon}_{\ell}\right)^{2}\hat{g}_{n}^{2}-\mathcal{V}_{L}\right)= & r_{L}\left(\sum_{n}\left(\sum_{\ell}e_{\ell}s_{\ell n}\varepsilon_{\ell}\right)^{2}\tilde{g}_{n}^{2}-\mathcal{V}_{L}\right)\nonumber \\
 & +r_{L}\sum_{n}\left(\left(\sum_{\ell}e_{\ell}s_{\ell n}\hat{\varepsilon}_{\ell}\right)^{2}-\left(\sum_{\ell}e_{\ell}s_{\ell n}\varepsilon_{\ell}\right)^{2}\right)\tilde{g}_{n}^{2}\nonumber \\
 & +r_{L}\sum_{n}\left(\sum_{\ell}e_{\ell}s_{\ell n}\hat{\varepsilon}_{\ell}\right)^{2}\left(\hat{g}_{n}^{2}-\tilde{g}_{n}^{2}\right).\label{eq:var_decomp}
\end{align}
\textcite{Adao} show that the second term of this expression is $o_{p}(1)$
under our assumptions, using the fact (their Lemma A.3, again generalized
to include importance weights) that for a triangular array $\left\{ A_{L1},\dots,A_{LL},B_{L1},\dots,B_{LL},C_{L1},\dots,C_{LN_{L}}\right\} _{L=1}^{\infty}$
with $\expec{A_{L\ell}^{4}\mid\left\{ \left\{ s_{\ell^{\prime}n}\right\} _{n},e_{\ell^{\prime}}\right\} _{\ell^{\prime}}}$,
$\expec{B_{L\ell}^{4}\mid\left\{ \left\{ s_{\ell^{\prime}n}\right\} _{n},e_{\ell^{\prime}}\right\} _{\ell^{\prime}}}$,
and $\expec{C_{Ln}^{2}\mid\left\{ \left\{ s_{\ell n^{\prime}}\right\} _{n^{\prime}},e_{\ell}\right\} _{\ell}}$
uniformly bounded, 
\begin{align}
r_{L}\sum_{\ell}\sum_{\ell^{\prime}}\sum_{n}e_{\ell}e_{\ell^{\prime}}s_{\ell n}s_{\ell^{\prime}n}A_{L\ell}B_{L\ell^{\prime}}C_{Ln} & =O_{p}(1).
\end{align}
Here with $D_{\ell}=(z_{\ell},w_{\ell}^{\prime})^{\prime}$, $\theta=(\beta,\gamma^{\prime})^{\prime}$,
and $\hat{\theta}=(\hat{\beta},\hat{\gamma}^{\prime})^{\prime}$ we
can write
\begin{align}
\left(\sum_{\ell}e_{\ell}s_{\ell n}\hat{\varepsilon}_{\ell}\right)^{2}= & \left(\sum_{\ell}e_{\ell}s_{\ell n}\varepsilon_{\ell}\right)^{2}+2\sum_{\ell}\sum_{\ell^{\prime}}e_{\ell}e_{\ell^{\prime}}s_{\ell n}s_{\ell^{\prime}n}D_{\ell}^{\prime}\left(\theta-\hat{\theta}\right)\varepsilon_{\ell^{\prime}}\nonumber \\
 & +\sum_{\ell}\sum_{\ell^{\prime}}e_{\ell}e_{\ell^{\prime}}D_{\ell}^{\prime}\left(\theta-\hat{\theta}\right)D_{\ell^{\prime}}^{\prime}\left(\theta-\hat{\theta}\right),
\end{align}
and both $D_{\ell}$ and $\varepsilon_{\ell}$ have bounded fourth
moments by the assumption of bounded fourth moments of $\epsilon_{\ell},$
$\eta_{\ell},$ $u_{\ell}$, and $q_{n}$, and $g_{n}$ in Assumption
B6. Thus by the lemma
\begin{align}
r_{L}\sum_{n}\left(\left(\sum_{\ell}e_{\ell}s_{\ell n}\hat{\varepsilon}_{\ell}\right)^{2}-\left(\sum_{\ell}e_{\ell}s_{\ell n}\varepsilon_{\ell}\right)^{2}\right)\tilde{g}_{n}^{2}= & 2\left(\theta-\hat{\theta}\right){}^{\prime}\left(r_{L}\sum_{\ell}\sum_{\ell^{\prime}}\sum_{n}e_{\ell}e_{\ell^{\prime}}s_{\ell n}s_{\ell^{\prime}n}\tilde{g}_{n}^{2}D_{\ell}\varepsilon_{\ell^{\prime}}\right)\nonumber \\
 & +\left(\theta-\hat{\theta}\right){}^{\prime}\left(r_{L}\sum_{\ell}\sum_{\ell^{\prime}}\sum_{n}e_{\ell}e_{\ell^{\prime}}s_{\ell n}s_{\ell^{\prime}n}\tilde{g}_{n}^{2}D_{\ell}D_{\ell^{\prime}}^{\prime}\right)\left(\theta-\hat{\theta}\right)\nonumber \\
= & \left(\theta-\hat{\theta}\right){}^{\prime}O_{p}(1)+\left(\theta-\hat{\theta}\right){}^{\prime}O_{p}(1)\left(\theta-\hat{\theta}\right),\label{eq:AKM_decomp}
\end{align}
which is $o_{p}(1)$ by the consistency of $\hat{\theta}$ (implied
by Assumptions B3-B6). \textcite{Adao} further show the first term
of equation (\ref{eq:var_decomp}) is $o_{p}(1)$, without using the
additional regularity conditions of their Proposition 5.

It thus remains for us to show the third term of (\ref{eq:var_decomp})
is also $o_{p}(1)$. Note that
\begin{align}
\hat{g}_{n}^{2} & =\left(g_{n}-q_{n}^{\prime}\hat{\mu}\right)^{2}=\tilde{g}_{n}^{2}+\left(q_{n}^{\prime}\left(\hat{\mu}-\mu\right)\right)^{2}-2\tilde{g}_{n}q_{n}^{\prime}\left(\hat{\mu}-\mu\right),
\end{align}
so that
\begin{align}
 & r_{L}\sum_{n}\left(\sum_{\ell}e_{\ell}s_{\ell n}\hat{\varepsilon}_{\ell}\right)^{2}\left(\hat{g}_{n}^{2}-\tilde{g}_{n}^{2}\right)\nonumber \\
= & r_{L}\sum_{n}\left(\sum_{\ell}e_{\ell}s_{\ell n}\hat{\varepsilon}_{\ell}\right)^{2}\left(q_{n}^{\prime}\left(\hat{\mu}-\mu\right)-2\tilde{g}_{n}\right)q_{n}^{\prime}\left(\hat{\mu}-\mu\right)\nonumber \\
= & r_{L}\sum_{n}\left(\sum_{\ell}e_{\ell}s_{\ell n}\varepsilon_{\ell}\right)^{2}\left(q_{n}^{\prime}\left(\hat{\mu}-\mu\right)-2\tilde{g}_{n}\right)q_{n}^{\prime}\left(\hat{\mu}-\mu\right)\nonumber \\
 & +r_{L}\sum_{n}\left(\left(\sum_{\ell}e_{\ell}s_{\ell n}\hat{\varepsilon}_{\ell}\right)^{2}-\left(\sum_{\ell}e_{\ell}s_{\ell n}\varepsilon_{\ell}\right)^{2}\right)\left(q_{n}^{\prime}\left(\hat{\mu}-\mu\right)-2\tilde{g}_{n}\right)q_{n}^{\prime}\left(\hat{\mu}-\mu\right).
\end{align}
Using the previous lemma, the first term of this expression is $O_{p}(1)\left(\hat{\mu}-\mu\right)$
since $\varepsilon_{\ell}$, $q_{n}$, and $\tilde{g}_{n}$ have bounded
fourth moments under Assumption B6. The second term is similarly
$O_{p}(1)\left(\hat{\mu}-\mu\right)$ by the lemma and the decomposition
used in equation (\ref{eq:AKM_decomp}). Noting that $\hat{\mu}-\mu=\left(\sum_{n}s_{n}q_{n}q_{n}^{\prime}\right)^{-1}\sum_{n}s_{n}q_{n}\tilde{g}_{n}\xrightarrow{p}0$
under the assumptions completes the proof.

\paragraph{The Case of No Controls}

We show that when there are no controls besides a constant, i.e. $w_{\ell}=g_{n}=1$,
the standard errors are numerically the same. To prove this, it suffices
to show that $\hat{g}_{n}=\ddot{g}_{n}$. Absent controls, $\hat{g}_{n}=g_{n}-\sum_{n}s_{n}g_{n}$
is the $s_{n}$-weighted demeaned shock. The $\ddot{g}_{n}$ are obtained
as the projection coefficient of $z_{\ell}^{\perp}=z_{\ell}-\sum_{\ell}e_{\ell}z_{\ell}$
on the $N$ shares. Note that
\begin{align}
\sum_{\ell}e_{\ell}z_{\ell} & =\sum_{\ell}e_{\ell}\sum_{n}s_{\ell n}g_{n}=\sum_{n}s_{n}g_{n},
\end{align}
so that, with $\sum_{n}s_{\ell n}=1$,
\begin{align}
z_{\ell}-\sum_{\ell}e_{\ell}z_{\ell} & =\sum_{n}s_{\ell n}g_{n}-\sum_{n}s_{n}g_{n}=\sum_{\ell}s_{\ell n}\hat{g}_{n}.
\end{align}
This means that the projection in \textcite{Adao} has exact fit and
produces $\ddot{g}_{n}=\hat{g}_{n}$.

\paragraph*{Relaxing Assumption B4}

We now show that the standard errors from our equivalent regression
in Proposition 5 are asymptotically conservative under a weaker assumption
on the structure of controls than Assumption B4:
\begin{lyxlist}{00.00.0000}
\item [{\textbf{Assumption}}] \noindent \textbf{B4}$^{\prime}$:\emph{
}There exists a $K$-dimensional vector $p_{n}$, with uniformly bounded
fourth moments, such that $w_{\ell}=\sum_{n}s_{\ell n}p_{n}+u_{\ell}$
for some $K$-dimensional vector $u_{\ell}$ and $\expec{g_{n}\mid\mathcal{I}_{L}}=p_{n}^{\prime}\mu$
for all $n$ and for $\mathcal{I}_{L}=\left\{ \left\{ p_{n}\right\} _{n},\left\{ u_{\ell},\epsilon_{\ell},\eta_{\ell},\left\{ s_{\ell n},\pi_{\ell n}\right\} _{n},e_{\ell}\right\} _{\ell}\right\} $.
\end{lyxlist}
This assumption requires that the controls $w_{\ell}$ can be represented
as noisy versions of some latent shift-share confounding variables
$\sum_{n}s_{\ell n}p_{n}$. Since the variance of $u_{\ell}$ is unrestricted,
this assumption relaxes not only our Assumption B4 but also the assumption
of approximate shift-share controls in \textcite{Adao}.

Specifically, we show that under Assumptions B3, B4$^{\prime}$, B5,
and B6 the shock-level regression from Proposition 5 that controls
for a subvector of confounders $q_{n}\subseteq p_{n}$ yields asymptotically
conservative standard errors. Consider
\begin{equation}
\widehat{\Delta}_{L}=r_{L}\sum_{n}\left(\sum_{\ell}e_{\ell}s_{\ell n}\hat{\varepsilon}_{\ell}\right)^{2}\hat{g}_{n}^{2}-r_{L}\mathcal{V}_{L}
\end{equation}
with $\hat{g}_{n}$ still denoting the $s_{n}$-weighted projection
of $g_{n}$ on $q_{n}$ (only), and $\mathcal{V}_{L}=\sum_{n}\left(\sum_{\ell}e_{\ell}s_{\ell n}\varepsilon_{\ell}\right)^{2}\var{g_{n}\mid\mathcal{I}_{L}}$
where $\mathcal{I}_{L}$ is the expanded set from Assumption B4$^{\prime}$.
Write
\begin{align}
\hat{g}_{n} & =g_{n}-p_{n}^{\prime}\hat{\mu}\nonumber \\
 & =\tilde{g}_{n}+p_{n}^{\prime}\left(\mu-\hat{\mu}\right),\label{eq:bad_ghat}
\end{align}
where the non-zero elements of $\hat{\mu}$ correspond to the $q_{n}$
subvector. We show that
\begin{equation}
\widehat{\Delta}_{L}-\Delta_{L}\xrightarrow{p}0\label{eq:conserv-result}
\end{equation}
for the non-negative
\begin{align}
\Delta_{L} & =r_{L}\sum_{n}\left(\sum_{\ell}e_{\ell}s_{\ell n}\hat{\varepsilon}_{\ell}\right)^{2}\left(p_{n}^{\prime}\left(\mu-\hat{\mu}\right)\right)^{2},
\end{align}
when $\hat{\mu}=O_{p}(1)$.\footnote{We note this is a weaker condition than convergence of the incorrect
shock-level projection (i.e. that $\hat{\mu}=\overline{\mu}+o_{p}(1)$
for some $\overline{\mu}$).}

First note by equation (\ref{eq:bad_ghat}) that, for $\tilde{g}_{n}=g_{n}-p_{n}^{\prime}\mu$,
\begin{align*}
\hat{g}_{n}^{2} & =\text{\ensuremath{\tilde{g}_{n}^{2}}}+2\text{\ensuremath{\tilde{g}_{n}}}p_{n}^{\prime}\left(\mu-\hat{\mu}\right)+\left(p_{n}^{\prime}\left(\mu-\hat{\mu}\right)\right)^{2}.
\end{align*}
Thus
\begin{align}
\widehat{\Delta}_{L}-\Delta_{L}= & r_{L}\sum_{n}\left(\sum_{\ell}e_{\ell}s_{\ell n}\hat{\varepsilon}_{\ell}\right)^{2}\text{\ensuremath{\tilde{g}_{n}^{2}}}-r_{L}\mathcal{V}_{L}\nonumber \\
 & +2r_{L}\sum_{n}\left(\sum_{\ell}e_{\ell}s_{\ell n}\hat{\varepsilon}_{\ell}\right)^{2}\text{\ensuremath{\tilde{g}_{n}}}p_{n}^{\prime}\left(\mu-\hat{\mu}\right).\label{eq:conserv}
\end{align}
We showed that the first term is $o_{p}(1)$ in the proof of Proposition
5. It remains to show the second term is also $o_{p}(1)$. To see
this, write
\begin{align}
r_{L}\sum_{n}\left(\sum_{\ell}e_{\ell}s_{\ell n}\hat{\varepsilon}_{\ell}\right)^{2}\text{\ensuremath{\tilde{g}_{n}}}p_{n}= & r_{L}\sum_{n}\left(\sum_{\ell}e_{\ell}s_{\ell n}\varepsilon_{\ell}\right)^{2}\text{\ensuremath{\tilde{g}_{n}}}p_{n}\label{eq:conserv_end}\\
 & +r_{L}\sum_{n}\left(\left(\sum_{\ell}e_{\ell}s_{\ell n}\hat{\varepsilon}_{\ell}\right)^{2}-\left(\sum_{\ell}e_{\ell}s_{\ell n}\varepsilon_{\ell}\right)^{2}\right)\text{\ensuremath{\tilde{g}_{n}}}p_{n}.
\end{align}
We have
\begin{align}
\expec{r_{L}\sum_{n}\left(\sum_{\ell}e_{\ell}s_{\ell n}\varepsilon_{\ell}\right)^{2}\tilde{g}_{n}p_{n}} & =\expec{r_{L}\sum_{n}\left(\sum_{\ell}e_{\ell}s_{\ell n}\varepsilon_{\ell}\right)^{2}\expec{\tilde{g}_{n}\mid\mathcal{I}_{L}}p_{n}}=0
\end{align}
since $\expec{\tilde{g}_{n}\mid\mathcal{I}_{L}}$. Furthermore, 
\begin{align}
\var{r_{L}\sum_{n}\left(\sum_{\ell}e_{\ell}s_{\ell n}\varepsilon_{\ell}\right)^{2}\tilde{g}_{n}p_{n}} & =r_{L}^{2}\sum_{n}\expec{\left(\sum_{\ell}e_{\ell}s_{\ell n}\varepsilon_{\ell}\right)^{4}\var{\tilde{g}_{n}\mid\mathcal{I}_{L}}p_{n}p_{n}^{\prime}}\rightarrow0,
\end{align}
implying the first term of (\ref{eq:conserv_end}) is $o_{p}(1)$.
The second term of this expression can also be shown to be $o_{p}(1)$
by applying the lemma from \textcite{Adao} and the representation
used in equation (\ref{eq:AKM_decomp}). Thus
\begin{align}
2r_{L}\sum_{n}\left(\sum_{\ell}e_{\ell}s_{\ell n}\hat{\varepsilon}_{\ell}\right)^{2}\text{\ensuremath{\tilde{g}_{n}}}p_{n}^{\prime}\left(\mu-\hat{\mu}\right) & =o_{p}(1)^{\prime}O_{p}(1),
\end{align}
completing the proof.

\paragraph{Comparison of Standard Errors under Assumption B4}

The characterization of the standard errors in equations (\ref{eq:se_equiv})\textendash (\ref{eq:akmSEs})
also offers insights into how these standard errors may differ in
presence of controls, when both standard error calculations are asymptotically
valid. We argue that under the conditions of Proposition 5, our standard
errors are likely smaller in finite samples. More precisely, we show
that the homoskedastic version of (\ref{eq:se_equiv}) is smaller
than the homoskedastic version of (\ref{eq:akmSEs}). This is suggestive
of the comparison under heteroskedasticity, but is not a proof.

To see this, consider versions of the two standard error formulas
obtained under shock homoskedasticity (i.e. $\var{g_{n}\mid\mathcal{I}_{L}}=\sigma_{g}^{2}$):
\begin{align}
\widehat{se}_{\text{equiv}}^{\text{homo}} & =\frac{\sqrt{\left(\sum_{n}s_{n}^{2}\hat{\varepsilon}_{n}^{2}\right)\left(\sum_{n}s_{n}\hat{g}_{n}^{2}\right)}}{\left|\sum_{n}s_{n}\bar{x}_{n}^{\perp}g_{n}\right|}\\
\widehat{se}_{\text{AKM}}^{\text{homo}} & =\frac{\sqrt{\left(\sum_{n}s_{n}^{2}\hat{\varepsilon}_{n}^{2}\right)\left(\sum_{n}s_{n}\ddot{g}_{n}^{2}\right)}}{\left|\sum_{\ell}e_{\ell}x_{\ell}^{\perp}z_{\ell}\right|},
\end{align}
which differ by a factor of $\sqrt{\sum_{n}s_{n}\hat{g}_{n}^{2}/\sum_{n}s_{n}\ddot{g}_{n}^{2}}$.

When the SSIV controls have an exact shift-share structure, $w_{\ell}=\sum_{n}s_{\ell n}q_{n}$,
the share projection producing $\ddot{g}_{n}$ has exact fit such
that one can represent $\ddot{g}_{n}=g_{n}-q_{n}^{\prime}\hat{\mu}_{AKM}$
for some $\hat{\mu}_{AKM}$. In this case the $s_{n}$-weighted sum
of squares of shock residuals is lower in our equivalent regression
by construction of $\hat{\mu}$: $\sum_{n}s_{n}\hat{g}_{n}^{2}\le\sum_{n}s_{n}\ddot{g}_{n}^{2}$
(with strict inequality when $\hat{\mu}_{\text{AKM}}\ne\hat{\mu}$).
Similarly, when $w_{\ell}$ instead contains controls that are included
for efficiency only and are independent of the shocks, projection
of $z_{\ell}$ on the shares produces a noisy estimate of $g_{n}-\sum_{n}s_{n}g_{n}$,
which again has a higher weighted sum of squares.

\paragraph{Null-Imposed Inference Procedure}

Finally, our shock-level equivalence provides a convenient implementation
for the alternative inference procedure that may have superior finite-sample
performance. \textcite{Adao} show how standard errors that impose
a given null hypothesis $\beta=\beta_{0}$ in estimating the residual
$\varepsilon_{\ell}$ can generate confidence intervals with better
coverage in situations with few shocks (and a similar argument can
be made in the case of shocks with a heavy-tailed distribution).\footnote{As explained by \textcite{Adao}, the problem that this ``AKM0''
confidence interval addresses generalizes the standard finite-sample
bias of cluster-robust standard errors with few clusters \parencite{ColinCameron2015}.
With few or heavy-tailed shocks, estimates of the residual variance
will tend to be biased downwards, leading to undercoverage of confidence
intervals based on standard errors that do not impose the null.} Building on Proposition 5, such confidence intervals can be constructed
in the same way as in any regular shock-level IV regression. To test
$\beta=\beta_{0}$, one regresses $\bar{y}_{n}^{\perp}-\beta_{0}\bar{x}_{n}^{\perp}$
on the shocks $g_{n}$ (weighting by $s_{n}$ and including any relevant
shock-level controls $q_{n}$) and uses a null-imposed residual variance
estimate. This procedure corresponds to the standard shock-level Lagrange
multiplier test for $\beta=\beta_{0}$ that can be implemented by
standard statistical software.\footnote{For example in Stata one can use the \emph{ivreg2} overidentification
test statistic from regressing $\bar{y}_{n}^{\perp}-\beta_{0}\bar{x}_{n}^{\perp}$
on $q_{n}$ with no endogenous variables and with $g_{n}$ specified
as the instrument (again with $s_{n}$ weights).} The confidence interval for $\beta$ is constructed by collecting
all candidate $\beta_{0}$ that are not rejected.

\subsection{Proposition A1\label{sec:A1_pf}}

We consider each expectation in equation (\ref{eq:hetfx_SSIV}) in
turn. For each $n$, write 
\begin{align}
\kappa_{n}(g_{-n},\varepsilon_{\ell},\eta_{\ell}) & =\lim_{g_{n}\rightarrow-\infty}y(x_{1}([g_{n};g_{-n}],\eta_{\ell1}),\dots,x_{R}([g_{n};g_{-n}],\eta_{\ell R}),\varepsilon_{\ell})
\end{align}
such that
\begin{align}
s_{\ell n}e_{\ell}g_{n}y_{\ell}= & s_{\ell n}e_{\ell}g_{n}\kappa_{n}(g_{-n},\varepsilon_{\ell},\eta_{\ell})\\
 & +s_{\ell n}e_{\ell}g_{n}\int_{-\infty}^{g_{n}}\frac{\partial}{\partial g_{n}}y(x_{1}([\gamma;g_{-n}],\eta_{\ell1}),\dots x_{R}([\gamma;g_{-n}],\eta_{\ell R}),\varepsilon_{\ell})d\gamma.\nonumber 
\end{align}
By as-good-as-random shock assignment, the expectation of the first
term is
\begin{align}
\expec{s_{\ell n}e_{\ell}g_{n}\kappa_{n}(g_{-n},\varepsilon_{\ell},\eta_{\ell})} & =\expec{s_{\ell n}e_{\ell}\expec{g_{n}\mid s,e,g_{-n},\varepsilon,\eta_{\ell}}\kappa_{n}(g_{-n},\varepsilon_{\ell},\eta_{\ell})}=0,
\end{align}
while the expectation of the second is
\begin{align}
 & \expec{s_{\ell n}e_{\ell}g_{n}\int_{-\infty}^{g_{n}}\frac{\partial}{\partial g_{n}}y(x_{1}([\gamma;g_{-n}],\eta_{\ell1}),\dots x_{R}([\gamma;g_{-n}],\eta_{\ell R}),\varepsilon_{\ell})d\gamma}\nonumber \\
 & =\expec{s_{\ell n}e_{\ell}\int_{-\infty}^{\infty}\int_{-\infty}^{g_{n}}g_{n}\frac{\partial}{\partial g_{n}}y(x_{1}([\gamma;g_{-n}],\eta_{\ell1}),\dots x_{R}([\gamma;g_{-n}],\eta_{\ell R}),\varepsilon_{\ell})d\gamma dF_{n}(g_{n}\mid\mathcal{I})}\nonumber \\
 & =\expec{s_{\ell n}e_{\ell}\int_{-\infty}^{\infty}\frac{\partial}{\partial g_{n}}y(x_{1}([\gamma;g_{-n}],\eta_{\ell1}),\dots x_{R}([\gamma;g_{-n}],\eta_{\ell R}),\varepsilon_{\ell})\int_{\gamma}^{\infty}g_{n}dF_{n}(g_{n}\mid\mathcal{I})d\gamma}\label{eq:hetFX-long}
\end{align}
where $F_{n}(\cdot\mid\mathcal{I})$ denotes the conditional distribution
of $g_{n}$. Thus
\begin{align}
\expec{s_{\ell n}e_{\ell}g_{n}y_{\ell}} & =\expec{s_{\ell n}e_{\ell}\int_{-\infty}^{\infty}\frac{\partial}{\partial g_{n}}y(x_{1}([\gamma;g_{-n}],\eta_{\ell}),\dots x_{R}([\gamma;g_{-n}],\eta_{\ell}),\varepsilon_{\ell})\mu_{n}(\gamma\mid\mathcal{I})d\gamma}\nonumber \\
 & =\sum_{r}\expec{\int_{-\infty}^{\infty}s_{\ell n}e_{\ell}\alpha_{\ell r}\pi_{\ell rn}([\gamma;g_{-n}])\mu_{n}(\gamma\mid\mathcal{I})\tilde{\beta}_{\ell rn}(\gamma)d\gamma}\label{eq:hetfx_rf}
\end{align}
where
\begin{align}
\mu_{n}(\gamma\mid\mathcal{I})\equiv & \int_{\gamma}^{\infty}g_{n}dF_{n}(g_{n}\mid\mathcal{I}).\nonumber \\
= & \left(\expec{g_{n}\mid g_{n}\ge\gamma,\mathcal{I}}-\expec{g_{n}\mid g_{n}<\gamma,\mathcal{I}}\right)Pr\left(g_{n}\ge\gamma\mid\mathcal{I}\right)\left(1-Pr\left(g_{n}\ge\gamma\mid\mathcal{I}\right)\right)\ge0\text{ }a.s.
\end{align}
Similarly
\begin{align}
\expec{s_{\ell n}e_{\ell}g_{n}x_{\ell}} & =\sum_{r}\expec{{\it \int_{-\infty}^{\infty}}s_{\ell n}e_{\ell}\alpha_{\ell r}\pi_{\ell rn}([\gamma;g_{-n}])\mu_{n}(\gamma\mid\mathcal{I})d\gamma}.\label{eq:hetfx_fs}
\end{align}
Combining equations (\ref{eq:hetfx_rf}) and (\ref{eq:hetfx_fs})
completes the proof, with 
\begin{align}
\omega_{\ell rn}(\gamma) & =s_{\ell n}e_{\ell}\alpha_{\ell r}\mu_{n}(\gamma\mid\mathcal{I})\pi_{\ell rn}([\gamma;g_{-n}])\ge0\text{ }\text{a.s.}
\end{align}

\subsection{Proposition A2\label{sec:A2_pf}}

By definition of $\bar{\varepsilon}_{n}$,
\begin{align}
\bar{\varepsilon}_{n} & =\frac{\sum_{\ell}e_{\ell}s_{\ell n}\left(\sum_{n^{\prime}}s_{\ell n^{\prime}}\nu_{n^{\prime}}+\check{\varepsilon}_{\ell}\right)}{\sum_{\ell}e_{\ell}s_{\ell n}}\nonumber \\
 & \equiv\sum_{n^{\prime}}\alpha_{nn^{\prime}}\nu_{n^{\prime}}+\bar{\check{\varepsilon}}_{n},
\end{align}
for $\alpha_{nn^{\prime}}=\frac{\sum_{\ell}e_{\ell}s_{\ell n}s_{\ell n^{\prime}}}{\sum_{\ell}e_{\ell}s_{\ell n}}$
and $\bar{\check{\varepsilon}}_{n}=\frac{\sum_{\ell}e_{\ell}s_{\ell n}\check{\varepsilon}_{\ell}}{\sum_{\ell}e_{\ell}s_{\ell n}}$.
Therefore,
\begin{align}
\var{\bar{\varepsilon}_{n}} & =\sum_{n^{\prime}}\sigma_{n^{\prime}}^{2}\alpha_{nn'}^{2}+\var{\bar{\check{\varepsilon}}_{n}}\nonumber \\
 & \ge\sigma_{\nu}^{2}\alpha_{nn}^{2},
\end{align}
and
\begin{equation}
\max_{n}\var{\bar{\varepsilon}_{n}}\ge\sigma_{\nu}^{2}\max_{n}\alpha_{nn}^{2}.
\end{equation}

To establish a lower bound on this quantity, observe that the $s_{n}$-weighted
average of $\alpha_{nn}$ satisfies:
\begin{align}
\sum_{n}s_{n}\alpha_{nn} & =\sum_{n}s_{n}\frac{\sum_{\ell}e_{\ell}s_{\ell n}^{2}}{s_{n}}\nonumber \\
 & =H_{L}.
\end{align}
Since $\sum_{n}s_{n}=1$, it follows that $\max_{n}\alpha_{nn}\ge H_{L}$
and therefore $\max_{n}\var{\bar{\varepsilon}_{n}}\ge\sigma_{\nu}^{2}H_{L}^{2}.$
Since $H_{L}\stackrel{}{\to}\bar{H}>0$, we conclude that, for sufficiently
large $L$, $\max_{n}\var{\bar{\varepsilon}_{n}}$ is bounded from
below by any positive $\delta<\sigma_{\nu}^{2}\bar{H}^{2}$.

\subsection{Proposition A3\label{sec:A3_pf}}

To prove (\ref{eq:spillovers_SSIV}), we aggregate (\ref{eq:spillover})
across industries within a region using $E_{\ell n}$ weights:
\begin{equation}
y_{\ell}=\left(\beta_{0}-\beta_{1}\right)x_{\ell}+\varepsilon_{\ell},
\end{equation}
where $\varepsilon_{\ell}=\sum_{n}s_{\ell n}\varepsilon_{\ell n}$.
The shift-share instrument $z_{\ell}$ is relevant because
\begin{align}
\expec{\sum_{\ell}e_{\ell}x_{\ell}z_{\ell}} & =\sum_{\ell}e_{\ell}\expec{\sum_{n}s_{\ell n}\left(\bar{\pi}g_{n}+\eta_{\ell n}\right)\cdot\sum_{n^{\prime}}s_{\ell n^{\prime}}g_{n^{\prime}}}\nonumber \\
 & =\sum_{\ell,n}e_{\ell}s_{\ell n}^{2}\bar{\pi}\sigma_{g}^{2}\nonumber \\
 & \ge\bar{H}_{L}\bar{\pi}\sigma_{g}^{2},
\end{align}
while exclusion holds because
\begin{align}
\expec{\sum_{\ell}e_{\ell}z_{\ell}\varepsilon_{\ell}} & =\sum_{\ell}e_{\ell}\expec{\sum_{n}s_{\ell n}\varepsilon_{\ell n}\cdot\sum_{n^{\prime}}s_{\ell n^{\prime}}g_{n^{\prime}}}\nonumber \\
 & =0.
\end{align}
Thus by an appropriate law of large numbers, $\hat{\beta}=\beta_{0}-\beta_{1}+o_{p}(1)$.

To study $\hat{\beta}_{\text{ind}}$, we aggregate (\ref{eq:spillover})
across regions (again with $E_{\ell n}$ weights):
\begin{equation}
y_{n}=\beta_{0}x_{n}-\beta_{1}\sum_{\ell}\omega_{\ell n}\sum_{n^{\prime}}s_{\ell n^{\prime}}x_{\ell n^{\prime}}+\varepsilon_{n},
\end{equation}
for $\varepsilon_{n}=\sum_{\ell}\omega_{\ell n}\varepsilon_{\ell n}$.
The resulting IV estimate yields
\begin{align}
\hat{\beta}_{\text{ind}}-\beta_{0} & =\frac{\sum_{n}s_{n}y_{n}g_{n}}{\sum_{n}s_{n}x_{n}g_{n}}-\beta_{0}\nonumber \\
 & =\frac{\sum_{n}s_{n}\left(-\beta_{1}\sum_{\ell}\omega_{\ell n}\sum_{n^{\prime}}s_{\ell n^{\prime}}x_{\ell n^{\prime}}+\varepsilon_{n}\right)g_{n}}{\sum_{n}s_{n}x_{n}g_{n}}.\label{eq:spillovers_ind_bias}
\end{align}
The expected denominator of $\hat{\beta}_{\text{ind}}$ is non-zero:
\begin{align}
\expec{\sum_{n}s_{n}x_{n}g_{n}} & =\sum_{n}s_{n}\expec{\sum_{\ell}\omega_{\ell n}\left(\bar{\pi}g_{n}+\eta_{\ell n}\right)g_{n}}\nonumber \\
 & =\sum_{n}s_{n}\omega_{\ell n}\bar{\pi}\sigma^{2}\nonumber \\
 & =\sum_{n}\frac{E_{n}}{E}\cdot\frac{E_{\ell n}}{E}\bar{\pi}\sigma^{2}\nonumber \\
 & =\bar{\pi}\sigma^{2},
\end{align}
while the expected numerator is
\begin{align}
\expec{\sum_{n}s_{n}\left(-\beta_{1}\sum_{\ell}\omega_{\ell n}\sum_{n^{\prime}}s_{\ell n^{\prime}}x_{\ell n^{\prime}}+\varepsilon_{n}\right)g_{n}} & =-\beta_{1}\sum_{n,\ell}s_{n}\omega_{\ell n}s_{\ell n}\bar{\pi}\sigma^{2}\nonumber \\
 & =-\beta_{1}H_{L}\bar{\pi}\sigma^{2},
\end{align}
where the last equality follows because 
\begin{align}
\sum_{n,\ell}s_{n}\omega_{\ell n}s_{\ell n} & =\sum_{n,\ell}\frac{E_{n}}{E}\frac{E_{\ell n}}{E_{n}}\frac{E_{\ell n}}{E_{\ell}}\nonumber \\
 & =\sum_{n,\ell}\frac{E_{\ell}}{E}\left(\frac{E_{\ell n}}{E_{\ell}}\right)^{2}\nonumber \\
 & =\sum_{n,\ell}e_{\ell}s_{\ell n}^{2}\nonumber \\
 & =H_{L}.
\end{align}
Thus by an appropriate law of large numbers,
\begin{align}
\hat{\beta}_{\text{ind}} & =\beta_{0}-\beta_{1}H_{L}+o_{p}(1).
\end{align}

\subsection{Proposition A4\label{sec:A4_pf}}

By appropriate laws of large numbers,
\begin{align}
\hat{\beta} & =\frac{\expec{\sum_{\ell}E_{\ell}\left(\sum_{n}s_{\ell n}y_{\ell n}\right)\left(\sum_{n^{\prime}}s_{\ell n^{\prime}}g_{n^{\prime}}\right)}}{\expec{\sum_{\ell}E_{\ell}\left(\sum_{n}s_{\ell n}x_{\ell n}\right)\left(\sum_{n^{\prime}}s_{\ell n^{\prime}}g_{n^{\prime}}\right)}}+o_{p}(1)\nonumber \\
 & =\frac{\sum_{\ell,n}E_{\ell}s_{\ell n}^{2}\pi_{\ell n}\sigma_{n}^{2}\beta_{\ell n}}{\sum_{\ell,n}E_{\ell}s_{\ell n}^{2}\pi_{\ell n}\sigma_{n}^{2}}+o_{p}(1)\nonumber \\
 & =\frac{\sum_{\ell,n}E_{\ell n}s_{\ell n}\pi_{\ell n}\sigma_{n}^{2}\beta_{\ell n}}{\sum_{\ell,n}E_{\ell n}s_{\ell n}\pi_{\ell n}\sigma_{n}^{2}}+o_{p}(1)
\end{align}
while
\begin{align}
\hat{\beta}_{\text{ind}} & =\frac{\sum_{n}E_{n}y_{n}g_{n}}{\sum_{n}E_{n}x_{n}g_{n}}\nonumber \\
 & =\frac{\expec{\sum_{n}E_{n}\left(\sum_{\ell}\omega_{\ell n}y_{\ell n}\right)g_{n}}}{\expec{\sum_{n}E_{n}\left(\sum_{\ell}\omega_{\ell n}x_{\ell n}\right)g_{n}}}+o_{p}(1)\nonumber \\
 & =\frac{\sum_{\ell,n}E_{n}\omega_{\ell n}\pi_{\ell n}\sigma_{n}^{2}\beta_{\ell n}}{\sum_{\ell,n}E_{n}\omega_{\ell n}\pi_{\ell n}\sigma_{n}^{2}}+o_{p}(1)\nonumber \\
 & =\frac{\sum_{\ell,n}E_{\ell n}\pi_{\ell n}\sigma_{n}^{2}\beta_{\ell n}}{\sum_{\ell,n}E_{\ell n}\pi_{\ell n}\sigma_{n}^{2}}+o_{p}(1).
\end{align}

\subsection{Proposition A5\label{sec:A5_pf}}

\noindent We prove each part of this proposition in turn.
\begin{enumerate}
\item Expanding the moment condition yields:
\begin{align}
\expec{\sum_{\ell}e_{\ell}\varepsilon_{\ell}\psi_{\ell,LOO}} & =\sum_{\ell}\expec{e_{\ell}\varepsilon_{\ell}\sum_{n}s_{\ell n}\frac{\sum_{\ell^{\prime}\ne\ell}\omega_{\ell^{\prime}n}\psi_{\ell^{\prime}n}}{\sum_{\ell^{\prime}\ne\ell}\omega_{\ell'n}}}\nonumber \\
 & =\sum_{\ell}e_{\ell}\sum_{n}s_{\ell n}\frac{\sum_{\ell^{\prime}\ne\ell}\omega_{\ell'n}\expec{\varepsilon_{\ell}\psi_{\ell^{\prime}n}}}{\sum_{\ell^{\prime}\ne\ell}\omega_{\ell^{\prime}n}}\nonumber \\
 & =0.
\end{align}
\item The assumption of part (1) is satisfied here, so $\expec{\sum_{\ell}e_{\ell}\varepsilon_{\ell}\psi_{\ell,LOO}}=0$.
We now establish that $\expec{\left(\sum_{\ell}e_{\ell}\varepsilon_{\ell}\psi_{\ell,LOO}\right)^{2}}\rightarrow0$,
which implies $\sum_{\ell}e_{\ell}\varepsilon_{\ell}\psi_{\ell,LOO}\stackrel{p}{\to}0$
and thus consistency of the LOO SSIV estimator provided it has a first
stage:
\begin{align}
\expec{\left(\sum_{\ell}e_{\ell}\varepsilon_{\ell}\psi_{\ell,LOO}\right)^{2}} & =\sum_{\substack{\ell_{1},\ell_{2},n_{1},n_{2},\\
\ell_{1}^{\prime}\ne\ell_{1},\ell_{2}^{\prime}\ne\ell_{2}
}
}e_{\ell_{1}}e_{\ell_{2}}s_{\ell_{1}n_{1}}s_{\ell_{2}n_{2}}\frac{\omega_{\ell_{1}^{\prime}n_{1}}}{\sum_{\ell\ne\ell_{1}}\omega_{\ell n_{1}}}\frac{\omega_{\ell_{2}^{\prime}n_{2}}}{\sum_{\ell\ne\ell_{2}}\omega_{\ell n_{2}}}\expec{\varepsilon_{\ell_{1}}\varepsilon_{\ell_{2}}\psi_{\ell_{1}^{\prime}n_{1}}\psi_{\ell_{2}^{\prime}n_{2}}}\nonumber \\
 & \le\sum_{\substack{\left(\ell_{1},\ell_{2},\ell_{1}^{\prime},\ell_{2}^{\prime}\right)\in\mathcal{J},\\
n_{1},n_{2}
}
}e_{\ell_{1}}e_{\ell_{2}}s_{\ell_{1}n_{1}}s_{\ell_{2}n_{2}}\frac{\omega_{\ell_{1}^{\prime}n_{1}}}{\sum_{\ell\ne\ell_{1}}\omega_{\ell n_{1}}}\frac{\omega_{\ell_{2}^{\prime}n_{2}}}{\sum_{\ell\ne\ell_{2}}\omega_{\ell n_{2}}}\cdot B\to0.
\end{align}
Here the second line used the first regularity condition, which implies
that $\expec{\varepsilon_{\ell_{1}}\varepsilon_{\ell_{2}}\psi_{\ell_{1}^{\prime}n_{1}}\psi_{\ell_{2}^{\prime}n_{2}}}=0$
whenever there is at least one index among $\left\{ \ell_{1},\ell_{2},\ell_{1}^{\prime},\ell_{2}^{\prime}\right\} $
which is not equal to any of the others, i.e. for all $\left(\ell_{1},\ell_{2},\ell_{1}^{\prime},\ell_{2}^{\prime}\right)\not\in\mathcal{J}$.
\item We show that under the given assumptions on $s_{\ell n}$, $e_{\ell}$,
and $\omega_{\ell n}$, the expression in (\ref{eq:LOO_sum}) is bounded
by $4N/L$:
\begin{align}
 & \sum_{\substack{\left(\ell_{1},\ell_{2},\ell_{1}^{\prime},\ell_{2}^{\prime}\right)\in\mathcal{J},\\
n_{1},n_{2}
}
}e_{\ell_{1}}e_{\ell_{2}}s_{\ell_{1}n_{1}}s_{\ell_{2}n_{2}}\frac{\omega_{\ell_{1}^{\prime}n_{1}}}{\sum_{\ell\ne\ell_{1}}\omega_{\ell n_{1}}}\frac{\omega_{\ell_{2}^{\prime}n_{2}}}{\sum_{\ell\ne\ell_{2}}\omega_{\ell n_{2}}}\nonumber \\
 & =\sum_{\left(\ell_{1},\ell_{2},\ell_{1}^{\prime},\ell_{2}^{\prime}\right)\in\mathcal{J}}\frac{1}{L^{2}}\frac{\omega_{\ell_{1}^{\prime}n\left(\ell_{1}\right)}}{\sum_{\ell\ne\ell_{1}}\omega_{\ell n\left(\ell_{1}\right)}}\frac{\omega_{\ell_{2}^{\prime}n\left(\ell_{2}\right)}}{\sum_{\ell\ne\ell_{2}}\omega_{\ell n\left(\ell_{2}\right)}}\nonumber \\
 & =\frac{1}{L^{2}}\sum_{\substack{\left(\ell_{1},\ell_{2},\ell_{1}^{\prime},\ell_{2}^{\prime}\right)\in\mathcal{J}\\
n(\ell_{1}^{\prime})=n(\ell_{1}),\\
n(\ell_{2}^{\prime})=n(\ell_{2})
}
}\frac{1}{L_{n(\ell_{1})}-1}\frac{1}{L_{n(\ell_{2})}-1}\nonumber \\
 & =\frac{1}{L^{2}}\sum_{n}\frac{2L_{n}\left(L_{n}-1\right)}{\left(L_{n}-1\right)^{2}}\stackrel{}{\le}4\frac{N}{L}.
\end{align}
Here the second line plugs in the expressions for $s_{\ell n}$ and
$e_{\ell}$, and the third line plugs in $\omega_{\ell n}$. The last
line uses the fact that any tuple $\left(\ell_{1},\ell_{2},\ell_{1}^{\prime},\ell_{2}^{\prime}\right)\in\mathcal{J}$
such that $n(\ell_{1}^{\prime})=n(\ell_{1})$ and $n(\ell_{2}^{\prime})=n(\ell_{2})$
has all four elements exposed to the same shock $n$. Moreover, it
is easily verified that all of these tuples have a structure $\left(\ell_{A},\ell_{B},\ell_{A},\ell_{B}\right)$
or $\left(\ell_{A},\ell_{B},\ell_{B},\ell_{A}\right)$ for any $\ell_{A}\ne\ell_{B}$
exposed to the same shock. Therefore, there are $2L_{n}\left(L_{n}-1\right)$
of them for each $n$. Finally, $\frac{L_{n}}{L_{n}-1}\le2$ as $L_{n}\ge2$.
\end{enumerate}

\subsection{Proposition A6\label{sec:A6_pf}}

National industry employment satisfies $E_{n}=\sum_{\ell}E_{\ell n}$;
log-linearizing this immediately implies (\ref{eq:natl-empl-growth}).
To solve for $g_{\ell n}$, log-linearize (\ref{eq:labor-supply}),
(\ref{eq:labor-demand}), and (\ref{eq:local-eqm}):
\begin{align}
\hat{E}_{\ell} & =\phi\hat{W}_{\ell}+\varepsilon_{\ell},\\
g_{\ell n} & =g_{n}^{\ast}+\hat{\xi}_{\ell n}-\sigma\hat{W}_{\ell},\\
\hat{E}_{\ell} & =\sum_{n}s_{\ell n}g_{\ell n}.
\end{align}
Solving this system of equations yields
\begin{equation}
\hat{W}_{\ell}=\frac{1}{\sigma+\phi}\left(\sum_{n}s_{\ell n}\left(g_{n}^{\ast}+\hat{\xi}_{\ell n}\right)-\varepsilon_{\ell}\right)
\end{equation}
and expression (\ref{eq:g_ln_model}).

\newpage{}

\section{Appendix Figures and Tables}

\vspace*{\fill}
\begin{center}
Figure C1: Industry-Level Variation in the \textcite{AutorDorn2001}
Setting
\par\end{center}

\vspace*{-0.2cm}

\vspace*{-0.2cm}
\begin{center}
{\small{}}%
\begin{tabular}{cc}
{\small{}First stage} & {\small{}Reduced form}\tabularnewline
{\small{}\includegraphics[width=0.35\paperwidth]{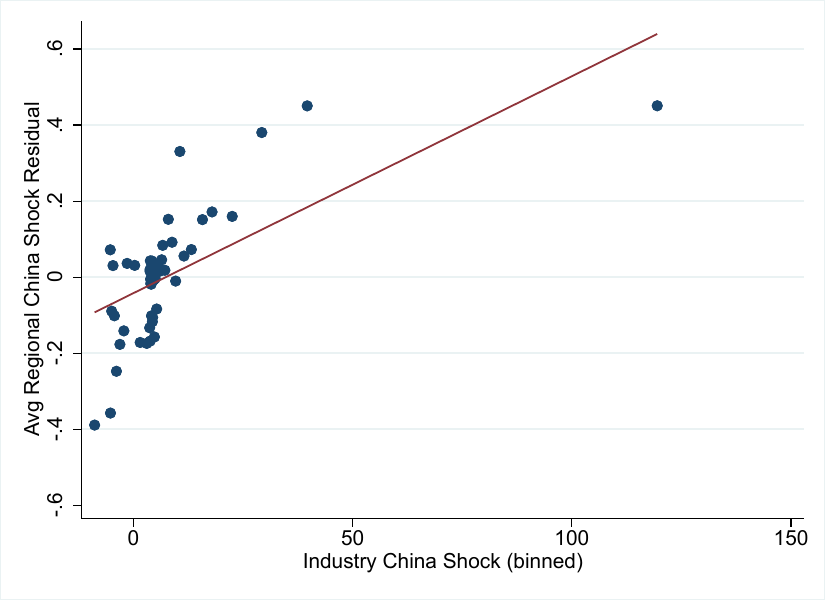}} & {\small{}\includegraphics[width=0.35\paperwidth]{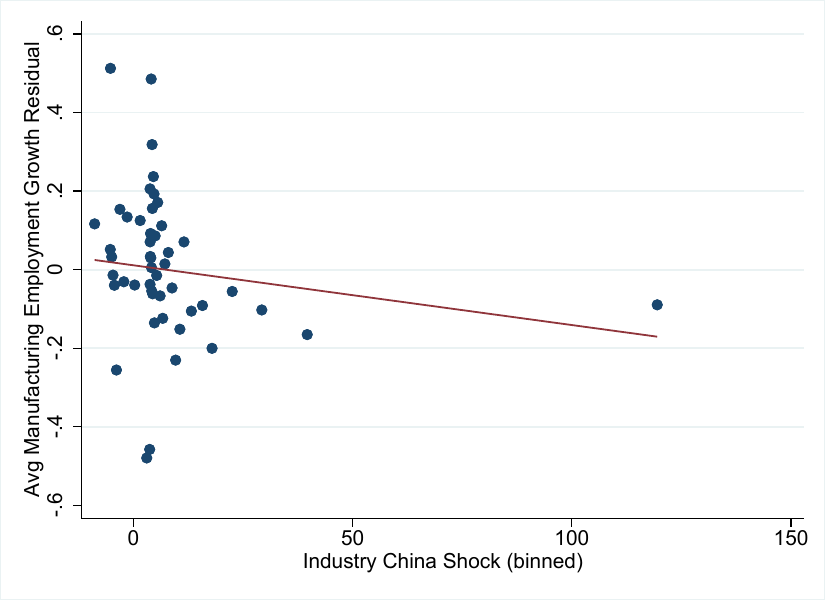}}\tabularnewline
\end{tabular}{\small\par}
\par\end{center}

\begin{singlespace}
\vspace*{-0.2cm}

{\footnotesize{}\noindent Notes: This figure shows binned scatterplots
of shock-level outcome and treatment residuals, $\bar{y}_{nt}^{\perp}$
and $\bar{x}_{nt}^{\perp}$, corresponding to the SSIV specification
in column 3 of Table 4. The manufacturing industry shocks, $g_{nt}$,
are residualized on period indicators (with the full-sample mean added
back) and grouped into fifty weighted bins, with each bin representing
around 2\% of total share weight $s_{nt}$. Lines of best fit, indicated
in red, are weighted by the same $s_{nt}$. The slope coefficients
equal $5.71\times10^{-3}$ and $-1.52\times10^{-3}$, respectively,
with the ratio (-0.267) equaling the SSIV coefficient in column 3
of Table 4.}{\footnotesize\par}
\end{singlespace}

\vspace*{\fill}

\newpage{}

\vspace*{\fill}
\begin{center}
Table C1: Shift-Share IV Estimates of the Effect of Chinese Imports
on Other Outcomes
\par\end{center}

\begin{center}
{\small{}}%
\begin{tabular}{lccccccc}
\toprule 
 & {\small{}(1)} & {\small{}(2)} & {\small{}(3)} & {\small{}(4)} & {\small{}(5)} & {\small{}(6)} & {\small{}(7)}\tabularnewline
\midrule
\midrule 
{\small{}Unemployment growth} & {\small{}0.221} & {\small{}0.217} & {\small{}0.063} & {\small{}-0.014} & {\small{}0.104} & {\small{}0.107} & {\small{}0.235}\tabularnewline
 & {\small{}(0.049)} & {\small{}(0.046)} & {\small{}(0.060)} & {\small{}(0.079)} & {\small{}(0.079)} & {\small{}(0.083)} & {\small{}(0.178)}\tabularnewline\addlinespace
{\small{}NILF growth} & {\small{}0.553} & {\small{}0.534} & {\small{}0.098} & {\small{}0.149} & {\small{}0.142} & {\small{}0.117} & {\small{}0.187}\tabularnewline
 & {\small{}(0.185)} & {\small{}(0.183)} & {\small{}(0.133)} & {\small{}(0.083)} & {\small{}(0.155)} & {\small{}(0.161)} & {\small{}(0.297)}\tabularnewline\addlinespace
{\small{}Log weekly wage growth} & {\small{}-0.759} & {\small{}-0.607} & {\small{}0.227} & {\small{}0.320} & {\small{}0.145} & {\small{}0.063} & {\small{}-0.211}\tabularnewline
 & {\small{}(0.258)} & {\small{}(0.226)} & {\small{}(0.242)} & {\small{}(0.209)} & {\small{}(0.264)} & {\small{}(0.260)} & {\small{}(0.651)}\tabularnewline\addlinespace[0.3cm]
{\small{}\# of industry-periods} & {\small{}796} & {\small{}794} & {\small{}794} & {\small{}794} & {\small{}794} & {\small{}794} & {\small{}794}\tabularnewline
{\small{}\# of region-periods} & {\small{}1,444} & {\small{}1,444} & {\small{}1,444} & {\small{}1,444} & {\small{}1,444} & {\small{}1,444} & {\small{}1,444}\tabularnewline
\bottomrule
\end{tabular}{\small\par}
\par\end{center}

\begin{singlespace}
\vspace*{-0.2cm}

{\footnotesize{}\noindent Notes: This table extends the analysis
of Table 4 to different regional outcomes in \textcite{AutorDorn2001}:
unemployment growth, labor force non-participation (NILF) growth,
and log average weekly wage growth. The specifications are otherwise
the same as in the corresponding columns of Table 4. SIC3-clustered
exposure-robust standard errors are computed using equivalent industry-level
IV regressions and reported in parentheses}.
\end{singlespace}

\vspace*{\fill}

\newpage{}
\begin{center}
\vspace*{\fill}
\par\end{center}

\begin{center}
Table C2: Alternative Standard Errors in the \textcite{AutorDorn2001}
Setting\vspace{2.5mm}
\par\end{center}

\begin{center}
{\small{}}%
\begin{tabular}{lccccccc}
\toprule 
 & {\small{}(1)} & {\small{}(2)} & {\small{}(3)} & {\small{}(4)} & {\small{}(5)} & {\small{}(6)} & {\small{}(7)}\tabularnewline
\midrule
\midrule 
{\small{}Coefficient} & {\small{}-0.596} & {\small{}-0.489} & {\small{}-0.267} & {\small{}-0.314} & {\small{}-0.310} & {\small{}-0.290} & {\small{}-0.432}\tabularnewline
{\small{}Table 4 SE} & {\small{}(0.114)} & {\small{}(0.100)} & {\small{}(0.099)} & {\small{}(0.107)} & {\small{}(0.134)} & {\small{}(0.129)} & {\small{}(0.205)}\tabularnewline\addlinespace[0.3cm]
{\small{}State-clustered SE} & {\small{}(0.099)} & {\small{}(0.086)} & {\small{}(0.086)} & {\small{}(0.097)} & {\small{}(0.104)} & {\small{}(0.101)} & {\small{}(0.193)}\tabularnewline\addlinespace[0.3cm]
{\small{}\textcite{Adao} SE} & {\small{}(0.126)} & {\small{}(0.116)} & {\small{}(0.113)} & {\small{}(0.107)} & {\small{}(0.143)} & {\small{}(0.140)} & {\small{}(0.192)}\tabularnewline\addlinespace[0.3cm]
{\small{}Confidence interval with} & {\small{}{[}-1.059,} & {\small{}{[}-0.832,\enskip{}} & {\small{}{[}-0.568,\enskip{}} & {\small{}{[}-0.637,\enskip{}} & {\small{}{[}-0.705,} & {\small{}{[}-0.699,\enskip{}} & {\small{}{[}-1.207,\enskip{}}\tabularnewline
{\small{}\qquad{}the null imposed} & {\small{}\enskip{}-0.396{]}} & {\small{}-0.309{]}} & {\small{}\enskip{}-0.028{]}} & {\small{}\enskip{}-0.018{]}} & {\small{}-0.002{]}} & {\small{}\enskip{}0.002{]}} & {\small{}\enskip{}0.122{]}}\tabularnewline
\bottomrule
\end{tabular}{\small\par}
\par\end{center}

\begin{singlespace}
\vspace*{-0.2cm}

{\footnotesize{}\noindent Notes: This table extends the analysis
of Table 4 by reporting conventional state-clustered standard errors,
the \textcite{Adao} SIC3-clustered standard errors, and confidence
intervals based on the equivalent industry-level IV regression with
the null imposed, as discussed in Section \ref{subsec:inference}.
The specifications are the same as those in the corresponding columns
of Table 4; for comparison we repeat the coefficient estimates and
exposure-robust standard errors from that table.}{\footnotesize\par}
\end{singlespace}

\vspace*{\fill}

\newpage{}

\vspace*{\fill}
\begin{center}
Table C3: Period-Specific Effects in the \textcite{AutorDorn2001}
Setting\vspace{2.5mm}
\par\end{center}

\begin{center}
{\small{}}%
\begin{tabular}{lcccc}
\toprule 
 & {\small{}(1)} & {\small{}(2)} & {\small{}(3)} & {\small{}(4)}\tabularnewline
 & {\small{}Mfg. emp.} & {\small{}Unemp.} & {\small{}NILF} & {\small{}Wages}\tabularnewline
\midrule
\midrule 
{\small{}Coefficient (1990s)} & {\small{}-0.491} & {\small{}0.329} & {\small{}1.209} & {\small{}-0.649}\tabularnewline
 & {\small{}(0.266)} & {\small{}(0.155)} & {\small{}(0.347)} & {\small{}(0.571)}\tabularnewline
{\small{}Coefficient (2000s)} & {\small{}-0.225} & {\small{}0.014} & {\small{}-0.109} & {\small{}0.391}\tabularnewline
 & {\small{}(0.103)} & {\small{}(0.083)} & {\small{}(0.123)} & {\small{}(0.288)}\tabularnewline
\bottomrule
\end{tabular}{\small\par}
\par\end{center}

\begin{singlespace}
\vspace*{-0.2cm}

{\footnotesize{}\noindent Notes: This table reports coefficient estimates
for versions of the shift-share IV specification in column 3 of Tables
4 and C1, allowing the treatment coefficient to vary by period. This
specification uses two endogenous treatment variables (treatment interacted
with period indicators) and two corresponding shift-share instruments.
The controls are the same as in column 3 of Table 4. SIC3-clustered
exposure-robust standard errors are obtained by the equivalent shock-level
regressions and reported in parentheses.}{\footnotesize\par}
\end{singlespace}

\vspace*{\fill}

\newpage{}
\begin{center}
\vspace*{\fill}
\par\end{center}

\begin{center}
Table C4: Robustness to \textcite{AADHP2016} Controls in the \textcite{AutorDorn2001}
Setting
\par\end{center}

\begin{center}
{\small{}}%
\begin{tabular}{lcccc}
\toprule 
 & {\small{}(1)} & {\small{}(2)} & {\small{}(3)} & {\small{}(4)}\tabularnewline
\midrule
\midrule 
{\small{}Coefficient} & {\small{}-0.200} & {\small{}-0.293} & {\small{}-0.241} & {\small{}-0.232}\tabularnewline
 & {\small{}(0.093)} & {\small{}(0.125)} & {\small{}(0.115)} & {\small{}(0.122)}\tabularnewline\addlinespace[0.3cm]
{\small{}\uline{Regional controls (\mbox{$w_{\ell t}$})}} &  &  &  & \tabularnewline
{\small{}\textcite{AutorDorn2001} controls} & {\small{}$\checked$} & {\small{}$\checked$} & {\small{}$\checked$} & {\small{}$\checked$}\tabularnewline
{\small{}Period-specific lagged mfg. share} & {\small{}$\checked$} & {\small{}$\checked$} & {\small{}$\checked$} & {\small{}$\checked$}\tabularnewline
{\small{}Lagged 10-sector shares} &  & {\small{}$\checked$} & {\small{}$\checked$} & {\small{}$\checked$}\tabularnewline
{\small{}Local \textcite{AADHP2016} controls} &  & {\small{}$\checked$} &  & {\small{}$\checked$}\tabularnewline
{\small{}Local \textcite{AADHP2016} pre-trends} & {\small{}$\checked$} &  & {\small{}$\checked$} & {\small{}$\checked$}\tabularnewline\addlinespace[0.3cm]
{\small{}SSIV first stage }\emph{\small{}F}{\small{}-stat.} & {\small{}118.9} & {\small{}53.3} & {\small{}65.9} & {\small{}56.6}\tabularnewline
{\small{}\# of region-periods} & \multicolumn{4}{c}{{\small{}1,444}}\tabularnewline
{\small{}\# of industry-periods} & \multicolumn{4}{c}{{\small{}794}}\tabularnewline
\bottomrule
\end{tabular}{\small\par}
\par\end{center}

\vspace*{-0.2cm}

\begin{singlespace}
{\footnotesize{}\noindent Notes: This table extends Table 4 by adding
exposure-weighted sums of the other industry-level controls in Table
3 of \textcite{AADHP2016}. Pre-trends controls refer to the changes
in industry log average wages and in the industry share of total U.S.
employment over 1976\textendash 91; see the notes to Table 4 notes
for details on the other controls and calculation of the SIC3-clustered
exposure-robust standard errors (in parentheses) and first-stage}\emph{\footnotesize{}
F-}{\footnotesize{}statistics.}{\footnotesize\par}
\end{singlespace}

\vspace*{\fill}

\newpage{}
\begin{center}
\vspace*{\fill}
\par\end{center}

\begin{singlespace}
\begin{center}
Table C5: Overidentified Shift-Share IV Estimates of the Effect\\
of Chinese Imports on Manufacturing Employment\vspace{2.5mm}
\par\end{center}
\end{singlespace}

\begin{center}
{\small{}}%
\begin{tabular}{lccc}
\toprule 
 & {\small{}(1)} & {\small{}(2)} & {\small{}(3)}\tabularnewline
\midrule
\midrule 
{\small{}Coefficient} & {\small{}-0.238} & {\small{}-0.247} & {\small{}-0.158}\tabularnewline
 & {\small{}(0.099)} & {\small{}(0.105)} & {\small{}(0.078)}\tabularnewline\addlinespace[0.3cm]
{\small{}Shock-level estimator} & {\small{}2SLS} & {\small{}LIML} & {\small{}GMM}\tabularnewline
{\small{}Effective first stage }\emph{\small{}F}{\small{}-statistic} & \multicolumn{3}{c}{{\small{}15.10}}\tabularnewline
{\small{}$\chi^{2}(7)$ overid. test stat. {[}}\emph{\small{}p}{\small{}-value{]}} & \multicolumn{3}{c}{{\small{}10.92 {[}0.142{]}}}\tabularnewline
\bottomrule
\end{tabular}{\small\par}
\par\end{center}

\begin{singlespace}
\vspace*{-0.2cm}

{\footnotesize{}\noindent Notes: Column 1 of this table reports an
overidentified estimate of the coefficient corresponding to column
3 of Table 4, obtained from a two-stage least squares regression of
shock-level average manufacturing employment growth residuals $\bar{y}_{nt}^{\perp}$
on shock-level average Chinese import competition growth residuals
$\bar{x}_{nt}^{\perp}$, instrumenting by the growth of imports (per
U.S. worker) in each of the eight non-U.S. countries from ADH, $g_{nk}$
for $k=1,\dots,8$, controlling for period fixed effects $q_{nt}$,
and weighting by average industry exposure $s_{nt}$. Column 2 reports
the corresponding limited information maximum likelihood estimate,
while column 3 reports a two-step optimal generalized method of moments
estimate. Standard errors, the optimal weight matrix, and the \textcite{hansen82}
$\chi^{2}$ test of overidentifying restrictions all allow for clustering
of shocks at the SIC3 industry group level. The first-stage }\emph{\footnotesize{}F}{\footnotesize{}-statistic
is computed by a shift-share version of the \textcite{MontielOleaPflueger2013}
method described in Appendix \ref{subsec:appx-multiple_shocks}.}{\footnotesize\par}
\end{singlespace}

\vspace*{\fill}

\newpage{}

\vspace*{\fill}
\begin{center}
Table C6: \textcite{Bartik1991} Application
\par\end{center}

\begin{center}
{\small{}}%
\begin{tabular}{lcc}
\toprule 
 & {\small{}(1)} & {\small{}(2)}\tabularnewline
\midrule
\midrule 
{\small{}Leave-one-out estimator} & {\small{}1.277} & {\small{}1.300}\tabularnewline
 & {\small{}(0.150)} & {\small{}(0.124)}\tabularnewline
{\small{}Conventional estimator} & {\small{}1.215} & {\small{}1.286}\tabularnewline
 & {\small{}(0.139)} & {\small{}(0.121)}\tabularnewline\addlinespace[0.3cm]
{\small{}$H$ heuristic} & {\small{}1.32} & {\small{}10.50}\tabularnewline
{\small{}Population weights} & {\small{}$\checked$} & \tabularnewline
{\small{}\# of region-periods} & \multicolumn{2}{c}{{\small{}2,166}}\tabularnewline
\bottomrule
\end{tabular}{\small\par}
\par\end{center}

\begin{singlespace}
\vspace*{-0.2cm}

{\footnotesize{}\noindent Notes: Column 1 replicates column 2 of
Table 3 from \textcite{GPSS}, reporting two SSIV estimators of the
inverse labor supply elasticity, with and without the leave-one-out
adjustment. Regions are U.S. commuting zones; periods are 1980s, 1990s,
and 2000s; all specifications include controls for 1980 regional characteristics
interacted with period indicators (see \textcite{GPSS} for more details).
Standard errors allow for clustering by commuting zones. Column 1
uses 1980 population weights, while column 2 repeats the same analysis
without population weights. The table also reports the $H$ heuristic
for the importance of the leave-one-out adjustment proposed in Appendix
\ref{subsec:appx-Estimated_shocks} (equation (\ref{eq:heuristic})).}{\footnotesize\par}
\end{singlespace}

\vspace*{\fill}

\vspace*{\fill}
\begin{center}
\textbf{\newpage}
\par\end{center}

\begin{center}
\vspace*{\fill}
\par\end{center}

\begin{center}
Table C7: Simulated 5\% Rejection Rates for Shift-Share and Conventional
Shock-Level IV
\par\end{center}

\begin{center}
{\small{}}%
\begin{tabular}{lllccccc}
\toprule 
 &  &  & \multicolumn{2}{c}{{\small{}SSIV}} &  & \multicolumn{2}{c}{{\small{}Shock-level IV}}\tabularnewline
 &  &  & \multicolumn{2}{c}{{\small{}Exposure-Robust SE}} &  & \multicolumn{2}{c}{{\small{}Robust SE}}\tabularnewline
\cmidrule{4-5} \cmidrule{5-5} \cmidrule{7-8} \cmidrule{8-8} 
 &  &  & {\small{}Null not} & {\small{}Null} &  & {\small{}Null not} & {\small{}Null}\tabularnewline
 &  &  & {\small{}Imposed} & {\small{}Imposed} &  & {\small{}Imposed} & {\small{}Imposed}\tabularnewline
 &  &  & {\small{}(1)} & {\small{}(2)} &  & {\small{}(3)} & {\small{}(4)}\tabularnewline
\midrule
\midrule 
\multicolumn{8}{c}{\textbf{\small{}Panel A: Benchmark Monte-Carlo Simulation}}\tabularnewline
{\small{}(a)} & {\small{}Normal shocks} &  & {\small{}7.6\%} & {\small{}5.2\%} &  & {\small{}6.8\%} & {\small{}5.0\%}\tabularnewline
{\small{}(b)} & {\small{}Wild bootstrap (benchmark)} &  & {\small{}8.0\%} & {\small{}4.9\%} &  & {\small{}14.2\%} & {\small{}4.0\%}\tabularnewline\addlinespace[0.3cm]
\multicolumn{8}{c}{\textbf{\small{}Panel B: Higher Industry Concentration}}\tabularnewline
{\small{}(c)} & {\small{}$1/HHI=50$} &  & {\small{}5.6\%} & {\small{}4.9\%} &  & {\small{}8.4\%} & {\small{}6.1\%}\tabularnewline
{\small{}(d)} & {\small{}$1/HHI=20$} &  & {\small{}7.3\%} & {\small{}5.5\%} &  & {\small{}7.0\%} & {\small{}10.7\%}\tabularnewline
{\small{}(e)} & {\small{}$1/HHI=10$} &  & {\small{}9.0\%} & {\small{}8.2\%} &  & {\small{}14.8\%} & {\small{}23.8\%}\tabularnewline\addlinespace[0.3cm]
\multicolumn{8}{c}{\textbf{\small{}Panel C: Smaller Numbers of Industries or Regions}}\tabularnewline
{\small{}(f)} & {\small{}$N=136$ (SIC3 industries)} &  & {\small{}5.4\%} & {\small{}4.5\%} &  & {\small{}7.7\%} & {\small{}4.3\%}\tabularnewline
{\small{}(g)} & {\small{}$N=20$ (SIC2 industries)} &  & {\small{}7.7\%} & {\small{}3.7\%} &  & {\small{}7.9\%} & {\small{}3.2\%}\tabularnewline
{\small{}(h)} & {\small{}$L=100$ (random regions)} &  & {\small{}9.7\%} & {\small{}4.5\%} &  & \multicolumn{2}{c}{{\small{}N/A}}\tabularnewline
{\small{}(i)} & {\small{}$L=25$ (random regions)} &  & {\small{}10.4\%} & {\small{}4.3\%} &  & \multicolumn{2}{c}{{\small{}N/A}}\tabularnewline
\bottomrule
\end{tabular}{\small\par}
\par\end{center}

\begin{singlespace}
\vspace*{-0.2cm}

{\footnotesize{}\noindent Notes: This table summarizes the results
of the Monte-Carlo analysis described in Appendix \ref{subsec:Finite-Sample-Performance},
reporting the rejection rates for a nominal 5\% level test of the
true null that $\beta^{*}=0$. In all panels, columns 1 and 2 are
simulated from the SSIV design based on \textcite{AutorDorn2001},
as in column 3 of Table 4, while columns 3 and 4 are based on the
conventional industry-level IV in \textcite{AADHP2016}. Column 1
uses exposure-robust standard errors from the equivalent industry-level
IV and column 2 implements the version with the null hypothesis imposed.
Columns 3 and 4 parallel columns 1 and 2 when applied to conventional
IV. In Panel A, the simulations approximate the data-generating process
using a normal distribution in row (a), with the variance matched
to the sample variance of the shocks in the data after de-meaning
by year, while wild bootstrap is used in row (b), following \textcite{Liu1988}.
Panel B documents the role of the Herfindahl concentration index across
industries, varying 1/HHI from 50 to 10 in rows (c) to (e), compared
with 191.6 for shift-share IV and 189.7 for conventional IV. Panel
C documents the role of the number of regions and industries. We aggregate
industries from 397 four-digit manufacturing SIC industries into 136
three-digit industries in row (f) and further into 20 two-digit industries
in row (g). In rows (h) and (i), we select a random subset of region
in each simulation. See Appendix \ref{subsec:Finite-Sample-Performance}
for a complete discussion.}{\footnotesize\par}
\end{singlespace}

\vspace*{\fill}
\begin{center}
\textbf{\newpage}\vspace*{\fill}
\par\end{center}

\begin{center}
Table C8: First Stage \emph{F}-statistics as a Rule of Thumb: Monte-Carlo
Evidence
\par\end{center}

\begin{center}
{\small{}}%
\begin{tabular}{lccccc}
\toprule 
 & \multicolumn{5}{c}{{\small{}Number of Instruments}}\tabularnewline
\cmidrule{2-6} \cmidrule{3-6} \cmidrule{4-6} \cmidrule{5-6} \cmidrule{6-6} 
 & {\small{}1} & {\small{}5} & {\small{}10} & {\small{}25} & {\small{}50}\tabularnewline
 & {\small{}(1)} & {\small{}(2)} & (3) & {\small{}(4)} & {\small{}(5)}\tabularnewline
\midrule 
\multicolumn{6}{c}{\textbf{\small{}Panel A: SSIV}}\tabularnewline
{\small{}5\% rejection rate} & {\small{}8.0\%} & {\small{}8.9\%} & {\small{}11.5\%} & {\small{}15.0\%} & {\small{}23.0\%}\tabularnewline
{\small{}Median bias, \% of std. dev.} & {\small{}0.3\%} & {\small{}14.6\%} & {\small{}28.3\%} & {\small{}43.2\%} & {\small{}72.1\%}\tabularnewline
{\small{}Median first-stage }\emph{\small{}F} & {\small{}54.3} & {\small{}14.8} & {\small{}9.1} & {\small{}6.4} & {\small{}7.7}\tabularnewline\addlinespace[0.3cm]
\multicolumn{6}{c}{\textbf{\small{}Panel B: Conventional Shock-Level IV}}\tabularnewline
{\small{}5\% rejection rate} & {\small{}13.6\%} & {\small{}13.9\%} & {\small{}14.9\%} & {\small{}17.7\%} & {\small{}22.0\%}\tabularnewline
{\small{}Median bias, \% of std. dev.} & {\small{}-0.3\%} & {\small{}10.1\%} & {\small{}27.1\%} & {\small{}57.0\%} & {\small{}80.2\%}\tabularnewline
{\small{}Median first-stage }\emph{\small{}F} & {\small{}59.4} & {\small{}19.4} & {\small{}13.2} & {\small{}10.0} & {\small{}11.2}\tabularnewline\addlinespace[0.3cm]
{\small{}Number of simulations} & {\small{}10,000} & {\small{}3,000} & {\small{}1,500} & {\small{}500} & {\small{}300}\tabularnewline
\bottomrule
\end{tabular}{\small\par}
\par\end{center}

\begin{singlespace}
\vspace*{-0.2cm}

{\footnotesize{}\noindent Notes: This table reports the results of
the Monte-Carlo analysis with many weak instruments, described in
Appendix \ref{subsec:Finite-Sample-Performance}. Panel A is simulated
from the SSIV design based on \textcite{AutorDorn2001}, as in column
3 of Table 4, while Panel B is based on the conventional industry-level
IV in \textcite{AADHP2016}. The five columns increase the number
of shocks $J=1,5,10,25,$ and $50$, with only one shock relevant
to treatment. The table reports the rejection rates corresponding
to a nominal 5\% level test of the true null that $\beta^{*}=0$,
the median bias of the estimator as a percentage of the simulated
standard deviation, and the median first-stage }\emph{\footnotesize{}F}{\footnotesize{}-statistic
obtained via the \textcite{MontielOleaPflueger2013} method (extended
to shift-share IV in Panel A, following Appendix \ref{subsec:appx-multiple_shocks}).
See Appendix \ref{subsec:Finite-Sample-Performance} for a complete
discussion.}{\footnotesize\par}
\end{singlespace}

\vspace*{\fill}

\newpage{}

\begin{singlespace}
\printbibliography
\end{singlespace}

\end{document}